\documentclass[final,3p,times]{elsarticle}

\usepackage{amssymb}
\usepackage{graphicx,multirow,caption,subcaption,multicol,setspace,amsmath,wrapfig}
\usepackage{commath}
\usepackage[linesnumbered,ruled]{algorithm2e}
\usepackage{algorithmic}
\usepackage{adjustbox}
\usepackage{booktabs,caption}
\usepackage[flushleft]{threeparttable}
\usepackage{lineno}
\usepackage{enumitem}
\setlist{nosep}
\usepackage{rotating,caption}
\usepackage[table]{xcolor}
\usepackage{soul,xcolor}
\usepackage{amsfonts}
\usepackage{makecell,eqnarray}

\usepackage{tikz}
\usetikzlibrary{shapes,arrows,decorations.markings}
\usepackage[EULERGREEK]{sansmath}
\usepackage{floatrow}
\DeclareFloatFont{henry}{\fontfamily{ccr}\fontseries{m}}
\usepackage{float}
\floatstyle{plaintop}
\restylefloat{table}

\biboptions{numbers,sort&compress}
\newcommand{\cmark}{\ding{51}}%
\newcommand{\xmark}{\ding{55}}%
\graphicspath{ {Images/} }

\makeatletter
\newcommand{\algorithmfootnote}[2][\footnotesize]{%
  \let\old@algocf@finish\@algocf@finish
  \def\@algocf@finish{\old@algocf@finish
    \leavevmode\rlap{\begin{minipage}{\linewidth}
    #1#2
    \end{minipage}}%
  }%
}

\journal{Neurocomputing}

\begin{document}
\begin{frontmatter}

\title{Multi-Objective Evolutionary Framework for Non-linear System Identification: A Comprehensive Investigation}

\author[label5]{Faizal Hafiz\corref{cor1}}
\address[label5]{Department of Electrical \& Computer Engineering, The University of Auckland, Auckland, New Zealand}
\ead{faizalhafiz@ieee.org}
\cortext[cor1]{Corresponding author}

\author[label5]{Akshya Swain}

\author[label1]{Eduardo Mendes}
\address[label1]{Department of Electronics Engineering, Federal University of Minas Gerais, Belo Horizonte, Brazil}

\begin{abstract}

The present study proposes a multi-objective framework for structure selection of nonlinear systems which are represented by polynomial NARX models. This framework integrates the key components of Multi-Criteria Decision Making (MCDM) which include preference handling, Multi-Objective Evolutionary Algorithms (MOEAs) and \textit{a posteriori} selection. To this end, three well-known MOEAs such as NSGA-II, SPEA-II and MOEA/D are thoroughly investigated to determine if there exists any significant difference in their search performance. The sensitivity of all these MOEAs to various \textit{qualitative} and \textit{quantitative} parameters, such as the choice of recombination mechanism, crossover and mutation probabilities, is also studied. These issues are critically analyzed considering seven discrete-time and a continuous-time benchmark nonlinear system as well as a practical case study of non-linear wave-force modeling. The results of this investigation demonstrate that MOEAs can be tailored to determine the correct structure of nonlinear systems. Further, it has been established through frequency domain analysis that it is possible to identify multiple valid discrete-time models for continuous-time systems. A rigorous statistical analysis of MOEAs via performance \textit{sweet spots} in the parameter space convincingly demonstrates that these algorithms are robust over a wide range of control parameters. 

\end{abstract}


\end{frontmatter}

\section{Introduction}
\label{s:Ch7Intro}

The system identification aims to encapsulate dynamical properties of the system under scrutiny in a mathematical \textit{descriptor/model}. Such identified models are not only critical to understand the system dynamics but also to the subsequent prediction and control applications. Given that the dynamic input-output behavior of most of the systems is predominantly nonlinear in nature, the identification of nonlinear systems has attracted significant attention over the past few decades~\cite{Soderstrom:Stoica:1989,Ljung:1999,ONelles:2001,Billings:2013,Haber:Unbehauen:1990,sjoberg:Zhang:1995,Hong:Mitchell:2008,Kar:Swain:2009}. The first step of the identification process is the selection of appropriate system representation, \textit{e.g.}, Volterra series, Wiener, Hammerstein, Neural Networks, polynomial Nonlinear Auto-Regressive Moving Average with eXogenous inputs (NARMAX) and others. Among these representations, the NARX models~\cite{Leontaritis:Billings:1985,LEONTARITIS:BILLINGS:1985b,Billings:2013,Chen:Billings:1989} are arguably the most popular due to various advantages, \textit{e.g.},  convenient \textit{linear-in-parameter} form and the availability of frequency domain analysis tools~\cite{Billings:Tsang:1989,Billings:2013}. The identification of non-linear systems in the form of a polynomial NARX models is, therefore, considered in this investigation.

The system identification is essentially a data-driven modelling process which begins by defining a search space of candidate \textit{terms/basis functions}. The subsequent model building procedure involves mainly two steps: 1) \textit{Structure Selection}: selection of significant \textit{terms/basis functions} and 2) \textit{Parameter Estimation}: estimation of associated \textit{parameters}. Due to \textit{linear-in-parameter} form of the NARX model, the parameters can be estimated using the least-squares based approaches, provided the terms to be included in the model are accurately determined. However, the selection of correct terms which captures the system dynamics is one of the major challenges of the system identification~\cite{sjoberg:Zhang:1995,Hong:Mitchell:2008}, and, therefore, is the main focus of this investigation. 

The exhaustive search to solve the structure selection problem is intractable even for moderate number of terms `$n$', as this would require an evaluation of $2^n$ \textit{term subsets/structures}. The structure selection problem is often `NP-Hard'~\cite{Mendes:1995} and therefore requires an efficient search strategy. Since the search for system structure is primarily dependent on the limited number of observed data, it involves \textit{bias-variance trade-off}~\cite{Ljung:1999,ONelles:2001,Hong:Mitchell:2008}, \textit{i.e.}, too sparse a structure may yield a high \textit{bias} error (\textit{under-fitting}) whereas a very complex structure may yield a high \textit{variance} error (\textit{over-fitting}). Further, earlier investigations~\cite{Aguirre:Billings:1995,AGUIRRE:MENDES:1996} show that the \textit{over-fitted} structures tend to induce `\textit{ghost}' dynamics which are not present in the actual system. Pursuing these arguments, it is clear that determination of the \textit{size} of the identified structure (\textit{i.e.}, \textit{number of terms} or `\textit{cardinality}') is a critical aspect of the structure selection process. However, this issue has received relatively less attention in the existing investigations.

Among the existing structure selection approaches, the Orthogonal Forward Regression (OFR) proposed by Billings and Korenberg~\cite{Korenberg:Billings:1988} has extensively been studied. It is a \textit{sequential} model building approach in which the most significant term is added in each step, which is determined on the basis of a statistical criterion. Over the years, this notion of the original OFR approach has been refined to further improve the search performance. These include OFR approaches with either new term evaluation criteria~\cite{Piroddi:Spinelli:2003,Wei:Billings:2006,Billings:Wei:2007,Guo:Guo:2016} or improved search strategies~\cite{Mao:Billings:1997,Li:Peng:2006,Guo:Billings:2015,Tang:Zhang:2018,Hafiz:Swain:Floating:2019}. A detailed treatment on this subject can be found in the recent investigation by the authors~\cite{Hafiz:Swain:Floating:2019}. It is worth noting that, while OFR and its extensions are quite effective in detecting the \textit{significant} terms, these approaches typically require an auxiliary procedure to estimate the \textit{number of terms} (\textit{cardinality}). Usually, an information theoretic criterion is used for this purpose, \textit{e.g.}, see~\cite{Billings:Wei:2008}. However, it is known that the core assumptions of the information criteria may not always hold. In such scenarios, the information criteria usually tend to \textit{over-fit} as suggested by the earlier investigation of Nakamura \textit{et al.}~\cite{Nakamura:Judd:2006}. Hence, the estimation of \textit{cardinality} still requires further investigations in OFR based approaches.

The other interesting approach to structure selection is based on stochastic sampling of the search space, \textit{e.g.}, Evolutionary Algorithms~\cite{Fonseca:Mendes:1993,Ho:French:1998,Rodriguez:Fonseca:1997,Madar:Abonyi:2005,Falsone:Piroddi:2015,Bianchi:Piroddi:2017,Avellina:Piroddi:2017} and Bayesian inference~\cite{Baldacchino:Kadirkamanathan:2012,Baldacchino:Kadirkamanathan:2013}.
In the earlier investigations, Genetic Algorithm (GA)~\cite{Fonseca:Mendes:1993,Ho:French:1998} and Genetic Programming (GP)~\cite{Rodriguez:Fonseca:1997,Madar:Abonyi:2005} have been proposed as an alternative to OFR. In these investigations, the structure selection has been posed as a single objective problem, where the search objective is chosen to be an aggregated function of predictive performance and number of terms to balance the bias-variance dilemma. Recently, a randomized structure selection approach is proposed in~\cite{Falsone:Piroddi:2015,Bianchi:Piroddi:2017,Avellina:Piroddi:2017}. In this approach, each candidate term is assigned with selection probability which is updated following a \textit{positive-feedback} mechanism similar to ant colony optimization~\cite{Dorigo:Maniezzo:1996}. In~\cite{Baldacchino:Kadirkamanathan:2013}, an alternative approach based on Bayesian framework is proposed wherein the posterior probabilities of both the terms and parameters are determined through Markov Chain Monte Carlo sampling.

Although the existing approaches have been shown to be effective, it is clear that their primary search focus is to determine the significance of the terms. The pursuit for parsimonious structure with a better predictive performance is essentially contradictory in nature. Therefore, the most challenging task of the structure selection is, arguably, to obtain a good balance of bias-variance dilemma. It is known that a unique solution to such problems may not exist. For instance, a discrete NARX model of a continuous-time system may not be unique~\cite{Billings:2013}. Pursuing these arguments, it is clear that the structure selection is, in essence, a \textit{multi-objective} problem. Note that over the past two decades, several Multi-Objective Evolutionary Algorithms (MOEAs) have specifically been developed to address such problems, \textit{e.g.}, see~\cite{Deb:Pratap:2002,Zitzler:Laumanns:2001,Zhang:Li:2007}. Hence, in this scenario, the following question arises: \textit{can the existing MOEAs be used to reach a better trade-off in bias-variance dilemma?}

The present study aims to answer this pertinent question. In particular, this study proposes a Multi-Objective Structure Selection (MOSS) framework which integrates all the key components involved in the Multi-Criterion Decision Making (MCDM) process to address both aspects of structure selection, \textit{i.e.}, \textit{term significance} and \textit{cardinality estimation}. 

It is worth noting that formulation of the structure selection problem as a multi-objective problem is not new, but it is under-explored. The research in this direction was pioneered in the early nineties by Fonseca et al.~\cite{Fonseca:Mendes:1993,Fonseca:Fleming:1996,Rodriguez:Fonseca:Fleming:1997,Rodriguez:Fleming:1998}. In~\cite{Fonseca:Mendes:1993,Fonseca:Fleming:1996}, the structure cardinality is pre-specified and the structure is identified to balance multiple prediction criteria. Subsequently, a multi-objective GP is developed for the structure selection on the basis of classical Multi-objective GA (MOGA)~\cite{Fonseca:Fleming:1993} which is extensively investigated in~\cite{Rodriguez:Fonseca:Fleming:1997,Rodriguez:Fleming:1998,Rodriguez:Fonseca:2004,Rodriguez:Fleming:2005}. Since then, several seminal MOEAs have been developed with significant improvement in the search performance (\textit{e.g.}, see~\cite{Deb:Pratap:2002,Zitzler:Laumanns:2001,Zhang:Li:2007}). However, in recent years, the application of MOEAs in structure selection has received relatively less attention except for a few preliminary investigations in~\cite{Lavinia:Patelli:2009a,Lavinia:Patelli:2009b,Zakaria:Jamaluddin:2012}. While the existing investigations have demonstrated the capabilities of MOEAs to address the structure selection problem and provided \textit{proof-of-concept}, several challenging issues are yet to be investigated. The focus of this study is, therefore, to address some of these issues,
\begin{itemize}
    \item Investigate the performance of well-known MOEAs such as NSGA-II, SPEA-II and MOEA/D, and to determine whether there exists any significant difference through rigorous statistical comparison.
    
    \item Determine the sensitivity of MOEAs to various \textit{qualitative} and \textit{quantitative} control parameters and determine the robustness through \textit{performance sweet spots}~\cite{Goldberg:1999,Purshouse:Fleming:2007} in the parameter space.

    \item Evaluate the \textit{a posteriori} MCDM techniques through Generalized Frequency Response Functions (GFRFs) in frequency domain validation. 
\end{itemize}
For this purpose, the performance of MOEAs is critically analyzed from the identification perspective considering seven discrete-time and a continuous-time benchmark system as well as a practical identification problem in structural hydro-dynamics.

It is worth noting that this study is a part of our broader investigation on alternative approaches for structure selection. In the first part of this investigation~\cite{Hafiz:Swain:Floating:2019}, the orthogonal floating search algorithms have been proposed to alleviate the \textit{nesting} effects of the classical OFR. Subsequently, a new two-dimensional structure selection approach was developed in~\cite{Hafiz:Swain:CEC:2018,Hafiz:Swain:2020a}, which explicitly integrates the information about subset cardinality (number of terms) into a probabilistic learning framework to balance the bias-variance dilemma. These approaches have been shown to be very effective on both simulated and practical nonlinear systems. These algorithms, however, have been designed to identify a unique structure which is optimum for certain criterion. The multi-objective framework (MOSS) proposed in this study can alleviate this issue. MOSS can identify multiple valid models for the given systems and thereby yield a better trade-off for the bias-variance dilemma. This, not only aids the decision making process but also instills confidence in the identified models. 

The rest of the article is organized as follows: the polynomial NARX model and the concept of Pareto dominance are discussed briefly in Section~\ref{s:Prelim}. This is followed by Multi-objective Structure Selection (MOSS) framework in Section~\ref{s:MOSS}, which describes, in detail, all the components involved in the multi-criteria decision making (MCDM) process. The framework of this investigation is introduced in Section~\ref{s:IF} which provides the detail about the benchmark nonlinear test systems, search environment and the performance metrics being used in the comparative evaluation. The results of this investigation are discussed in three stages. First, the MOEAs are critically analyzed from the identification perspective on the benchmark systems in Section~\ref{s:resIDP}. Next, the search performance of MOEAs is compared through rigorous statistical analysis in Section~\ref{s:CompMOEA}. Finally, the robustness of MOEAs to the search parameters is analyzed at length in Section~\ref{s:resRobust}, followed by the conclusions in Section~\ref{s:conclusion}.

\section{Preliminaries}
\label{s:Prelim}

\subsection{The Polynomial NARX Model}
\label{s:NARX}
The NARX model represents a non-linear system as a function of recursive lagged input and output terms as follows:
\begin{linenomath*}
\begin{align*}
y(k) & = F^{n_l} \ \{ \ y(k-1),\ldots,y(k-n_y),u(k-1),\ldots, u(k-n_u) \ \}+e(k)
\end{align*}
\end{linenomath*}
where $y(k)$ and $u(k)$ respectively represent the output and input at time intervals $k$, $n_y$ and $n_u$ are corresponding lags and $F^{n_l}\{ \cdotp \}$ is some nonlinear function of degree $n_l$. 

The \textit{total number of possible terms} or \textit{model size} ($n$) of the NARX model is given by,
\begin{linenomath*}
\begin{align}
\label{eq:n}
n & = \sum_{i=0}^{n_l} n_i, \ n_0=1 \ \textit{and \ } n_i = \frac{n_{i-1}(n_y+n_u+i-1)}{i}, \ i=1,\ldots, n_l 
\end{align}
\end{linenomath*}
This model is essentially linear-in-parameters and can be expressed as:
\begin{linenomath*}
\begin{align}
\label{eq:NARXmodel}
    y(k) & = \sum_{i=1}^{n} \theta_i x_i(k) + e(k), \ \ k = 1,2,\dots \mathcal{N}\\
    \text{where, } x_1(k) & = 1, \ \ \text{and \ } x_i(k)  = \prod_{j=1}^{p_y}y(k-n_{y_j})\prod_{k=1}^{q_u}u(k-n_{u_k}) \, \ i=2,\ldots, n, \nonumber
\end{align}
\end{linenomath*}
$p_y,q_u \geq 0$; $1\leq p_y+q_u \leq n_l$; $1 \leq n_{y_j}\leq n_y$;$1 \leq n_{u_k}\leq n_u$; $n_l$ is the degree of polynomial expansion; `$\mathcal{N}$' denotes the total number of data points.

\subsection{Pareto Dominance}
\label{s:PD}
For multi-objective problems, the ideal solution which would yield optimum values across all the objectives may not exist; rather there may exist a set of \textit{non-dominated} or \textit{Pareto Optimal} solutions~\cite{Deb:Book:2001}. Such non-dominated solutions may not necessarily be optimum for each objective, however, they are better than the other solution when all the objectives are simultaneously considered. 

To understand the concept of the Pareto dominance, consider a problem with `$n$' decision variables and having a total of `$m$' number of objectives. Let $\mathcal{X}_1$ and $\mathcal{X}_2$ denote the candidate solutions of this problem. Further, denote the corresponding objective function vectors by $\vec{J}_1$ and $\vec{J_2}$, as follows:
\begin{align*}
    \vec{J}_1 = \{ J_{1}(\mathcal{X}_1), \ J_{2}(\mathcal{X}_1), \dots J_{m}(\mathcal{X}_1)\} \qquad \vec{J}_2 = \{ J_{1}(\mathcal{X}_2), \ J_{2}(\mathcal{X}_2), \dots J_{m}(\mathcal{X}_2)\}
\end{align*}
The Pareto dominance of these solutions can be determined on the basis of the objective values as follows: $\mathcal{X}_1$ dominates $\mathcal{X}_2$ (denoted as $\mathcal{X}_1 \preceq \mathcal{X}_2$),
\begin{align}
  if{f} \ \ &  \forall{p} \in \{1 \dots m\} : \, J_{p}(\mathcal{X}_1) \leq J_{p}(\mathcal{X}_2) \ \wedge \: \exists{p} \in \{1 \dots m\} : J_{p}(\mathcal{X}_1) < J_{p}(\mathcal{X}_2)
\end{align}
In this study, the set of such non-dominated solutions is denoted by `$\Gamma$' and the corresponding set of objective function vectors by `$\Lambda$'.

\begin{figure}[!t]
\centering
\begin{adjustbox}{max width=0.6\textwidth}
\tikzstyle{block} = [rectangle, draw, fill=white, 
    text width=7em, text centered, rounded corners, minimum height=4em,line width=0.04cm]
\tikzstyle{line} = [draw, -latex']
\tikzstyle{cloud} = [draw, ellipse,fill=white, node distance=6cm,
    minimum height=3em, line width=0.04cm]

\begin{tikzpicture}[node distance = 4cm, auto, scale=0.1]
    \centering
    \node [cloud] (data) {Identification Data};
    \node [cloud, right of=data] (DM) {DM's Preferences};
    \node [block, below of=data] (search){MOEA + Goals};
    \node[block, left of=search](alg){NSGA-II, SPEA-II, MOEA/D};
    \node[block, below of=DM](prefArt){Preference Articulation};
    \node[block,below of=search](MCDM){\textit{a posteriori} MCDM};
    \node[block, left of=MCDM](mcdmApp){MMD, MTD};
    \node[block, below of=MCDM](final){System Structure};
    \path [line,ultra thick] (data) -- (search);
    \path[line,ultra thick] (DM)--(prefArt);
    \path [line, ultra thick] (prefArt) -- node [align=center, above]{\textit{goals}} node [below]{(\textit{a priori})}(search); %
    \path [line, ultra thick] (alg) -- (search);
    \path [line, ultra thick] (search) -- node [align=center, near start]{APS} (MCDM);
    \path [line, ultra thick] (prefArt) |-node [near end, above]{\textit{weights}} node [near end, below]{(\textit{a posteriori})} (MCDM); 
    \path [line, ultra thick] (mcdmApp) -- (MCDM);
    \path [line, line width=0.05cm] (MCDM) -- (final);
\end{tikzpicture}  
\end{adjustbox}
\caption{Multi-Objective Structure Selection (MOSS) framework. Approximate Pareto Set (APS) denotes the set of identified non-dominated structures. Minimum Manhattan Distance (MMD) and Multi-criteria Tournament Decision (MTD) denote the \textit{a posteriori} decision-making process.}
\label{f:MOSSFrame}
\end{figure}
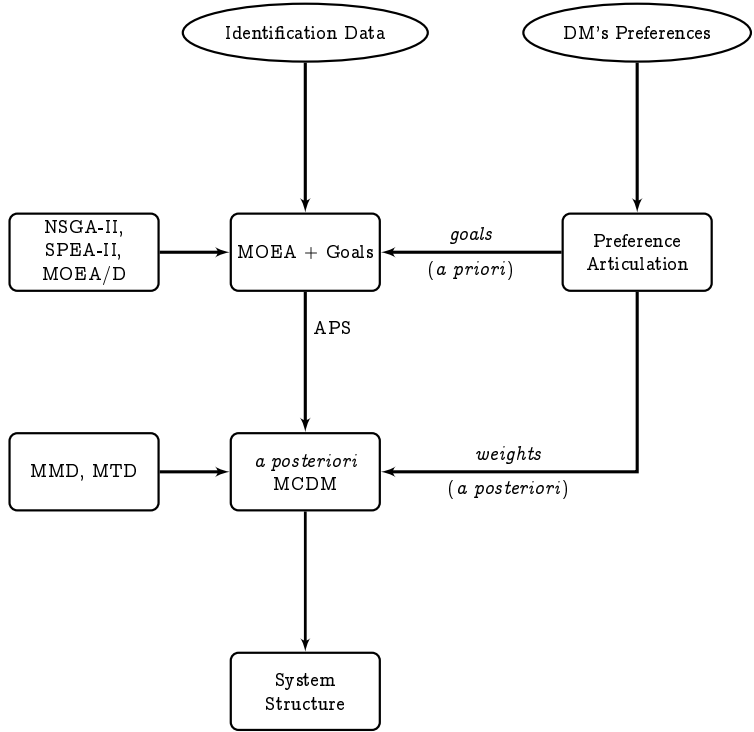
\section{Multi-objective Structure Selection Framework}
\label{s:MOSS}

In this study, the structure selection problem is posed as a Multi-Criteria Decision Making (MCDM) problem to balance the \textit{bias-variance} dilemma. The objective here is to minimize both the \textit{predictive error} as well as the \textit{number of terms} (\textit{cardinality}). Without loss of generality, this study takes an \textit{a posteriori} approach to the multi-criteria decision making, where a set of non-dominated structures are first identified by an MOEA. Next, an \textit{a posteriori} decision method is used to select the final structure from the pool of identified structures. Such approach decouples the search and decision making process. Since the search process is unbiased to any of the objective, it is possible to accommodate distinct preferences \textit{a posteriori} without the need to reiterate the search process.

To this end, a Multi-objective Structure Selection (MOSS) framework is proposed which consists mainly of three procedures: 1) Preference Articulation 2) Search for Non-dominated Structures and 3) \textit{a posteriori} MCDM. 

The search objectives of the proposed MOSS are discussed in Section~\ref{s:MOCrF}. To solve this problem, three seminal Multi-objective Evolutionary Algorithms (MOEAs) are being considered: NSGA-II~\cite{Deb:Pratap:2002}, SPEA-II~\cite{Zitzler:Laumanns:2001} and MOEA/D~\cite{Zhang:Li:2007}. These are briefly discussed in Section~\ref{s:MOEAs}. In all the MOEAs, the binary string representation is used to encode system structures, as discussed in Section~\ref{s:SolRep}. Further, in system identification, the Decision Maker (DM) is often interested in a particular region of the Pareto front. This information is incorporated to focus the search in the region of interest by introducing \textit{a priori goal} point, which is discussed in Section~\ref{s:PrefArt}. The search process yields a set of non-dominated structures, which is referred to as Approximate Pareto Set (APS). For the final selection, the identified non-dominated structures in APS are \textit{ranked} using either Minimum Manhattan Distance (MMD) or Multi-criteria Tournament Decision (MTD) approach, which is discussed at length in Section~\ref{s:MCDM}.

All the steps involved in the proposed MOSS are outlined in Fig.~\ref{f:MOSSFrame}. In the following, these steps are discussed in detail from the identification perspective.

\subsection{Multi-objective Structure Selection Problem}
\label{s:MOCrF}

To understand the structure selection problem, consider the identification of a system from a total `$\mathcal{N}$' number of input-output data-points, \textit{i.e.}, $\{ u_i, y_i \}, i= 1,2, \dots, \mathcal{N}$. Denote the given super-set of NARX terms by `$\mathcal{X}_{model}$', as follows:
\begin{linenomath*}
\begin{align}
    \label{eq:Xmodel}
    \mathcal{X}_{model}=\begin{Bmatrix} x_1 & x_2 & \dots & x_{n} \end{Bmatrix}
\end{align}
\end{linenomath*}
where, $x_i \in \mathcal{X}_{model}$ could represent any linear or non-linear term of the NARX model in (\ref{eq:NARXmodel}); `$n$' denotes the total number of NARX terms which is expressed in ~(\ref{eq:n}). It is dependent on the maximum lag of inputs and outputs as well as on the degree of non-linearity.

For the given identification dataset and a model-set $\mathcal{X}_{model}$, the goal of the structure selection is to identify the subset of \textit{system/significant} terms, $\mathcal{X}^{\star} \subset \mathcal{X}_{model}$. This can be formulated as an optimization problem as:
\begin{linenomath*}
\begin{align}
    \label{eq:soss}
    \mathcal{X}^{\star} & = \Big\{ \mathcal{X} \subset \mathcal{X}_{model} \ | \    J(\mathcal{X}) = \min \limits_{\forall{\mathcal{X}_i} \subset \mathcal{X}_{model}} J(\mathcal{X}_i) \Big\}
\end{align}
\end{linenomath*}
where, `$J(\cdot)$' denotes a suitable criterion function. Most of the existing structure selection approaches often use the predictive performance of the model as the criterion function~\cite{Fonseca:Mendes:1993,Ho:French:1998,Rodriguez:Fonseca:1997,Madar:Abonyi:2005,Falsone:Piroddi:2015,Bianchi:Piroddi:2017,Avellina:Piroddi:2017}. These can, therefore, be categorized as the \textit{single objective} approach to structure selection. In contrast, in this study, the structure selection is formulated as a \textit{multi-objective} optimization problem with the following search objectives: 
\begin{linenomath*}
\begin{align}
    \label{eq:moss}
    \vec{J}(\mathcal{X}_i) & = \min \begin{cases} J_1(\mathcal{X}_i): \xi_i\\
                                                J_2(\mathcal{X}_i): \mathcal{E}_i
                            \end{cases}
\end{align}
\end{linenomath*}
\begin{linenomath*}
\begin{align}
    \label{eq:NMSE}
    \text{where, } \xi_i & = |\mathcal{X}_i|, \quad \text{and, \ } \mathcal{E}_i = \sqrt{\frac{\sum_{k=1}^{\mathcal{N}_v} \big[ y(k) - \hat{y}(k) \big]^2}{\sum_{k=1}^{\mathcal{N}_v} \big[ y(k) - \Bar{y} \big]^2} } \times 100
\end{align}
\end{linenomath*}
In (\ref{eq:NMSE}), `$\mathcal{X}_i$' denotes the $i^{th}$ structure under consideration; `$\xi_{i}$' denotes the cardinality (\textit{number of terms}); `$\mathcal{E}_i$' denotes the Normalized Mean Squared Error (NMSE) obtained over the validation data; `$\hat{y}$' denotes the model predicted output; and `$\mathcal{N}_v$' denotes the length of the validation data. 

It is easier to follow that the search objectives, $\{ \xi, \ \mathcal{E}\}$, have been selected to identify parsimonious models which would give better predictive performance. For such multi-objective formulation, there may exist several non-dominated structures instead of single optimum structure ($\mathcal{X}^{\star}$) (as discussed in Section~\ref{s:PD}). Hence, the goal of the multi-objective structure selection is to identify the set of non-dominated structures, $\Gamma^\star$, which are Pareto optimal for the criteria function, $\vec{J}(\cdot)$ in~(\ref{eq:moss}):  
\begin{linenomath*}
\begin{align}
    \label{eq:}
    \Gamma^\star = \Big\{ \mathcal{X} \subset \mathcal{X}_{model} \ | \ \nexists{\mathcal{X}_j} \subset \mathcal{X}_{model} : \mathcal{X}_j \preceq \mathcal{X}  \Big\}
\end{align}
\end{linenomath*}

It is worth noting that the formulation of the structure selection as a multi-objective optimization problem is not new. The pioneer investigation in this direction has been carried out by Fonseca et al. in~\cite{Fonseca:Mendes:1993,Fonseca:Fleming:1996,Rodriguez:Fleming:1998,Rodriguez:Fonseca:2004}. In most of these investigations, the structure selection problem is essentially formulated as a many-objective problem with $5$ or more search objectives, \textit{i.e.}, $m\geq 5$. For instance, in~\cite{Rodriguez:Fonseca:2004}, to reduce the model complexity, maximum lags and degree of nonlinearity are also included as the search objectives. Further, linear and higher order cross-correlation functions between input and error residuals are also framed as the search objectives to embed the model validity tests~\cite{Billings:Voon:1986} into the search process. While the inclusion of these additional objectives can narrow the feasible search space, in this study, we have limited to number of search objectives to two (\textit{i.e.}, $m=2$) for the following reasons:
\begin{itemize}
    \item The identification of the Pareto optimal structures is essentially the penultimate step. In the last step, the decision maker (DM) has to select the final structure from the pool of non-dominated structures which is often referred to as \textit{a posteriori} multi-criteria decision making (MCDM). By limiting the search objectives to $\{ \xi, \ \mathcal{E}\}$, \textit{a posteriori} MCDM can be reduced to a model order selection problem. 
    \item The dominance relations tend to lose the comparative power for a higher number of objectives. This could deteriorate the search performance of the dominance based MOEAs such as NSGA-II and SPEA-II (which are being used in this study)~\cite{Deb:Book:2001,Purshouse:Fleming:2007}. 
    \item The complexity of preference articulation is directly dependent on the number of objectives. For instance, the specification of multiplicative preference relations to select the final structure becomes increasingly difficult as the number of objectives increase. 
\end{itemize}

\subsection{Solution Representation}
\label{s:SolRep}

Let the set containing all model terms be denoted as $\mathcal{X}_{model}=\begin{Bmatrix} x_1 & x_2 & \dots & x_{n} \end{Bmatrix}$. For this problem, each member of an MOEA ($\beta$) encodes a \textit{candidate structure} or \textit{set of terms} in an $n$ - dimensional binary vector. For example, the $i^{th}$ member, $\beta_i$, is represented as:
\begin{linenomath*}
\begin{align}
\label{eq:binaryrepresentation}
    \beta_i & = \begin{Bmatrix} \beta_{i,1} & \beta_{i,2} & \dots & \beta_{i,n} \end{Bmatrix}, \ \ \beta_{i,m} \in \{0,1\}, \ \ m=1,2,\dots n
\end{align}
\end{linenomath*}
The $m^{th}$ term ($x_m$) from $\mathcal{X}_{model}$ is included into the candidate structure provided the corresponding bit in the member, `$\beta_{i,m}$' is set to `$1$'.

For sake of clarity, consider a simple NARX model with a total of $5$ terms ($n=5$) as follows:
\begin{linenomath*}
\begin{align}
\label{eq:solRep1}
	\mathcal{X}_{model} & = \begin{Bmatrix} x_1 & x_2 & x_3 & x_4 & x_5 \end{Bmatrix} \\
			  & =\begin{Bmatrix} y(k-1) & u(k-2) & y(k-3) & y(k-2)u(k-2) & u(k-3)^3 \end{Bmatrix} \nonumber	  
\end{align}
\end{linenomath*}
For this problem, assume that the $i^{th}$ member is given by,
\begin{linenomath*}
\begin{align}
\label{eq:solRep2}
    \beta_i & = \begin{bmatrix} 1 & 0 & 1 & 1 & 0 \end{bmatrix}
\end{align}
\end{linenomath*}
This implies that only the \textit{first, third} and \textit{fourth} terms from the set $\mathcal{X}_{model}$ are included into the \textit{structure/term subset}. Thus, the structure `$\mathcal{X}_i$' encoded by the particle $\beta_i$ is given by,
\begin{linenomath*}
\begin{align}
\label{eq:solRep3}
    \mathcal{X}_i & = \begin{Bmatrix} x_1 & x_3 & x_4 \end{Bmatrix} = \begin{Bmatrix} y(k-1) & y(k-3) & y(k-2)u(k-2) \end{Bmatrix}
\end{align}
\end{linenomath*}
\subsection{Multi-objective Evolutionary Algorithms}
\label{s:MOEAs}

The Multi-Objective Evolutionary Algorithms (MOEAs) are designed to address the multi-criteria decision making (MCDM) process \textit{a posteriori}. It is, therefore, vital that the search process identifies well distributed solutions which are closer to the true Pareto front. This provides a better insight to the Decision Maker (DM).
These two main goals of the search process are, however, contradictory in nature. The existing MOEAs can be categorized based on their approach to balance the convergence with the solution diversity of the approximated Pareto front. In this study, the following three seminal MOEAs are selected to represent two distinct categories of MOEAs: Pareto dominance based (NSGA-II~\cite{Deb:Pratap:2002}  and SPEA-II~\cite{Zitzler:Laumanns:2001}) and decomposition based (MOEA/D~\cite{Zhang:Li:2007}). 

In the following subsections, first the solution representation used to encode term subset/structure is discussed. Next, the compared MOEAs are briefly summarized.
\subsubsection{Pareto Dominance based MOEA}
\label{s:DomMOEA}

Most of the existing Pareto Dominance based MOEAs are based on Goldberg's~\cite{Goldberg:Book:1989} earlier proposal of the direct integration of dominance relations into the search process. This integration can loosely be thought of as a pseudo fitness approach, where a scalar value derived from the dominance relations is assigned to the members to accommodate multiple search objectives. However, since the Pareto optimality defines only a partial order~\cite{Deb:Book:2001,Zhou:Qu:2011}, several members of the population may become incomparable due to similar value of the pseudo fitness, \textit{e.g.}, all the non-dominated members are incomparable. Such a scenario may lead to genetic drift phenomenon~\cite{Goldberg:Book:1989}, \textit{i.e.}, even though all the non-dominated members are equivalent, the population tend to converge only to the part of the Pareto front. An additional distinguishing metric is, therefore, required. To this end, different measures are introduced which essentially encapsulates the diversity information, \textit{e.g.}, fitness sharing, crowding distance, density estimate. The idea here is to assign more importance to the members in an under-represented part of the Pareto front and thereby promote the solution diversity. Furthermore, it has been shown that the introduction of elitism can lead to improved convergence properties~\cite{Zitzler:Deb:2000}.

In this study, the following two seminal dominance based MOEAs are considered: Non-dominated Sorting Genetic Algorithm-II (NSGA-II)~\cite{Deb:Pratap:2002} and Strength Pareto Evolutionary Algorithm-2 (SPEA-II)~\cite{Zitzler:Laumanns:2001}. While both NSGA-II and SPEA-II are based on the Goldberg's proposal, they differ in terms of how these search goals are achieved. 

NSGA-II can be thought of as a direct extension of the Goldberg's proposal. In NSGA-II, the dominance relations are used to rank and segregate the entire population into successive fronts using a well-known \textit{non-dominated sorting} operation. Each front contains the members with a similar dominance rank. To promote the solution diversity, a \textit{crowding-distance} metric is proposed which essentially determines the average distance to neighbors in the objective space. The mating pool is determined using the Crowded Tournament Selection (CTS) which is similar to binary tournaments except with a modified selection operator to balance convergence with diversity, as follows. The member with a lower rank is preferred, however, when the ranks are incomparable, the tie is broken by selecting a member with smaller crowding distance. The offspring population is generated by usual recombination and mutation operators. The elitism is maintained by combining population of `$ps$' parents with the population of `$ps$' offspring to form the population union with $(2 \times ps)$ members. The elitism is ensured by allowing only `$ps$' members of the union population to survive where the non-dominated members are given first priority followed by other members with the lower rank and smaller crowding distance.

In contrast, SPEA-II maintains elitism by keeping an external population (referred to as \textit{archive}) of non-dominated solutions which is continually updated throughout the search process. Although, the archive size may differ from the population size, in this study both are fixed to the same value, and is denoted by `$ps$', for the sake of simplicity. Further, in SPEA-II, dominance relations are incorporated by introducing the concept of `\textit{strength value}'. For each member in the population and in the archive, the strength value is determined from the total number of dominated members. Further, the Euclidean distance in objective space to the $k^{th}$ nearest neighbor is used as the density estimate, where `$k$' is the function of the population size and the archive size. To balance convergence with diversity, the pseudo fitness in SPEA-II is composed of two components: `\textit{raw}' fitness derived from strength value and the density estimate. The mating pool is generated by the binary tournament selection among the members of the archive. The offspring population is generated by usual recombination and mutation operators. Consequently, the external archive is updated to maintain elitism. To this end, in each iteration, the archive and the offspring population are combined to form the union population of size $(2 \times ps)$. The elitism is maintained by including all the non-dominated members from the population union into the archive. Note that in SPEA-II, the archive size is fixed \textit{i.e.}, the archive is always filled. Accordingly, two different scenarios may exist based on the number of non-dominated solutions. If the size of the archive, $ps$, is higher than the number of non-dominated members, the remaining positions in the archive are filled by selecting the members with the lowest pseudo-fitness from the population union. On the other hand, if the non-dominated members are more than `$ps$', then the additional members from the crowded region of the Pareto front are removed using the density estimate metric. 

\begin{algorithm}[!t]
    \small
    \SetKwInOut{Input}{Input}
    \SetKwInOut{Output}{Output}
    \SetKwComment{Comment}{*/ \ \ \ }{}
    \Input{Population/Archive of `$ps$' parents, $\beta_1, \ \beta_2, \ \dots, \ \beta_{ps}$}
    \Output{Population of `$ps$' offspring, $\hat{\beta}_1, \ \hat{\beta}_2, \ \dots, \ \hat{\beta}_{ps}$}
    \algorithmfootnote{`CTS' denotes the \textit{Crowded Tournament Selection}, see~\cite{Deb:Pratap:2002};\\ \smallskip `BTS' denotes the \textit{Binary Tournament Selection}, see~\cite{Zitzler:Laumanns:2001}}
    \BlankLine
    \Comment*[h] {Selection \& Crossover}\\
    \For{i = 1 to $\frac{ps}{2}$}
    {
        \BlankLine
        \Comment*[h] {NSGA-II: Crowded Tournament Selection}\\
        \BlankLine
        $\{ \beta_p, \ \beta_q \} = CTS (population, \ ranks, \ crowding \ distance)$
        \BlankLine
        \BlankLine
        \Comment*[h] {SPEA-II: Binary Tournament Selection}\\
        \BlankLine
        $\{ \beta_p, \ \beta_q \} = BTS (archive, \ pseudo \ fitness)$
        \BlankLine
        \BlankLine
        \Comment*[h] {Parameterized Uniform Crossover}\\
        \BlankLine
        $\hat{\beta}_p \leftarrow \beta_p$, $\hat{\beta}_{q} \leftarrow \beta_q$\\ 
        \BlankLine
        \If{$p_c>rand$}
        {
         \BlankLine
         \For{j = 1 to $n$}
         {
                \BlankLine
                \If{$0.5>rand$}
                    {\BlankLine
                     $\hat{\beta}_{p,j} \leftarrow \beta_{q,j}, \quad \hat{\beta}_{q,j} \leftarrow \beta_{p,j}$}
         } 
        }
    } 
    \BlankLine
    \BlankLine
    \Comment*[h] {Mutation}\\
    \For{i = 1 to $ps$}
    {
       \BlankLine
       \For{j = 1 to $n$}
        {
         \BlankLine
        \If{$p_m>rand$} 
         {
           $\hat{\beta}_{i,j} = 1-\hat{\beta}_{i,j}$ 
         } 
        }
    } 
   
\caption{Reproduction procedures being used for NSGA-II and SPEA-II}
\label{al:moea}
\end{algorithm}

Since NSGA-II and SPEA-II have extensively been studied from both theoretical and empirical perspective, further details and pseudo codes for these algorithms are omitted here, and can be found elsewhere~\cite{Deb:Pratap:2002,Deb:Book:2001,Zitzler:Laumanns:2001}. Nevertheless, in the following the necessary implementation details are discussed. Since one of the objectives of this study is to compare the search performance of MOEAs, the reproduction is performed using similar recombination and mutation operators in all the compared MOEAs. Hence, the first step is to select the appropriate reproduction operators for the binary string representation which is used to encode the candidate term subsets/structures (see Section~\ref{s:SolRep}). To this end, two recombination operators are considered in this study: \textit{single-point} and \textit{uniform} crossover. A detailed investigation to study the effects of recombination operator will be discussed in Section~\ref{s:SingleOrUniform}. Based on this investigation, the uniform crossover has been selected in this study, unless specified otherwise. Further, the mutation is induced through \textit{flip-bit} mutation. 

The common reproduction operators used in this study for NSGA-II and SPEA-II are shown in Algorithm~\ref{al:moea}, where `$\beta_i$' and `$\hat{\beta}_i$' respectively denote the $i^{th}$ parent and the corresponding offspring. Note that the selection procedure in NSGA-II and SPEA-II are different. NSGA-II uses Crowded Tournament Selection (CTS) which is based on ranks and the crowding distance as discussed earlier in this section. In contrast, SPEA-II uses Binary Tournament Selection (BTS) on the basis of the pseudo fitness function. Further issues regarding the parameter settings such as crossover and mutation probabilities will be discussed at length in Section~\ref{s:SweetSpot}.

\subsubsection{Decomposition based MOEA}
\label{s:DecompMOEA}

MOEA/D takes a different approach to balance convergence with diversity. In MOEA/D, each member of the search population is tasked with solving a unique scalar sub-problem of a decomposed multi-objective problem. To this end, a multi-objective problem is decomposed into several scalar sub-problems by a set of uniformly spaced weight vectors, `$\lambda$'. For instance, with a population of `$ps$' members, an $m$-dimensional problem is decomposed into a total of `$ps$' sub-problems. Each sub-problem corresponds to a unique weight vector, \textit{e.g.}, the $i^{th}$ weight vector, $\lambda^i$, corresponds to $i^{th}$ sub-problem, where, $\lambda^i=\{ \lambda_1^i, \ \lambda_2^i, \ \dots, \ \lambda_m^i \}$ and $\sum \limits_{p=1}^m \lambda_p^i =1$. The candidate solution for each sub-problem is evaluated using the corresponding weight vector and an aggregation function such as weighted sum function or Tchebycheff function~\cite{Zhang:Li:2007}. 

Further, prior to the search process, for each sub-problem, a neighborhood of `$T$' sub-problems is determined such that the corresponding distance in the weight vectors is minimized. The core idea of MOEA/D is to exploit the neighborhood relations for the benefit of the search process, \textit{i.e.}, if the $j^{th}$ sub-problem is in the neighborhood of the $i^{th}$ sub-problem, then a good candidate solution for the $i^{th}$ sub-problem may be promising for the $j^{th}$ sub-problem as well. In each iteration, a new candidate solution is determined by reproduction operations on two randomly selected members of the same neighborhood. The new solution is allowed to replace a total of `$nr$' ($nr<T$) worse solutions in the neighborhood. Note that the population in MOEA/D contains the best solution found hitherto for each sub-problem. In addition, an external archive of non-dominated solutions is also maintained and updated in each iteration.

For the sake of fair comparison, the recombination and the mutation operator being used with MOEA/D are similar to NSGA-II and SPEA-II, and are shown in Algorithm~\ref{al:moea}. However, the selection procedure in MOEA/D is different, where two parents are selected at random from the neighborhood of the parent under consideration. 
\subsection{Preference Articulation}
\label{s:PrefArt}

The preferences of the Decision Maker (DM), if incorporated correctly, can simplify the MCDM process. To this end, the preferences can be formulated \textit{a priori}, \textit{a posteriori} or \textit{progressively} with respect to the search process~\cite{Fonseca:Fleming:1993,Fonseca:Fleming:1998,Branke:2008,Coello:2000,Cvetkovic:Parmee:2002,Branke:Deb:2005,Thiele:Miettinen:2009}. This study takes a two-pronged approach to the preference articulation wherein the \textit{a priori} preferences are formulated as \textit{goals} in the objective space~\cite{Fonseca:Fleming:1993,Fonseca:Fleming:1998} and the \textit{a posteriori} preferences are formulated as \textit{weights}~\cite{Zhang:Chen:2004}. This distinction arises from the role being played by the preferences with respect to the search process. The key role of the \textit{a priori} preferences is to focus on the subset of actual Pareto front to better suit the needs of the system identification (to be discussed in Section~\ref{s:aprioripref}). In contrast, the \textit{a posteriori} preferences are being used to select the final structure from the set of identified non-dominated structures (to be discussed in Section~\ref{s:aposterioripref} and~\ref{s:MTD}).

It is worth emphasizing that the preference articulation is not critical to the search performance of the MOEAs. Instead, they are crucial to simplify the \textit{a posteriori} MCDM process by focusing on a particular region of the actual Pareto front which is of interest to the DM.

\subsubsection{A priori Preferences}
\label{s:aprioripref}

Most of the existing MOEAs are designed to support MCDM in an \textit{a posteriori} manner~\cite{Deb:Pratap:2002,Deb:Book:2001,Zitzler:Laumanns:2001,Zhang:Li:2007}, \textit{i.e.}, the goal is to balance diversity in the non-dominated solutions with the convergence to the optimal Pareto front. For instance, in NSGA-II~\cite{Deb:Pratap:2002}, the crowding distance operator has specifically been designed to encourage diversity in the identified non-dominated solutions. This is primarily based on the assumption that the DM does not have any \textit{a priori} preferences. 

However, this may not be the case in system identification applications, wherein it is possible to formulate partial \textit{a priori} preferences. For instance, earlier investigation suggests that most of the non-linear dynamics can sufficiently be described by $20$ or fewer NARX terms~\cite{Chen:Billings:1989}, \textit{i.e.}, $\xi \leq 20$. Further, it is easier to follow that a similar upper limit can be determined for the other objective, \textit{i.e.}, NMSE ($\mathcal{E}$). In practice, the structures with NMSE higher than $30\%$ are usually not recommended, as this indicates a high bias error~\cite{ONelles:2001}. This, in essence, specifies the region of interest for most of the identification applications.

If such information can be embedded in the search process, then MOEAs can focus only on the portion of the Pareto front which is of interest to the Decision Maker (DM). While this exercise has several benefits (\textit{e.g.}, \textit{speed, focus})~\cite{Branke:2008}, in this study particular interest is on `\textit{focus}'. By focusing the search only to regions of interest to the DM, irrelevant structures in the approximate Pareto front can be reduced which can simplify the \textit{a posteriori} MCDM process. This has been the main motivation for including preference articulation in this study.

The integration of DM preferences in MOEA has been investigated in several existing investigation, \textit{e.g.}, see~\cite{Fonseca:Fleming:1993,Fonseca:Fleming:1998,Branke:2008,Coello:2000,Cvetkovic:Parmee:2002,Branke:Deb:2005,Thiele:Miettinen:2009}. The preferences can be formulated as weights, priorities or goals~\cite{Fonseca:Fleming:1998,Coello:2000,Branke:2008}. The weights and priority essentially assign the importance to each objective as per preference of the DM. In goal formulation, the preferences are specified in the objective space as an utopian or minimum levels to be attained by the objectives. For the system identification, articulation of \textit{a priori} preferences as \textit{goals} is more appropriate for the following reasons:
\begin{itemize}
    \item Specification of goal in objective space is relatively easier than determining the priority or weight for each objective \textit{a priori}.
    \item The approximate Pareto set obtained with the goal formulation is essentially subset of the original Pareto set, and therefore not biased towards any objective.
\end{itemize}  

Hence, without the loss of generality, it is possible to formulate \textit{a priori} preferences for the structure selection problem as the following goals: $\xi \leq 20$ and $\mathcal{E} \leq 30$, \textit{i.e.}, the `\textit{goal}' point for the structure selection problem can be given by: $\{\xi_{lim}, \mathcal{E}_{lim} \} = \{ 20, 30 \}$, where `$\xi_{lim}$' and `$\mathcal{E}_{lim}$' denote respectively the maximum allowable limit for the cardinality and the prediction error. 

There exist several approaches to focus the search efforts in the specific region which is determined by the specified goal. For example, Fonseca and Fleming~\cite{Fonseca:Fleming:1993,Fonseca:Fleming:1998} proposed new dominance relations wherein a higher priority is assigned to the objective in which the goal is not attained. While this approach can easily be integrated with the dominance based MOEAs such as NSGA-II and SPEA-II, it may be challenging for non-dominance based MOEAs such as MOEA/D. Hence, in this study the other approach to goal attainment is considered which casts goals as constraints in the objective space as follows: 
\begin{linenomath*}
\begin{align}
    \label{eq:mossUpdated}
    \vec{J}(\mathcal{X}_i) & = \min \begin{cases} J_1(\mathcal{X}_i): \xi_i + \mathcal{P}\\
                                                J_2(\mathcal{X}_i): \mathcal{E}_i + \mathcal{P}
                            \end{cases}\\
    \label{eq:penalty}
    \text{with, } \mathcal{P} & = 10 \times \Big(\langle \mathcal{E}_{lim} - \mathcal{E} \rangle + \langle \xi_{lim} - \xi \rangle \Big)
\end{align}
\end{linenomath*}
where, the bracket-operator, `$\langle \cdot \rangle$' in~(\ref{eq:penalty}) returns the absolute value if the operand is negative else it returns zero. The penalty function, $\mathcal{P}$, has been designed to discourage exploration beyond the specified goal point, $\{\xi_{lim}, \mathcal{E}_{lim} \}$. 

The effects of \textit{a priori} preferences is illustrated through the following numerical example:
\begin{small}
\begin{align}
    \label{eq:numExample}
    \mathcal{S}_1 : y(k) = & \ 0.2 y(k-1)^3 + 0.7 y(k-1)u(k-1) + 0.6 u(k-2)^2\\\nonumber
                           & - 0.7 y(k-2)u(k-2)^2 - 0.5 y(k-2) + e(k) 
\end{align}
\end{small}
For the purpose of identification, a total of $1000$ input-output data-points are generated by exciting the system $\mathcal{S}_1$ with zero-mean uniform white noise sequence, $u \sim WUN(-1,1)$ and Gaussian white noise, $e \sim WGN(0,0.004)$. The model set of $165$ NARX terms is obtained with the following specifications: $[n_u, n_y, n_l] = [4,4,3]$.
\begin{figure}[!t]
    \centering
    \includegraphics[width=0.4\textwidth]{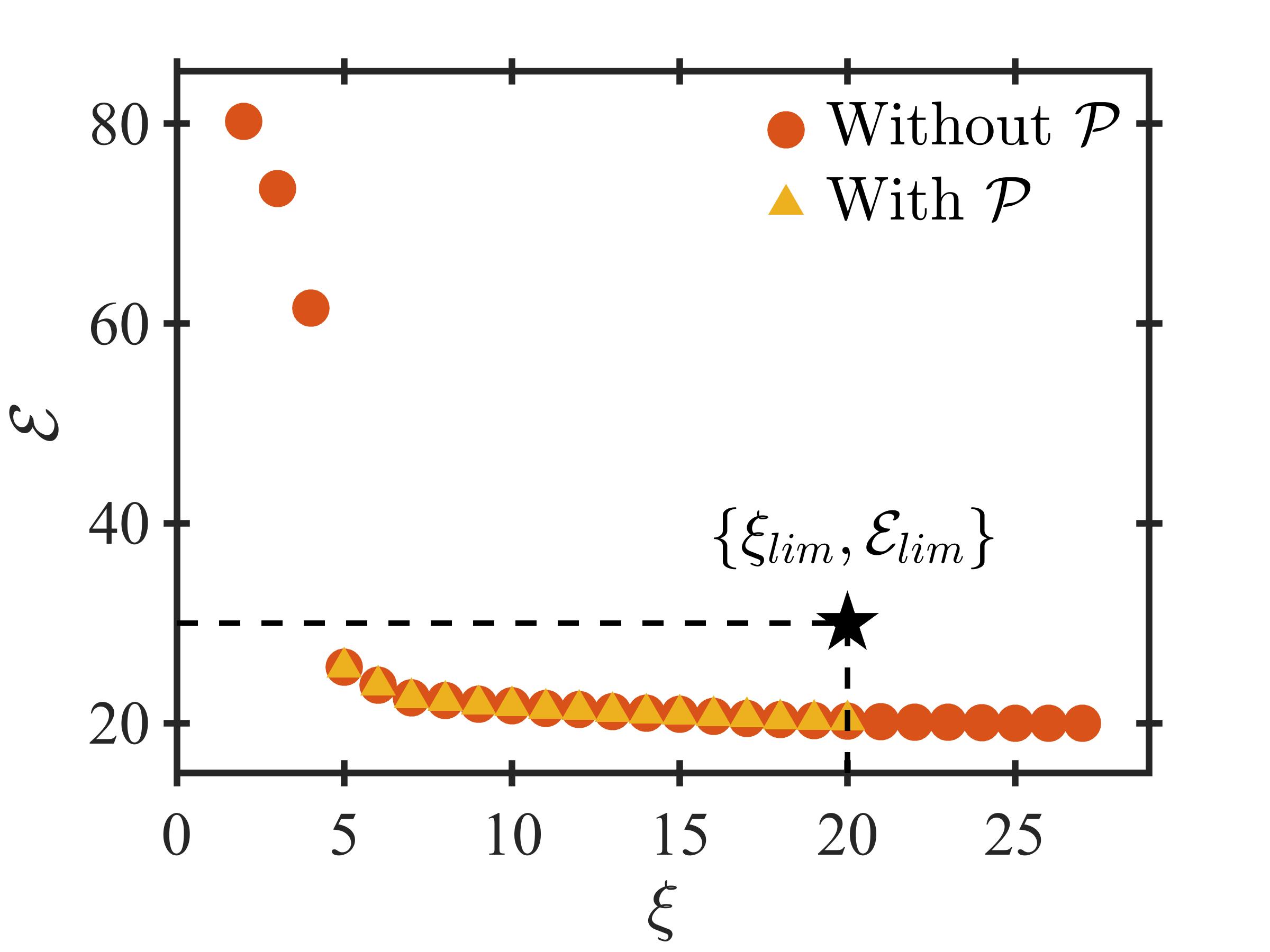}
    \caption{Effects of Preference Articulation}
    \label{f:effPref}
\end{figure}

To investigate the effects of goal formulation, NSGA-II is applied to obtain the approximate Pareto fronts in two scenarios: 1) Without any \textit{a priori} preferences, \textit{i.e.}, using the criteria function in~(\ref{eq:moss}) and 2) With \textit{a priori} preferences, \textit{i.e.}, using the criteria function in~(\ref{eq:mossUpdated}). For the second scenario, the goal point is specified as $\{\xi_{lim}, \mathcal{E}_{lim} \} = \{ 20, 30 \}$. The approximate Pareto front obtained over $40$ independent runs of NSGA-II for both the scenarios are shown in Fig~\ref{f:effPref}.  It is clear that with goal formulation, the search is focused in the region of interest. Due to the penalty function, $\mathcal{P}$, the structure outside the region of interest `appear' less rewarding. Hence, the search is focused in the boundaries defined by the specified goal point. Further, note that the front identified with the goal formulation is the subset of the approximate Pareto front which is obtained without preference articulation, \textit{i.e.}, no bias is introduced towards any of the search objective.

This procedure of accommodating the goal formulation into the objective space is outlined Algorithm~\ref{al:crf}. 

It is worth noting that, without the loss of generality, the goal point $\{\xi_{lim}, \mathcal{E}_{lim} \}$ is fixed to $\{ 20, 30 \}$ on the basis of prevailing common knowledge in system identification. While this can be considered as a rule of thumb, some exception may exist. A simple approach to identify such exceptions and the remedial procedure are discussed in Section~\ref{s:commentPH}.

\begin{algorithm}[!t]
    \small
    \SetKwInOut{Input}{Input}
    \SetKwInOut{Output}{Output}
    \SetKwComment{Comment}{*/ \ \ \ }{}
    \Input{Search Agent, $\beta_i$}
    \Output{Criterion Function, $\vec{J}(\mathcal{X}_i)=\begin{Bmatrix} J_1(\mathcal{X}_i) & J_2(\mathcal{X}_i) \end{Bmatrix}$}
    \BlankLine
    Set the $i^{th}$ structure to null vector, \textit{i.e.},
    $\mathcal{X}_i \leftarrow \varnothing$ and $\xi_i \leftarrow 0$\\
    
    \BlankLine
    \Comment*[h] {Decode the Parent}\\\nllabel{line:crf1}
    \BlankLine
    \For{m = 1 to n} 
        { \BlankLine
           \If{$\beta_{i,m}=1$}
           {\BlankLine
             $\mathcal{X}_i \leftarrow \{ \mathcal{X}_i \cup x_m \}$  \Comment*[h] {add the $m^{th}$ term}\\
             $\xi_i \leftarrow \xi_i+1$
            \BlankLine
           }
          \BlankLine
        } 
    \BlankLine\nllabel{line:crf2}
    Estimate Coefficients, `$\Theta$', corresponding to the terms in $\mathcal{X}_i$
    \BlankLine
    \BlankLine
    \Comment*[h] {Evaluate the Criterion Function}\\
    \BlankLine
    Determine the dynamic prediction error $\mathcal{E}_i$ using~(\ref{eq:NMSE})\\
    Determine the penalty assignment, $\mathcal{P}$ using~(\ref{eq:penalty})
    \BlankLine
    $J_1(\mathcal{X}_i) \leftarrow \xi_i + \mathcal{P}$, \ \ $J_2(\mathcal{X}_i) \leftarrow \mathcal{E}_i + \mathcal{P}$
\caption{Evaluation of the Multi-Objective Criterion Function, $\vec{J}(\cdotp)$}
\label{al:crf}
\end{algorithm}

\subsubsection{A Posteriori Preferences}
\label{s:aposterioripref}

Once the search process is complete and the approximate Pareto set is obtained, the next step is to select the system structure from the pool of non-dominated structures, as shown in Fig~\ref{f:MOSSFrame}. To this end, the DM may be impartial towards the objectives or may have particular preferences. The selection of an appropriate \textit{a posteriori} MCDM approach is therefore dependent on the prevailing scenario, and will be discussed in detail in Section~\ref{s:MCDM}. Further, the human preferences are often partial and/or abstract~\cite{Cvetkovic:Parmee:2002,Branke:2008,Zhang:Chen:2004}. Hence, for the scenario where the DM is biased towards a particular objective, such partial preferences have to be quantified before they can be used with a suitable MCDM approach. While there exist several approaches for this purpose (\textit{e.g.}, see~\cite{Coello:2000,Branke:2008,Zhang:Chen:2004}), in this study, \textit{a posteriori} preferences are formulated as weights using the preference ordering approach~\cite{Zhang:Chen:2004,Parreiras:Vasconcelos:2009} due to its simplicity. In the following, this approach is discussed briefly in the context of system identification.

In the preference ordering approach~\cite{Zhang:Chen:2004,Parreiras:Vasconcelos:2009}, as the name suggests, each objective is ranked based on the order of its importance by the DM. For instance, the DM may prefer parsimonious structures over predictive performance, \textit{i.e.}, `$\xi$' is preferred over `$\mathcal{E}$'. Consequently, the objectives are ranked as follows: $[O_\xi, O_\mathcal{E}] = [1,2]$, where `$O$' denotes the objective rank. It is easier to follow that, for a total of $m$ number of objectives, an $m$-dimensional ranking vector is required. 

Next, the intensity of the preference, denoted by $\mathcal{I}$, is fixed on a scale from `$1$' to `$9$'; where $\mathcal{I}=1$ denotes equal importance to the objectives and $\mathcal{I}=9$ denotes extreme prejudice. Subsequently, the multiplicative preference relations (denoted by `$a$') are obtained using the objective rankings ($O$) and the preference intensity ($\mathcal{I}$) as follows: 
\begin{align}
    \label{eq:mprelation}
    a_{i,j} & = \mathcal{I}^{\delta_O}, \quad i,j \in [1,m] \\ 
    \text{where, } \delta_O & = \frac{O_j - O_i}{m-1} \nonumber
\end{align}
and `$a_{i,j}$' denotes the multiplicative preference over objectives $i$ and $j$. This implies that the $i^{th}$ objective is $a_{i,j}$ times more important than the $j^{th}$ objective.

Finally, the preference weights (denoted by $w$) are determined from the aggregated multiplicative preferences:
\begin{align}
    \label{eq:postw}
    \vec{w} & = \frac{\begin{bmatrix} w_1 & w_2 & \dots & w_m \end{bmatrix}}{\sum_{p=1}^{m} w_p}, \qquad \text{where, \ } w_i = \Big( \prod \limits_{p=1}^{m} a_{i,j} \Big)^{1/m}
\end{align}
For illustration purposes, consider a particular identification problem. Let the objective ranks and the preference intensity specified by the DM be given by: $[O_\xi, O_\mathcal{E}] = [1,2]$ and $\mathcal{I}=5$. Using~(\ref{eq:mprelation}), the multiplicative preference relations can be determined as:
\begin{align}
    \label{eq:prefExample1}
    \begin{bmatrix} a_{\xi,\xi} & a_{\xi,\mathcal{E}} \\ a_{\mathcal{E},\xi} & a_{\mathcal{E},\mathcal{E}} \end{bmatrix} = \begin{bmatrix} 1 & 5 \\ \frac{1}{5} & 1 \end{bmatrix}
\end{align}
Consequently, the preference weights determined using~(\ref{eq:postw}) are as follows:
\begin{align}
    \label{eq:prefExample2}
    w_\xi & = 2.2361, \qquad w_\mathcal{E}=0.4472, \qquad  \text{which gives,} \quad \vec{w} = \begin{bmatrix} 0.83 & 0.17 \end{bmatrix}
\end{align}
The complexity of this procedure increases with the number of objectives. It is worth emphasizing that, by limiting number of objectives to two (\textit{i.e.}, $m=2$), the determination of preference weights is significantly simplified. This, in part, is the motivation behind limiting the of objectives to cardinality ($\xi$) and prediction error ($\mathcal{E}$).

\subsection{A Posteriori Multi-Criteria Decision Making}
\label{s:MCDM}

Given a set of non-dominated structures, the objective of the \textit{a posteriori} Multi-Criteria Decision Making (MCDM) methods is to rank these structures to facilitate the final selection by a Decision Maker (DM). Note that the DM may or may not have particular preferences. To accommodate these two distinct scenarios, the following two MCDM methods have been selected in this study: 1) Minimum Manhattan Distance (MMD)~\cite{Chiu:Yen:2016} and 2) Multi-criteria Tournament Decision (MTD)~\cite{Parreiras:Vasconcelos:2009}.

Note that both MMD and MTD are independent of the search algorithm. Hence, they can easily be combined with any MOEA for \textit{a posteriori} decision making. In the following subsections, these methods are briefly discussed from the perspective of system identification.

\subsubsection{Minimum Manhattan Distance}
\label{s:MMD}

If the identified Pareto front is \textit{bent}, then the solutions in the `\textit{knee-region}' of the front represents the best overall compromise. Therefore, many existing \textit{a posteriori} MCDM methods have been designed to select the \textit{knee-point} solutions. Hence, in the absence of any preferences, the \textit{knee-point} structures can be considered for further analysis. To this end, Minimum Manhattan Distance (MMD) approach has been selected, which is similar to Divide $\&$ Conquer approach to identify the knee-point solution, albeit MMD is computationally more efficient~\cite{Chiu:Yen:2016}.

For the given set of non-dominated structures, $\Gamma$, MMD begins by identifying the hypothetical ideal point, $\vec{J}^\star$, in the corresponding objective space, as follows:
\begin{align}
\vec{J}^{\star} & = \begin{Bmatrix} J_1^{min} & J_2^{min} & \dots &  J_m^{min} \end{Bmatrix}, \ \text{where,} \ J_p^{min} = \min \limits_{\forall{\mathcal{X}_i} \in \Gamma}  J_p(\mathcal{X}_i), \quad p\in[1,m]
\end{align}
Subsequently, for each non-dominated structure $\mathcal{X}_i \in \Gamma$, the Manhattan distance, $\mathcal{D}(\cdot)$, is evaluated with respect to $\vec{J}^\star$:
\begin{align}
\label{eq:MMD}
\mathcal{D}(\mathcal{X}_i) & = \sum \limits_{p=1}^m \abs{\frac{J_p(\mathcal{X}_i)- J_p^{min}}{J_p^{max}- J_p^{min}}}, \ \text{where, \ } J_p^{max} = \max \limits_{\forall{\mathcal{X}_i} \in \Gamma}  J_p(\mathcal{X}_i)
\end{align}
Note that the Manhattan distance $\mathcal{D}(\cdot)$ is determined in the normalized objective space to avoid scaling issues. In MMD, as the name suggests, the structure with the minimum $\mathcal{D}(\cdot)$ is selected as the final choice. However, a slightly different approach is followed in this study, where each structure is ranked in the ascending order of $\mathcal{D}(\cdot)$, and the top five structures are selected for further analysis.

It is worth to note that MMD has explicitly been designed to discourage any preference specification from the DM. Further, the knee-point selection approach of MMD may not necessarily align with the preferences of the DM. Hence, MMD is not recommended when DM can specify preferences. For such scenario, Multi-criteria Tournament Decision (MTD)~\cite{Parreiras:Vasconcelos:2009} may be more appropriate, which is discussed in the following. 

\subsubsection{Multi-criteria Tournament Decision}
\label{s:MTD}

In practice, DM may have specific preferences which can somehow be embedded in the MCDM process. For instance, in system identification, the DM can assign higher priority to the \textit{number of terms}, `$\xi$', as a parsimonious description of the given system is usually preferred. Such preferences can easily be converted into a \textit{priority weight}, `$w$', which highlights the importance of a particular search objective as per DM's preference. A detailed investigation for determining the \textit{priority weights} from abstract DM preferences can be found in~\cite{Zhang:Chen:2004}. The priority weights for this study are determined following the guidelines in~\cite{Zhang:Chen:2004,Parreiras:Vasconcelos:2009}, as discussed in Section~\ref{s:aposterioripref}. 

The Multi-criteria Tournament Decision (MTD) approach essentially ranks the given non-dominated structures based on priority weights specified by the DM~\cite{Parreiras:Vasconcelos:2009}. For this purpose, for each structure $\mathcal{X}_i \in \Gamma$, a pairwise comparison with the remaining structures is carried out, considering only the $p^{th}$ search objective, $p \in [1,m]$. This is accomplished by the following tournament function:
\begin{align}
\label{eq:MTD1}
T_p(\mathcal{X}_i,\Gamma) & = \sum \limits_{\forall{\mathcal{X}_j} \in \Gamma \wedge \mathcal{X}_j \neq \mathcal{X}_i} \frac{t_p(\mathcal{X}_i,\mathcal{X}_j)}{|\Gamma|-1}, \quad p \in [1,m]\\
\text{where,} \ t_p(\mathcal{X}_i,\mathcal{X}_j) & = \begin{cases} 1, \quad \textit{if} \ J_{p}(\mathcal{X}_j)-J_{p}(\mathcal{X}_i)>0 \\ 0, \quad \textit{otherwise} \end{cases} \nonumber 
\end{align}
Thus, for the $p^{th}$ objective, the tournament function $T_p(\mathcal{X}_i,\Gamma)$ determines the total number of structures compared to which $\mathcal{X}_i$ yields better objective value. For the structure under consideration, the procedure is repeated to determine the tournament function across all the `$m$' objectives. A similar procedure is repeated for each of the structure $\mathcal{X}_i \in \Gamma$. Subsequently, the following geometric aggregation function is used to derive a global rank which considers the tournament function for all the `$m$' objectives as well as the priority weights which are specified by DM, $[w_1, w_2, \dots, w_m]$: 
\begin{align}
\label{eq:MTDRank}
\mathcal{R}(\mathcal{X}_i) = \Big( \prod \limits_{p=1}^{m} T_p(\mathcal{X}_i,\Gamma)^{w_p} \Big)^\frac{1}{m}
\end{align}
It is easier to follow that a higher value of the global rank `$\mathcal{R}(\cdot)$' is desirable. In this study, the structures are ranked in the descending order of $\mathcal{R}(\cdot)$. On the basis of this rankings, the top five structures are selected for further analysis. 
\begin{algorithm}[!t]
    \small
    \SetKwInOut{Input}{Input}
    \SetKwInOut{Output}{Output}
    \SetKwComment{Comment}{*/ \ \ \ }{}
    \Input{Input-output Data, $(u,y)$}
    \Output{Identified Structure and Coefficients}
    \BlankLine
    \Comment*[h] {Data Pre-processing}\\
    \BlankLine
    Split the input-output data into the \textit{estimation} and \textit{validation} sets\\
    Generate set of model terms by specifying $n_u, n_y$ and $n_l$ of the NARX model \\
    \BlankLine
    \BlankLine
    \Comment*[h] {Search for the non-dominated system structures}\\
    \BlankLine
    Select a meta-heuristic search algorithm (NSGA-II, SPEA-II or MOEA/D in this study)\\
    Perform $R$ independent runs of the selected search algorithm \nllabel{line:1}\\
    \BlankLine
    $\Gamma \leftarrow \varnothing$, $\Lambda \leftarrow \varnothing$\\
    \For{k = 1 to $R$} 
        { \BlankLine
           Identify and record the \textit{non-dominated} structures, \textit{i.e.},\\
          \BlankLine
          $\Gamma \leftarrow \Gamma \cup \begin{Bmatrix} \mathcal{X}_1 & \mathcal{X}_2 & \dots \end{Bmatrix}$, $\Lambda \leftarrow \Lambda \cup \begin{Bmatrix} \vec{J}(\mathcal{X}_1) & \vec{J}(\mathcal{X}_2) & \dots \end{Bmatrix}$\nllabel{line:moss1}
          \BlankLine 
        } 
     \BlankLine
     Keep only the non-dominated structures, \textit{i.e.}, $\Gamma^* \prec \Gamma$\\
     \BlankLine
     \Comment*[h] {A posteriori MCDM}\\
     Rank the identified structures using MMD and/or MTD\\
\caption{Multi-Objective Evolutionary Structure Selection}
\label{al:moss}
\end{algorithm}

\section{Investigation Framework}
\label{s:IF}

In the following, the framework of this investigation is discussed. First, the search environment of this investigation is discussed in Section~\ref{s:searchSetup}. The test non-linear systems being used to benchmark the search performance are discussed next in Section~\ref{s:Data}. Finally, the performance metrics being used to evaluate search from the quantitative and qualitative perspectives are discussed in Section~\ref{s:QuantMetric} and~\ref{s:QualMetric}. 

\subsection{Search Environment}
\label{s:searchSetup}

The overall procedure followed for the multi-objective structure selection is outlined in Algorithm~\ref{al:moss}. The search process begins by formulating \textit{a priori} preferences as goals. All the results in this study are obtained with the following goal specification: $\{ \xi_{limit}, \mathcal{E}_{limit} \} = \{ 20, 30 \}$. Subsequently, the non-dominated structures are gathered over multiple independent runs of the MOEA due to stochastic nature of the MOEAs. The dominance of the ensuing pool of the structures is again determined and only non-dominated structures are considered for further analysis. For the \textit{a posteriori} decision making, the identified non-dominated structures are ranked using the Manhattan distance $D(\cdot)$ in MMD (see Section~\ref{s:MMD}) and the rank $\mathcal{R}(\cdot)$ in MTD (see Section~\ref{s:MTD}). The priority weights required for MTD approach are determined using preference ordering approach (see Section~\ref{s:aposterioripref}). 

All of the compared MOEAs are implemented in MATLAB. For fair comparison, each MOEAs is terminated after 25,000 Function Evaluations (FEs). The population size, `$ps$' is fixed to $50$ for all the MOEAs. The size of external archives in SPEA-II and MOEA/D is also set to $50$. In SPEA-II, the density estimate is determined with respect to $k^{th}$ nearest neighbor, where $k=10$. In MOEA/D, the sub-problems are created by uniformly spaced weight vectors in $[0,1]$. The neighborhood size, `$T$', is fixed to $10\%$ of the population size. The new solution is allowed to replace a total of two solutions in the neighborhood, \textit{i.e.}, $nr=2$. To accommodate the stochastic nature of the MOEAs, the non-dominated structures over $40$ independent runs are selected.

Further, the \textit{uniform} crossover is selected as the recombination mechanism in all the MOEAs on the basis of detailed comparative analysis (\textit{see} Section~\ref{s:SingleOrUniform}). The following crossover and mutation probabilities for MOEAs are selected on the basis the performance \textit{sweet spot} analysis discussed in Section~\ref{s:SweetSpot}: NSGA-II: $\{ p_c, \ p_m \} = \{0.9, 0.006 \}$; SPEA-II: $\{ p_c, \ p_m \} = \{0.7, 0.008 \}$;  MOEA/D: $\{ p_c, \ p_m \} = \{0.8, 0.008 \}$.

\subsection{Test Non-linear Systems}
\label{s:Data}

In this study, a total of eight known benchmark nonlinear systems have been included. These systems have been used by various researchers to study the efficacy of the structure selection approaches~\cite{Mendes:1995,Mao:Billings:1997,Bonin:Pirrodi:2010,Piroddi:Spinelli:2003b,Baldacchino:Kadirkamanathan:2013,Falsone:Piroddi:2015}. The first six systems, shown in Table~\ref{t:sys}, are being used for comparative evaluation of MOEAs. The other known system has been included to highlight the exception to the specification of \textit{a priori} goal point in Section~\ref{s:commentPH}. 

For the comparative evaluation, a total of $1000$ data-points are collected for each system shown in Table~\ref{t:sys}. Each system is excited by a white noise input with either Gaussian (\textit{denoted by} `WGN') or uniform (\textit{denoted by} `WUN') distribution. The model set ($\mathcal{X}_{model}$) of $165$ NARX terms is obtained by the following specifications in NARX model~(\ref{eq:NARXmodel}): $[n_u,n_y,n_l]=[4,4,3]$. For structure selection purposes, $700$ data-points are used for the parameter estimation and the remaining data-points are used for the validation, \textit{i.e.}, $\mathcal{N}_v = 300$.

In addition to these known systems, two case studies are also included: the identification of 1) Duffing's oscillator and 2) nonlinear wave-forces acting on fixed structures. These case studies are discussed respectively in Section~\ref{s:ResDuff} and~\ref{s:wav}.

\begin{table}[!t]
  \centering
  \small
  \caption{Test Non-linear Systems}
  \label{t:sys} 
  \begin{adjustbox}{max width=0.9\textwidth}
  \begin{threeparttable}
    \begin{tabular}{ccccccccc}
    \toprule
    
    \textbf{System} & \textbf{Known Structure} & \textbf{Input}(\boldmath$u$)$^\dagger$ & \textbf{Noise} (\boldmath$e$)$^\dagger$  \\
    \midrule

    $\mathcal{S}_1$    &  $y(k) =$  \makecell{$0.2 y(k-1)^3 + 0.7 y(k-1)u(k-1) + 0.6 u(k-2)^2$\\[0.7ex]
                    $- 0.7 y(k-2)u(k-2)^2 - 0.5 y(k-2) + e(k)$} &  WUN$(-1,1)$ & WGN$(0,0.004)$\\[4ex]
    $\mathcal{S}_2$    &  $y(k) = 0.5 + 0.5y(k-1) + 0.8u(k-2) + u(k-1)^2 - 0.05y(k-2)^2 + e(k)$     &  WUN$(0,1)$ & WGN$(0,0.05)$\\[2ex]
    $\mathcal{S}_3$    &  $y(k) = 0.8y(k-1) + 0.4u(k-1) + 0.4u(k-1)^2 + 0.4u(k-1)^3 + e(k) $         &  WGN$(0,1)$ & WGN$(0,0.33^2)$\\[3ex]
    $\mathcal{S}_4$    &  $y(k) =$  \makecell{$ 0.1586 y(k-1) + 0.6777 u(k-1) + 0.3037 y(k-2)^2$\\[0.7ex]
                                   $ -0.2566 y(k-2) u(k-1)^2 - 0.0339 u(k-3)^3 + e(k)$} &  WUN$(0,1)$ & WGN$(0,0.002)$\\[5ex]
                                   
    $\mathcal{S}_5$    &  $y(k) = $  \makecell{$0.7 y(k-1)u(k-1) - 0.5 y(k-2)$\\[0.7ex]
    $+ 0.6 u(k-2)^2 - 0.7 y(k-2)u(k-2)^2 + e(k)$} &  WUN$(-1,1)$ & WGN$(0,0.004)$\\[5ex]
    $\mathcal{S}_6$    &  $y(k) = 0.5y(k-1) + 0.3u(k-1) + 0.3u(k-1)y(k-1) + 0.5u(k-1)^2 + e(k)$     &  WUN$(0,1)$ & WGN$(0,0.002)$\\ [2ex]
    
    \bottomrule
    \end{tabular}%
    \begin{tablenotes}
      \scriptsize
       \item $\dagger$ WUN$(a,b)$ denotes white uniform noise sequence in the interval $[a,b]$; WGN$(\mu,\sigma)$ denotes white Gaussian noise sequence with the mean `$\mu$' and the variance `$\sigma$'.  
    \end{tablenotes}
  \end{threeparttable}
  \end{adjustbox}
\end{table}%
\subsection{Quantitative Performance Metrics}
\label{s:QuantMetric}

The part of this study aims to compare the search performance of MOEAs on the identification problems, which requires a quantitative performance metric to infer the differences in the identified non-dominated structures. Further, a quantitative metric is essential to investigate the effects of search parameters and the choice of the crossover operator which will be discussed in Section~\ref{s:ExpSetupPcPm}. Usually, MOEAs are evaluated on the basis of the following two criteria: 1) proximity of the identified solutions to the optimal Pareto front and 2) diversity of the identified solutions. Most of the existing performance metrics have been designed to integrate the evaluation of these criteria to yield a singular value~\cite{Deb:Book:2001,Zitzler:Thiele:2003}. In this study, two of such performance metrics are being used to evaluate the performance of MOEAs: \textit{Set Coverage Metric} and \textit{Hyper-volume Indicator}~\cite{Zitzler:Thiele:1998}, which are discussed briefly in the following:

Consider a structure selection problem with `$m$' number of objectives being solved by two MOEAs, `$A$' and `$B$'. Let the set of non-dominated structures and the corresponding set of objective function vectors identified by these algorithms be denoted by $\Gamma_A$, $\Gamma_B$, $\Lambda_A$ and $\Lambda_B$, \textit{i.e.},
\begin{align}
    \label{eq:exampleAB}
    \Gamma_A & = \{ \mathcal{X}_{A,1}, \ \mathcal{X}_{A,2}, \dots \}, \qquad \Lambda_A = \{ \vec{J}_{A,1}, \ \vec{J}_{A,2}, \dots \}\\
    \Gamma_B & = \{ \mathcal{X}_{B,1}, \ \mathcal{X}_{B,2}, \dots \}, \qquad \Lambda_B = \{ \vec{J}_{B,1}, \ \vec{J}_{B,2}, \dots \}
\end{align}
The set coverage metric, $\mathcal{C}(\cdotp)$, is designed to compare two sets of non-dominated structures. For instance, for $\Gamma_A$ and $\Gamma_B$, it essentially determines the number of structures in $\Gamma_B$ which are dominated by the structures in $\Gamma_A$, as follows~\cite{Zitzler:Thiele:1998}:
\begin{align}
    \label{eq:coverageMetric}
    \mathcal{C}(A,B) & = \frac{\Big| \Big\{ \mathcal{X}_{B,i} \in \Gamma_B | \ \exists{\mathcal{X}_{A,j}}\in \Gamma_A :  \mathcal{X}_{A,j} \preceq \mathcal{X}_{B,i}   \Big\}\Big|}{|\Gamma_B|}
\end{align}
It is easier to follow that this metric is bounded between $[0,1]$. It is worth emphasizing that the $\mathcal{C}$-metric may not be symmetric~\cite{Zitzler:Thiele:1998}. Hence, the comparison of $\Gamma_A$ and $\Gamma_B$ requires evaluation of both $\mathcal{C}(A,B)$ and $\mathcal{C}(B,A)$. It is obvious that $\mathcal{C}(A,B)>\mathcal{C}(B,A)$ implies the search performance of algorithm `$A$' is better than that of Algorithm `$B$', \textit{i.e.}, $\Gamma_A \preceq \Gamma_B$.        

In contrast to $\mathcal{C}$-metric, the hyper-volume (HV) indicator is designed to independently evaluate the given set of non-dominated structures. In essence, the HV-indicator measures the region in the objective space which is dominated by the given set of non-dominated structures, say $\Gamma_A$, and bounded by a reference point, $\vec{r} \in \mathbb{R}^m$, which is given by,
\begin{align}
\vec{r} & \geq \begin{Bmatrix} J_{A,1}^{max} & J_{A,2}^{max} & \dots & J_{A,m}^{max}\end{Bmatrix}, \ \text{where,} \ J_{A,p}^{max} = \max \limits_{\forall{\mathcal{X}_{A,i}} \in \Gamma_A} J_p(\mathcal{X}_{A,i}) , \ p=1, 2, \dots,m.
\end{align}
A higher value of this metric denotes a better performance. In this study, HV-indicator is evaluated following the dimension sweep algorithm~\cite{Fonseca:Paquete:2006,Beume:Fonseca:2009}. Further, to avoid scaling issues, it is determined using the normalized objective values. Note that HV-indicator may be sensitive to the specification of the reference point, $\vec{r}$. To avoid such issues, we follow the guidelines in~\cite{Ishibuchi:Imada:2017} to determine $\vec{r}$.

\subsection{Qualitative Evaluation of Search Outcomes}
\label{s:QualMetric}

It is worth emphasizing that a discrete NARX model of a continuous-time system is not unique, \textit{i.e.}, there may exist several valid NARX structures for the system under consideration~\cite{Billings:2013}. This issue is further explored later through the case studies in Section~\ref{s:ResDuff} and~\ref{s:wav}. However, this is not the case for the known discrete non-linear system shown in Table~\ref{t:sys}, which are being used for the comparative evaluation. Given that the identification data are directly generated from these systems, there exists a unique and known structure for each system in Table~\ref{t:sys}. Hence, for these systems, the non-dominated structures identified by MOEAs can easily be evaluated. To this end, each non-dominated structure is `\textit{qualitatively}' evaluated following the \textit{search outcome} definitions developed in our previous investigations~\cite{Hafiz:Swain:Floating:2019,Hafiz:Swain:2020a}.

Consider the following subsets of NARX terms for this purpose,
\begin{itemize}
    \item $\varnothing$ : the null set
    \item $\mathcal{X}_{model}$ : the super-set containing all the NARX terms
    \item $\mathcal{X}^{\star}$ : the optimum structure or the set of system terms, $\mathcal{X}^{\star} \subset \mathcal{X}_{model}$
    \item $\mathcal{X}_{k}$ : $k^{th}$ non-dominated structure identified by an MOEA, $\mathcal{X}_{k} \in \Gamma^{\ast}$ 
    \item $\mathcal{X}_k^{spur}$ : set of \textit{spurious} terms which are not present in the actual system but are included in $\mathcal{X}_{k}$, \textit{i.e.}, $\mathcal{X}_k^{spur} = \mathcal{X}_{k} \setminus \mathcal{X}^{\star}$
\end{itemize}
\smallskip

Following these definitions, each non-dominated structure can be categorized into one of the following search outcome~\cite{Hafiz:Swain:Floating:2019}:
\begin{enumerate}
\item \textbf{Identification of the Correct Structure} (\textit{\textbf{Exact Fitting}}) :\\
\textit{In this scenario the identified model contains all system terms and does not include any spurious terms, i.e.}, $\mathcal{X}_{k}=\mathcal{X}^{\star}$ and $\mathcal{X}_k^{spur} = \varnothing$
\smallskip
\smallskip
\item \textit{\textbf{Over Fitting}} :\\
\textit{The identified model contains all system terms; however spurious terms are also selected}, \textit{i.e.}, $\mathcal{X}_{k} \supset \mathcal{X}^{\star}$ and $\mathcal{X}_k^{spur} \neq \varnothing$
\smallskip
\smallskip
\item \textit{\textbf{Under Fitting-1}} :\\
\textit{The algorithm fails to identify all system terms; though it does not include any spurious terms}, \textit{i.e.}, $\mathcal{X}_{k} \subset \mathcal{X}^{\star}$ and $\mathcal{X}_k^{spur} = \varnothing$
\smallskip
\smallskip
\item \textit{\textbf{Under Fitting-2}} :\\
\textit{The algorithm fails to identify all system terms; however spurious terms are selected}, \textit{i.e.}, $\mathcal{X}_{k} \not \supset \mathcal{X}^{\star}$ and $\mathcal{X}_{spur} \neq \varnothing$
\smallskip
\end{enumerate}

\section{Results: Identification Perspective}
\label{s:resIDP}

This study aims to establish the multi-objective evolutionary algorithms (MOEAs) as viable alternatives for structure selection. To this end, several issues encountered in multi-objective structure selection are investigated. First, the identification of a set of benchmark nonlinear systems is discussed in Section~\ref{s:CompEval2}. Next, the issues related to the specification of goal point are discussed through an illustrative example in Section~\ref{s:commentPH}. Finally, two case studies are included in Section~\ref{s:ResDuff} and~\ref{s:wav}. The identification of discrete-time model of the continuous-time Duffing's oscillator is taken as the first case study. Next, the practical problem of non-linear wave-force identification is considered. The objective of these case studies is to highlight key distinctive ability of MOEAs, \textit{i.e.}, identification of multiple valid system structures. 

\subsection{Identification of Discrete-time Systems}
\label{s:CompEval2}
The first phase of the investigation begins by considering several discrete-time nonlinear systems shown in Table~\ref{t:sys}. The non-dominated structures for these systems are identified using all the MOEAs and as per the search environment described in Section~\ref{s:searchSetup} and Algorithm~\ref{al:moss}. It is worth emphasizing that, since the structure of these systems are known, the correctness of the proposed approach can be easily validated. As discussed and defined in Section~\ref{s:QualMetric}, the non-dominated structures identified by MOEAs are categorized using qualitative search outcomes. Since the MOEAs are designed to obtain diverse solutions across the entire Pareto front, the approximate Pareto set may contain several \textit{under-fitted} and/or \textit{over-fitted} structures.

From the system identification perspective, the ideal search outcome is `\textit{exact-fitting}'. However, the \textit{over-fitted} structures can also be tolerated provided the spurious terms can be removed through some refinement procedure. To investigate this further, the identified non-dominated structures are refined through a simple null-hypothesis test~\cite{Wei:Billings:2006} on the coefficients. Subsequently, the refined structures are categorized into qualitative search outcomes as per the definitions in Section~\ref{s:QualMetric}. Following these steps, the frequency of each search outcome is determined for the approximated Pareto set, as shown in Table~\ref{t:outcome}. For instance, the approximate Pareto set identified by NSGA-II for the system $\mathcal{S}_2$ contains a total of $17$ non-dominated structures. Each of these structures is refined using null-hypothesis test and categorized into one of the four possible search outcomes. Following these steps, $16$ non-dominated structures are categorized as `\textit{exact-fitting}' whereas only one non-dominated structure is classified as `\textit{under-fitting}' (as seen in Table~\ref{t:outcome}). 

It can be argued that `\textit{success}' of the given MOEA can be determined on the basis of total number of `\textit{exact-fitting}' outcomes. Pursuing this argument, it is interesting to see that most of the identified non-dominated structures are categorized as \textit{exact-fitting} (see Table~\ref{t:outcome}), which implies that these structures could include all the system terms. It is, therefore, safe to conclude that all the MOEAs have been \textit{successful} in identifying the correct structure for all the benchmark discrete-time systems in Table~\ref{t:sys}.

\begin{table}[!t]
  \centering
  \small
  \caption{Qualitative Evaluation of the Non-dominated Structures}
  \label{t:outcome}%
  \begin{adjustbox}{max width=0.6\textwidth}
  \begin{threeparttable}
  \begin{tabular}{ccccc}
    \toprule
    \textbf{System} & \textbf{Outcome} & \textbf{NSGA-II} & \textbf{SPEA-II} & \textbf{MOEA/D} \\
    \midrule
    \multirow{5}{*}{$\mathcal{S}_1$} & \textit{Exact-Fitting} & 16    & 16    & 16 \\
          & \textit{Over-fitting} & 0     & 0     & 0 \\
          & \textit{Under-fitting 1} & 0     & 0     & 0 \\
          & \textit{Under-fitting 2} & 0     & 0     & 0 \\
    \cmidrule{2-5}          & \textit{Total Structures} & 16    & 16    & 16 \\
    \midrule
    \multirow{5}{*}{$\mathcal{S}_2$} & \textit{Exact-Fitting} & 16    & 16    & 16 \\
          & \textit{Over-fitting} & 0     & 0     & 0 \\
          & \textit{Under-fitting 1} & 0     & 0     & 0 \\
          & \textit{Under-fitting 2} & 1     & 1     & 0 \\
    \cmidrule{2-5}          & \textit{Total Structures} & 17    & 17    & 16 \\
    \midrule
    \multirow{5}{*}{$\mathcal{S}_3$} & \textit{Exact-Fitting} & 17    & 17    & 17 \\
          & \textit{Over-fitting} & 0     & 0     & 0 \\
          & \textit{Under-fitting 1} & 1     & 1     & 1 \\
          & \textit{Under-fitting 2} & 0     & 0     & 0 \\
    \cmidrule{2-5}          & \textit{Total Structures} & 18    & 18    & 18 \\
    \midrule
    \multirow{5}{*}{$\mathcal{S}_4$} & \textit{Exact-Fitting} & 16    & 16    & 16 \\
          & \textit{Over-fitting} & 0     & 0     & 0 \\
          & \textit{Under-fitting 1} & 1     & 1     & 1 \\
          & \textit{Under-fitting 2} & 0     & 0     & 0 \\
    \cmidrule{2-5}          & \textit{Total Structures} & 17    & 17    & 17 \\
    \midrule
    \multirow{5}{*}{$\mathcal{S}_5$} & \textit{Exact-Fitting} & 17    & 17    & 17 \\
          & \textit{Over-fitting} & 0     & 0     & 0 \\
          & \textit{Under-fitting 1} & 0     & 0     & 0 \\
          & \textit{Under-fitting 2} & 0     & 0     & 0 \\
    \cmidrule{2-5}          & \textit{Total Structures} & 17    & 17    & 17 \\
    \midrule
    \multirow{5}{*}{$\mathcal{S}_6$} & \textit{Exact-Fitting} & 17    & 17    & 17 \\
          & \textit{Over-fitting} & 0     & 0     & 0 \\
          & \textit{Under-fitting 1} & 1     & 1     & 1 \\
          & \textit{Under-fitting 2} & 0     & 0     & 0 \\
    \cmidrule{2-5}          & \textit{Total Structures} & 18    & 18    & 18 \\
    \bottomrule
    \end{tabular}%
    \begin{tablenotes}
      \scriptsize
       \item $\dagger$ see Section~\ref{s:QualMetric} for the definition of search outcomes
    \end{tablenotes}
  \end{threeparttable}
  \end{adjustbox}
\end{table}%

\subsection{Comments on the a priori Goal Point}
\label{s:commentPH}

This study takes a pragmatical approach to set the \textit{a priori} goal point, \textit{i.e.}, $\{ \xi_{limit}, \mathcal{E}_{limit} \} = \{ 20, 30 \}$, as discussed in Section~\ref{s:aprioripref}. In particular, these specifications have been selected to guide the search towards structures with fewer terms. However, it is important to note that these specifications are ultimately system-dependent. In the following, a numerical example is considered to highlight this issue along with a simple approach to handle such exceptions.

Consider the following non-linear system with $23$ number of terms: 
\begin{linenomath*}
\begin{small}
\begin{align}
	\label{eq:S4}
    \mathcal{S}_7 : y(k)  = & \ 0.8833u(k-1) + 0.0393u(k-2) + 0.8546u(k-3) + 0.8528u(k-1)^2 + 0.7582u(k-1)u(k-2) \nonumber\\ 
            & + 0.1750u(k-1)u(k-3) + 0.0864u(k-2)^2 + 0.4916u(k-2)u(k-3) + 0.0711u(k-3)^2 - 0.0375y(k-1) \nonumber\\ 
            & - 0.0598y(k-2) - 0.0370y(k-3) - 0.0468y(k-4) - 0.0476y(k-1)^2 - 0.0781y(k-1)y(k-2) \nonumber\\ 
            & - 0.0189y(k-1)y(k-3) - 0.0626y(k-1)y(k-4) - 0.0221y(k-2)^2 - 0.0617y(k-2)y(k-3) \nonumber \\ 
            & - 0.0378y(k-2)y(k-4) - 0.0041y(k-3)^2 - 0.0543y(k-3)y(k-4) - 0.0603y(k-4)^2 + e(k)
\end{align}
\end{small}
\end{linenomath*}
A total of  $1000$ input-output data points are collected by exciting the system with $u \sim WUN(0,1)$ and $e \sim WGN(0,0.01^2)$. The model set ($\mathcal{X}_{model}$) of $286$ NARX terms is obtained with the following specification in~(\ref{eq:NARXmodel}): $[n_u,n_y,n_l]=[5,5,3]$. 

\begin{figure*}[!t]
\centering
\small
\begin{subfigure}{.33\textwidth}
  \centering
  \includegraphics[width=\textwidth]{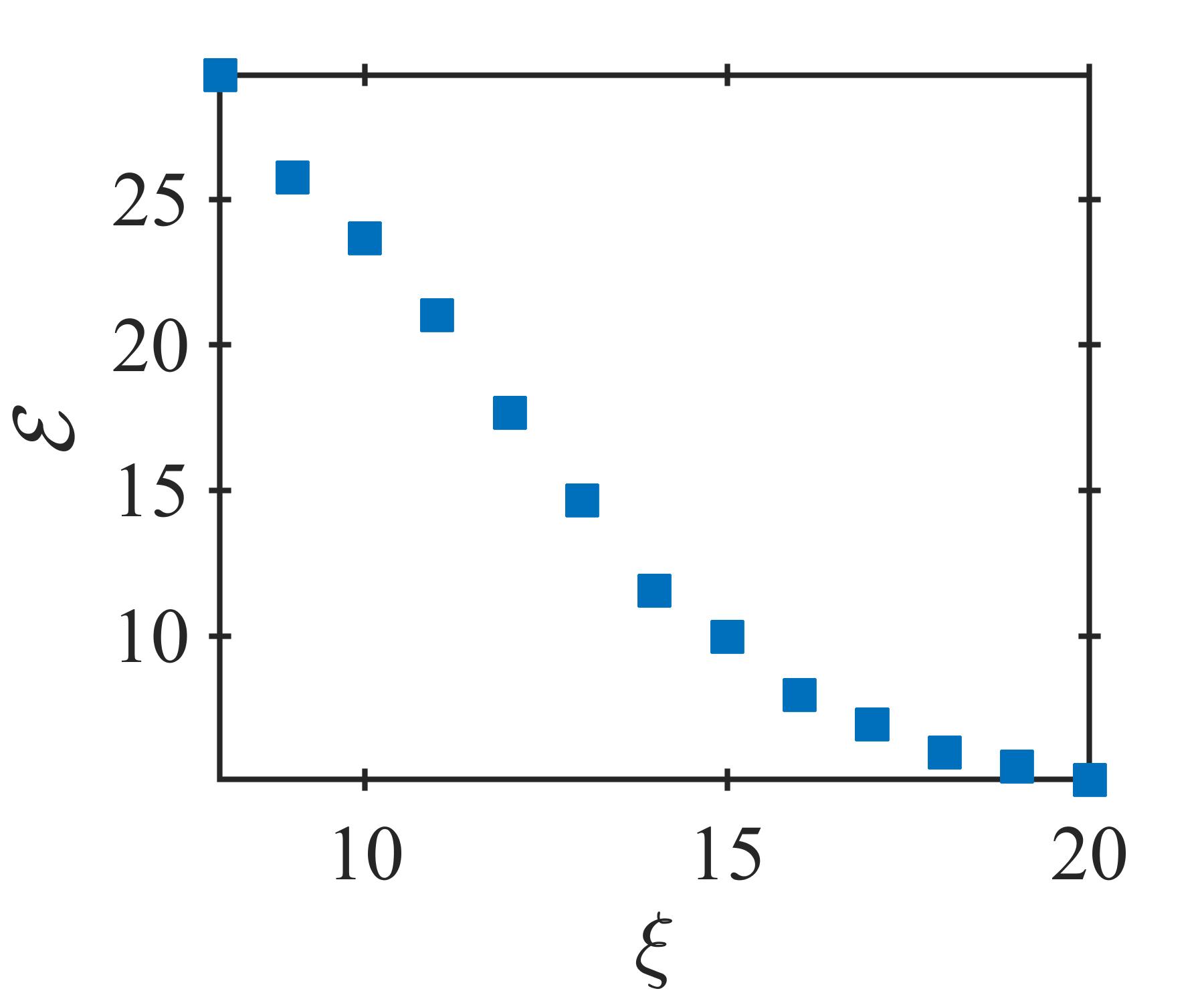}
  \caption{The Pareto front obtained by NSGA-II with $\xi_{limit} = 20$}
  \label{f:S4_P1}
\end{subfigure}%
\hspace{0.1\textwidth}
\begin{subfigure}{.33\textwidth}
  \centering
  \includegraphics[width=\textwidth]{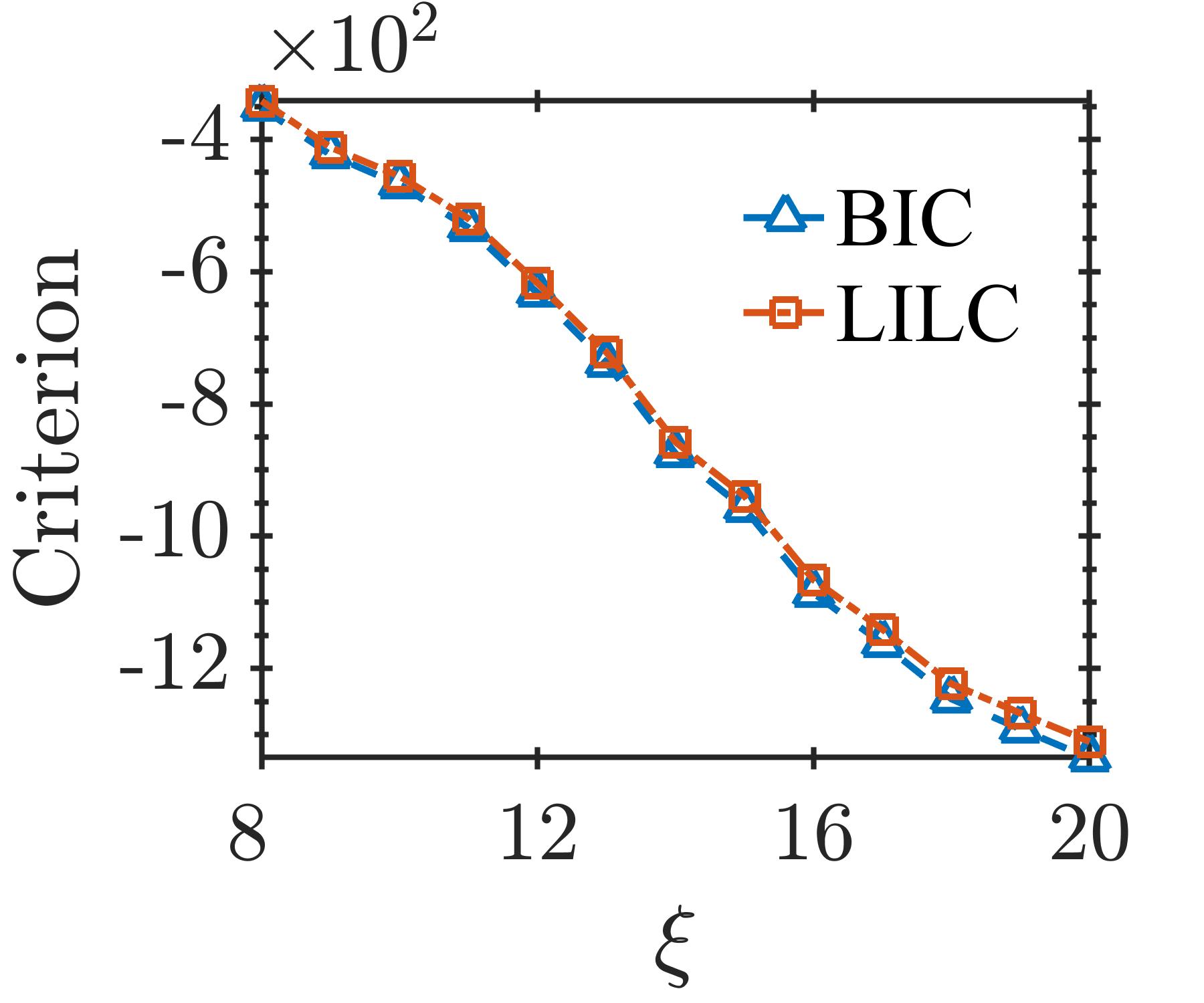}
  \caption{Information Criteria for ND structures obtained with $ \xi_{limit} = 20$}
  \label{f:S4_1}
\end{subfigure}
\smallskip
\begin{subfigure}{.33\textwidth}
  \centering
  \includegraphics[width=\textwidth]{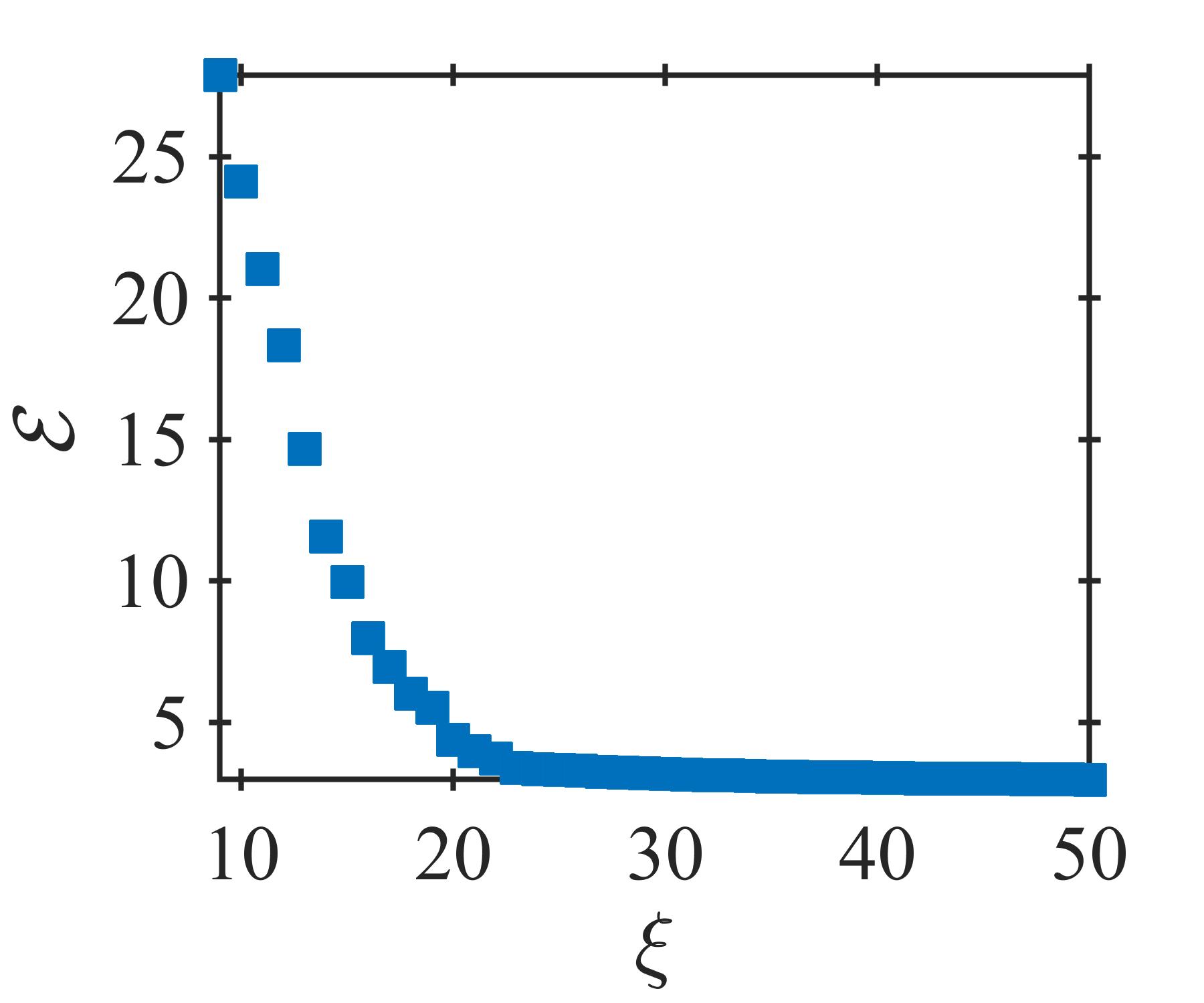}
  \caption{The Pareto front obtained by NSGA-II with $\xi_{limit} = 50$}
  \label{f:S4_P2}
\end{subfigure}%
\hspace{0.1\textwidth}
\begin{subfigure}{.33\textwidth}
  \centering
  \includegraphics[width=\textwidth]{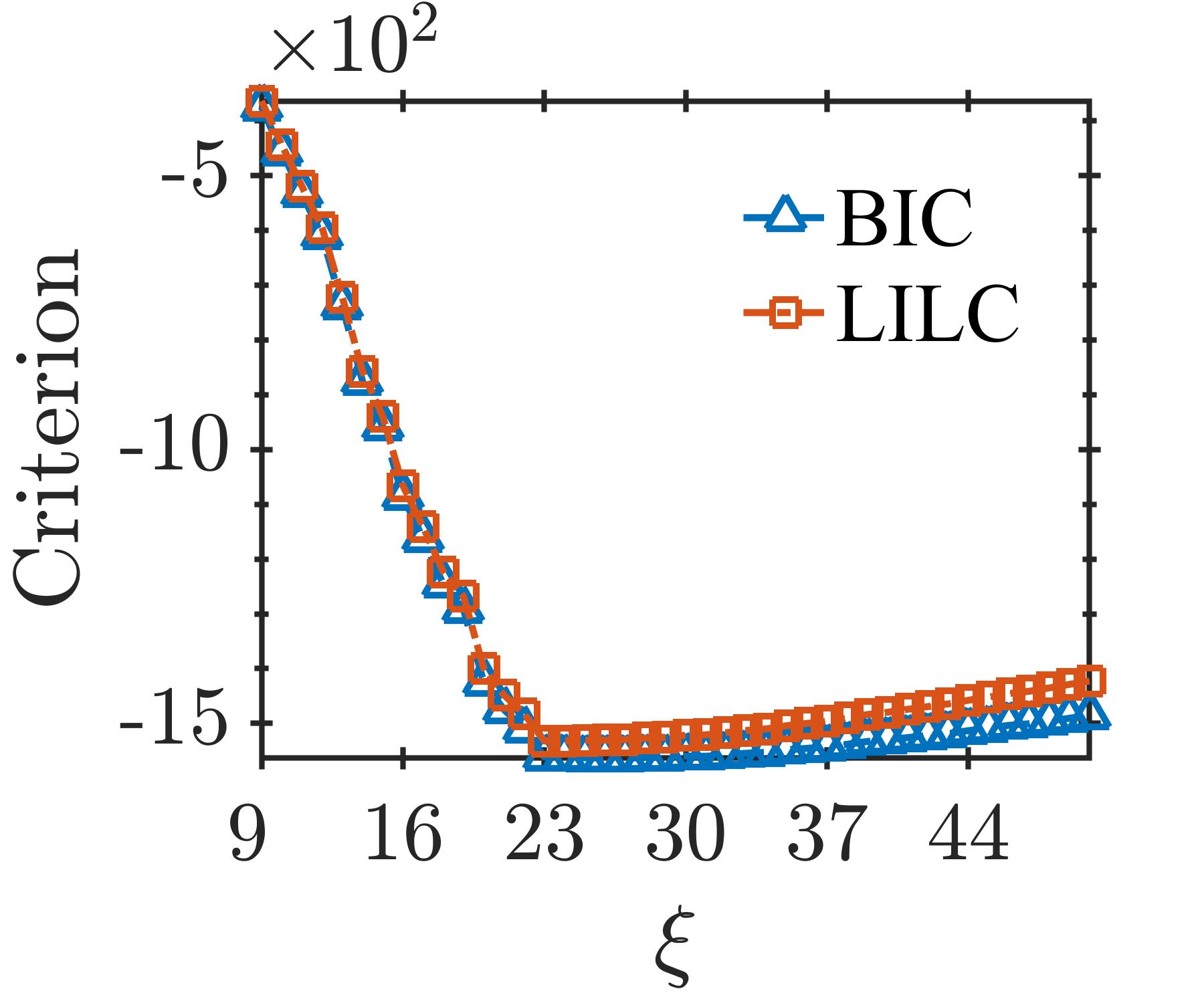}
  \caption{Information Criteria for ND structures obtained with $\xi_{limit} = 50$}
  \label{f:S4_2}
\end{subfigure}
\caption{An example to illustrate the Preference Update}
\label{f:S4Interval}
\end{figure*}

NSGA-II is applied to identify this system with $\{ \xi_{limit}, \mathcal{E}_{limit} \} = \{ 20, 30 \}$, as per the procedure outlined in Section~\ref{s:searchSetup} and Algorithm~\ref{al:moss}. The approximate Pareto front (APF) for this scenario is shown in Fig.~\ref{f:S4_P1}. Note that the specified cardinality limit in this case is less than the actual system cardinality, \textit{i.e.}, $\xi_{limit}<\xi^\ast$. Hence, the search is directed towards only the \textit{under-fitted} structures (see Section~\ref{s:QualMetric}). This is not apparent in APF shown in Fig.~\ref{f:S4_P1}. However, this can easily be detected by observing the behavior of classical Information Criteria (IC) (\textit{e.g.}, BIC, LILC) for the non-dominated structures. For instance, Fig.~\ref{f:S4_1} shows the IC obtained with the non-dominated structures. It is clear that the minimum value for both BIC and LILC is obtained at $\xi=20$. However, it is not clear whether this is the actual knee-point/plateau of these criteria. Hence, the cardinality preference is updated to $\xi_{limit}=50$ and the identification procedure is repeated. The consequent Pareto front and IC are shown respectively in Fig.~\ref{f:S4_P2} and~\ref{f:S4_2}. With the updated preference, a continual increase in information criteria is observed for $\xi \geq 24$. This, therefore, confirms that the plateau observed in the information criteria for $\xi \in [23,27]$ indeed points towards the true system cardinality. Hence, no further increase in $\xi_{limit}$ is required.

As illustrated by this example, evaluation of the specified \textit{a priori} preferences and any required updates can easily be accomplished by observing the behavior of the information criteria of the non-dominated structures. 
\subsection{Discrete Models for Nonlinear Continuous-Time Systems}
\label{s:ResDuff}

\begin{table}[!t]
\centering
\small
  \caption{Selected Models for the Duffing's Oscillator}%
  \label{t:duffmodel}%
  \begin{adjustbox}{max width=0.65\textwidth}
  \begin{threeparttable}
    \begin{tabular}{cccccc}
    \toprule
    \textbf{Model} & \makecell{$^\dagger$ \textbf{MMD} \\ \boldmath{$\mathcal{D}(\cdot)$}} & \makecell{$^\ddagger$ \textbf{MTD} \\ \boldmath{$\mathcal{R}(\cdot)$}} & \makecell{\textbf{Number of}\\ \textbf{Terms} (\boldmath$\xi$)} & \textbf{NMSE} (\boldmath$\mathcal{E}$) \\
    \midrule
     $\mathcal{M}_{D_1}$    & 0.07 & 3.07   & 5     & $1.98\times10^{-2}$ \\[1ex]
     $\mathcal{M}_{D_2}$    & 0.13 & 3.17   & 6     & $1.62\times10^{-2}$ \\[1ex]
     $\mathcal{M}_{D_3}$    & 0.19 & 3.18   & 7     & $3.97\times10^{-4}$ \\[1ex]
    \bottomrule
    \end{tabular}%
    \begin{tablenotes}
      \small
       \item $\dagger$ `$\mathcal{D}(\cdot)$' is determined following~(\ref{eq:MMD}) from the hypothetical ideal point: $\vec{J}^{\star}=\begin{Bmatrix} 4 & 2.16\times10^{-6} \end{Bmatrix}$  
       \item $\ddagger$ `$\mathcal{R}(\cdot)$' is determined following~(\ref{eq:MTDRank}) with $\vec{w}=[0.83, \ 0.17]$
    \end{tablenotes}
  \end{threeparttable}
 \end{adjustbox}
\end{table}

In this part of the investigation, identification of a discrete-time model for a continuous-time system is considered as a case study to evaluate the multi-objective framework further. In particular, the NARX model is fitted to identify the dynamics of \textit{Duffing}'s oscillator, which is given by,
\begin{linenomath*}
\begin{align}
	\label{eq:duffing}
	\ddot{y}(t) + 2\zeta\omega_n\dot{y}(t)+\omega_n^2y(t)+\omega_n^2\varepsilon y(t)^3-u(t)=0
\end{align}
\end{linenomath*}
For this purpose, a total of $1000$ data points are generated by exciting the system with a uniform white noise sequence, \textit{i.e.}, $u(\cdotp)\sim WUN(0,1)$, and sampling the output at $500 \ Hz$. The system parameters are set as follows: $\omega_n = 45\pi$, $\zeta=0.01$ and $\varepsilon=3$. The model set ($\mathcal{X}_{model}$) with a total of $286$ terms is obtained by specifying $[n_u, \ n_y, \ n_l] = [5, \ 5, \ 3]$ in~(\ref{eq:NARXmodel}). A total of $700$ data points is used for estimation purposes and the remainder is used for validation.

The NARX models for the Duffing's oscillator are identified following the procedure outlined in Section~\ref{s:searchSetup} and Algorithm~\ref{al:moss}. Note that, in the following, only the results obtained by SPEA-II are discussed, as it yields comparatively better approximation of the Pareto set (\textit{to be discussed in} Section~\ref{s:CompMOEA}). A total of $17$ non-dominated structures are identified by SPEA-II. Subsequently, the identified non-dominated structures are ranked based on Manhattan distance `$\mathcal{D}(\cdot)$' (see Section~\ref{s:MMD}) and the global rank `$\mathcal{R}(\cdot)$' (see Section~\ref{s:MTD}). On the basis of these rankings, three non-dominated structures are selected for further analysis. Table~\ref{t:duffmodel} shows the objective functions of the selected non-dominated structures along with the corresponding Manhattan distance and the global rank. Note that predictive performance, shown in Table~\ref{t:duffmodel}, is determined over the model-predictive output. The selected structures and the corresponding coefficients are given by the following models:
\begin{linenomath*}
\begin{small}                                     
\begin{align}
    \label{eq:md1}                                  
    \mathcal{M}_{D_1}: \ y(k) = & \ 1.9152 \, y(k-1) - 0.99436 \, y(k-2) +                                1.983 \times 10^{-6} \, u(k-1) + 1.9792\times 10^{-6} \, u(k-2) -0.23154 \, y(k-1)^3\\ 
   \label{eq:md2}                                  
   \mathcal{M}_{D_2}: \ y(k) = & \ 1.9152 \, y(k-1) - 0.99436 \, y(k-2) +                            1.983\times 10^{-6} \, u(k-1) + 1.9792\times 10^{-6} \, u(k-2) - 0.22981 \,y(k-1)^3 \nonumber \\
                        & \ - 3.4686\times 10^{-3} \, y(k-3)^3 \\
   \label{eq:md3}  
   \mathcal{M}_{D_3}: \ y(k) = & \ 1.9152 \, y(k-1) - 0.99436 \, y(k-2) +                                1.983 \times 10^{-6} \, u(k-1) + 1.9792\times 10^{-6} \, u(k-2) - 0.25637 \, y(k-1)^3 \nonumber\\
                        & \ + 5.4467\times 10^{-2} \, y(k-3)y(k-1)^2 - 3.191\times 10^{-2} \, y(k-1)y(k-3)^2
\end{align} 
\end{small} 
\end{linenomath*}
\begin{figure*}[!b]
\centering
\small
\begin{subfigure}{.248\textwidth}
  \centering
  \includegraphics[width=\textwidth]{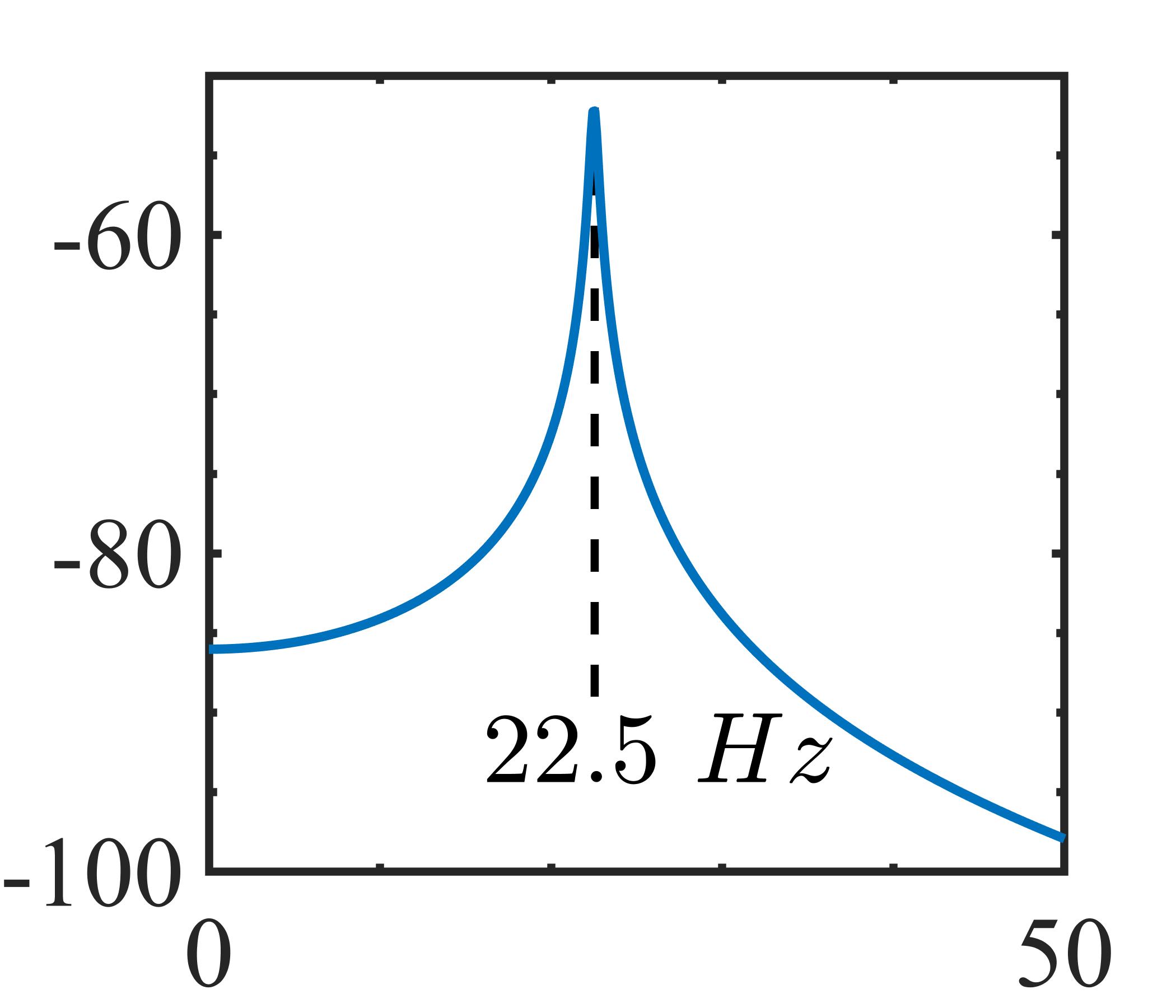}
  \caption{Duffing's Oscillator}
  \label{f:frfDuffC1}
\end{subfigure}%
\begin{subfigure}{.248\textwidth}
  \centering
  \includegraphics[width=\textwidth]{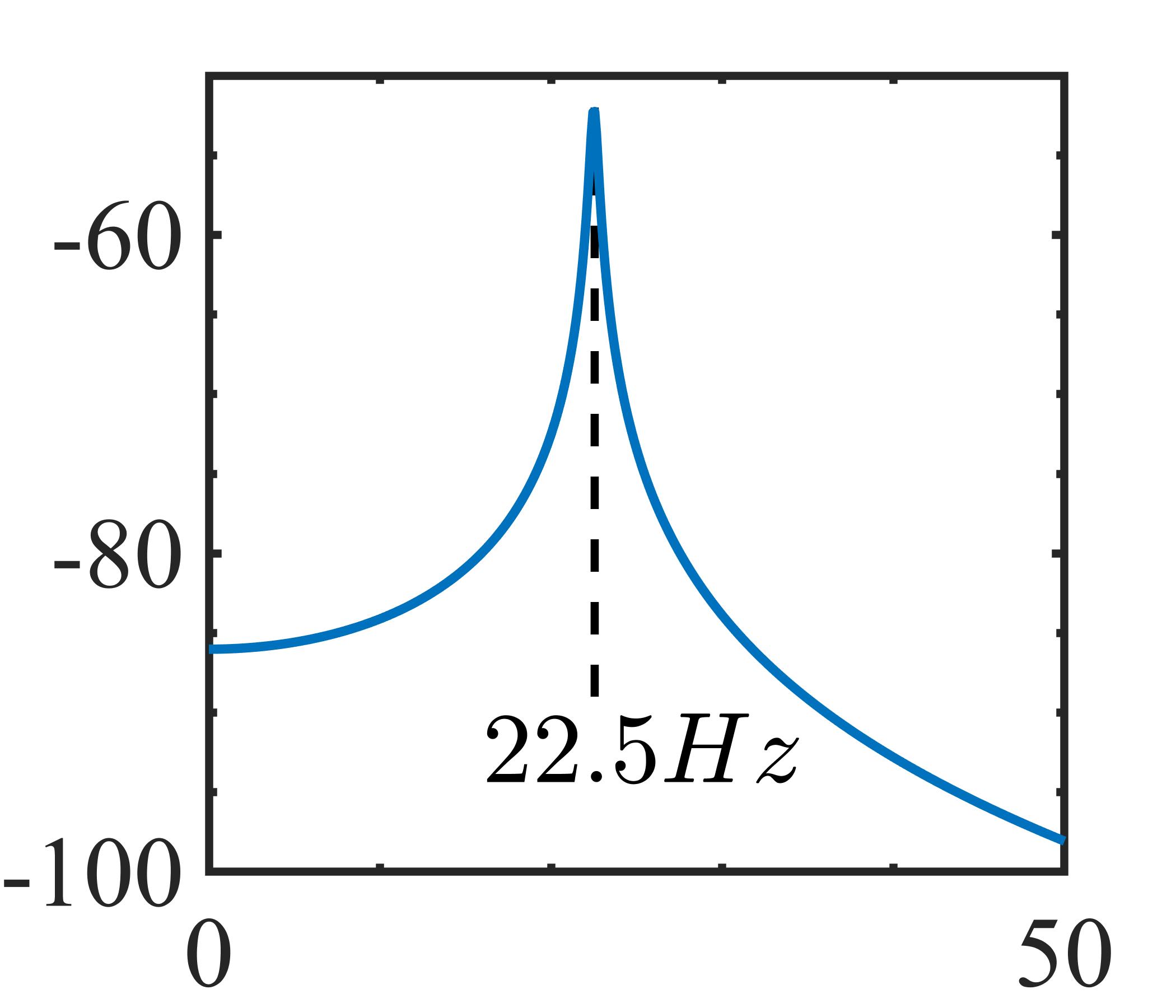}
  \caption{$\mathcal{M}_{D_1}$}
  \label{f:frfDuffD1M1}
\end{subfigure}
\begin{subfigure}{.248\textwidth}
  \centering
  \includegraphics[width=\textwidth]{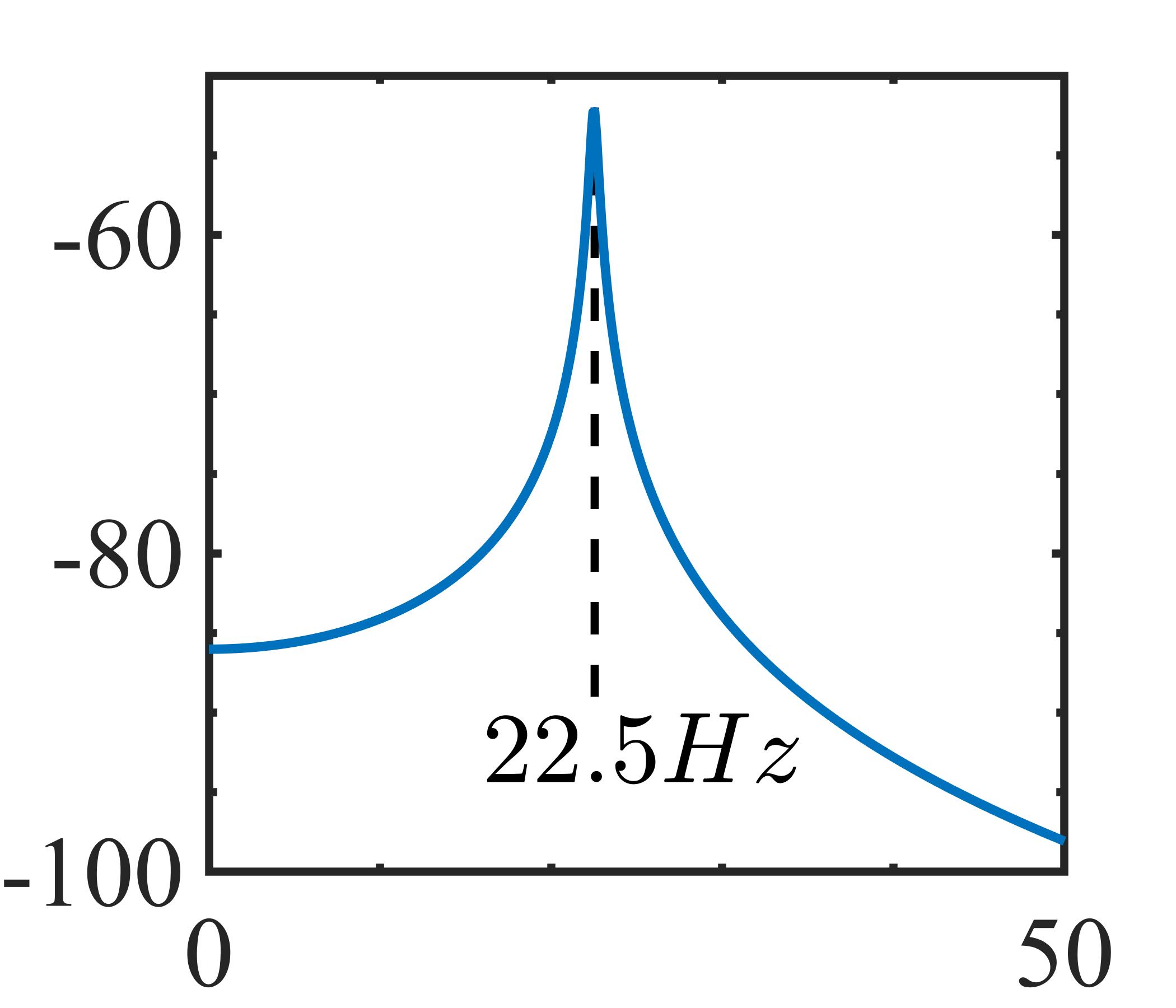}
  \caption{$\mathcal{M}_{D_2}$}
  \label{f:frfDuffD1M2}
\end{subfigure}%
\begin{subfigure}{.248\textwidth}
  \centering
  \includegraphics[width=\textwidth]{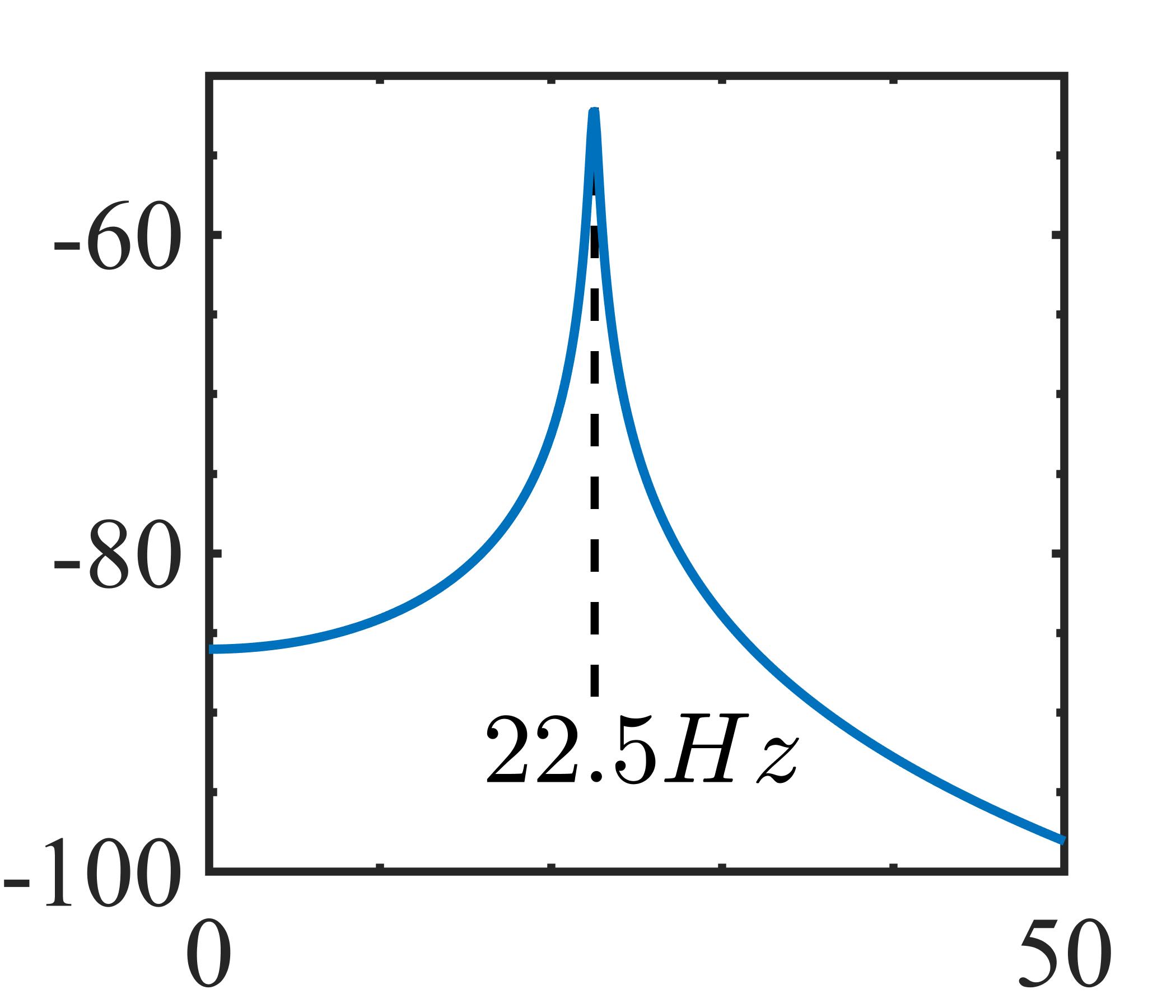}
  \caption{$\mathcal{M}_{D_3}$}
  \label{f:frfDuffD1M3}
\end{subfigure}
\caption{Linear Frequency Response of the Duffing's oscillator and the identified models. X-axis and Y-axis respectively denote frequency in `$Hz$' and magnitude in `$dB$'.}
\label{f:gfrfduff1}
\end{figure*}
\begin{figure*}[!t]
\centering
\begin{subfigure}{.38\textwidth}
  \centering
  \includegraphics[width=\textwidth]{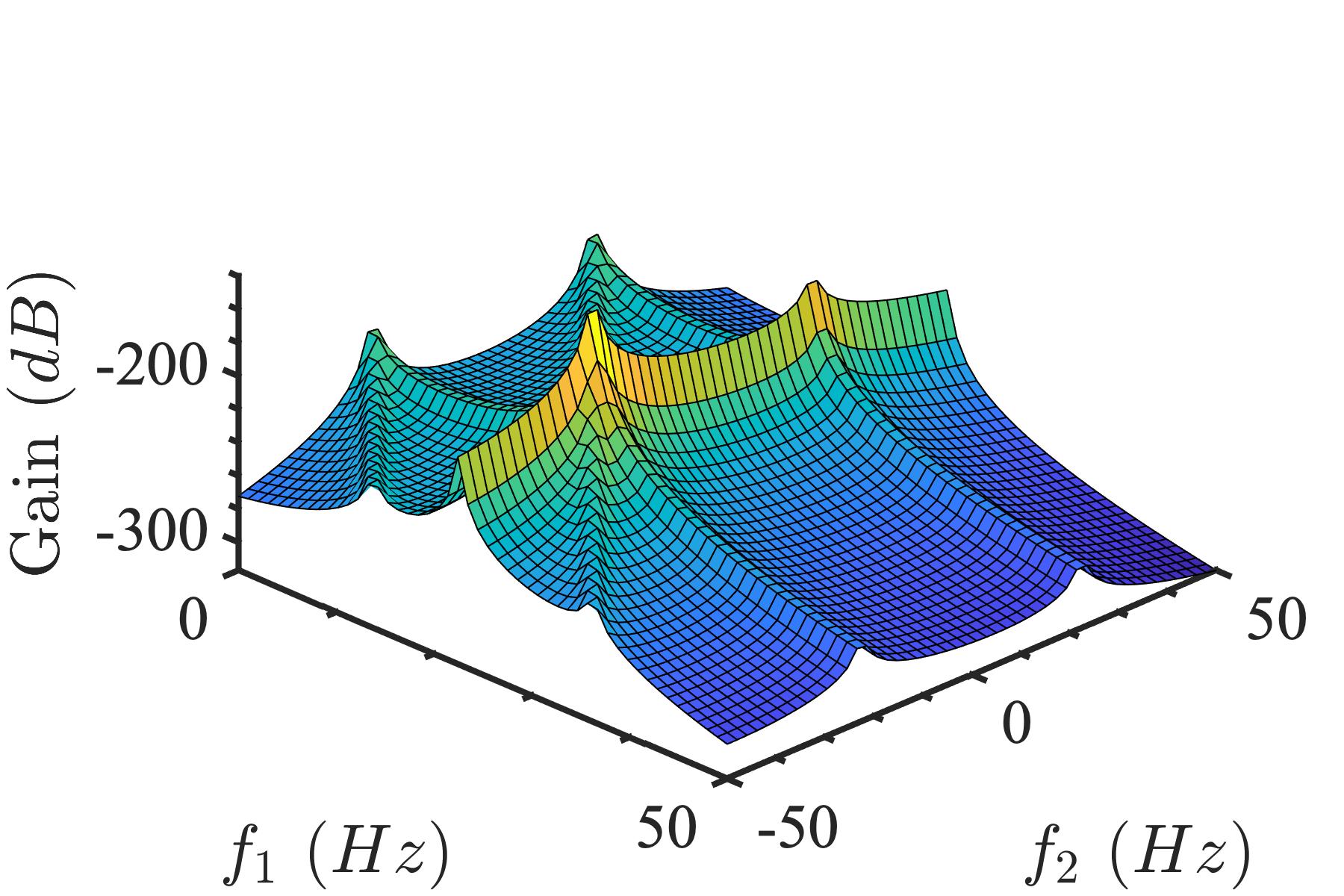}
  \caption{Duffing's Oscillator}
  \label{f:frfDuffC3}
\end{subfigure}%
\hspace{0.1\textwidth}
\begin{subfigure}{.38\textwidth}
  \centering
  \includegraphics[width=\textwidth]{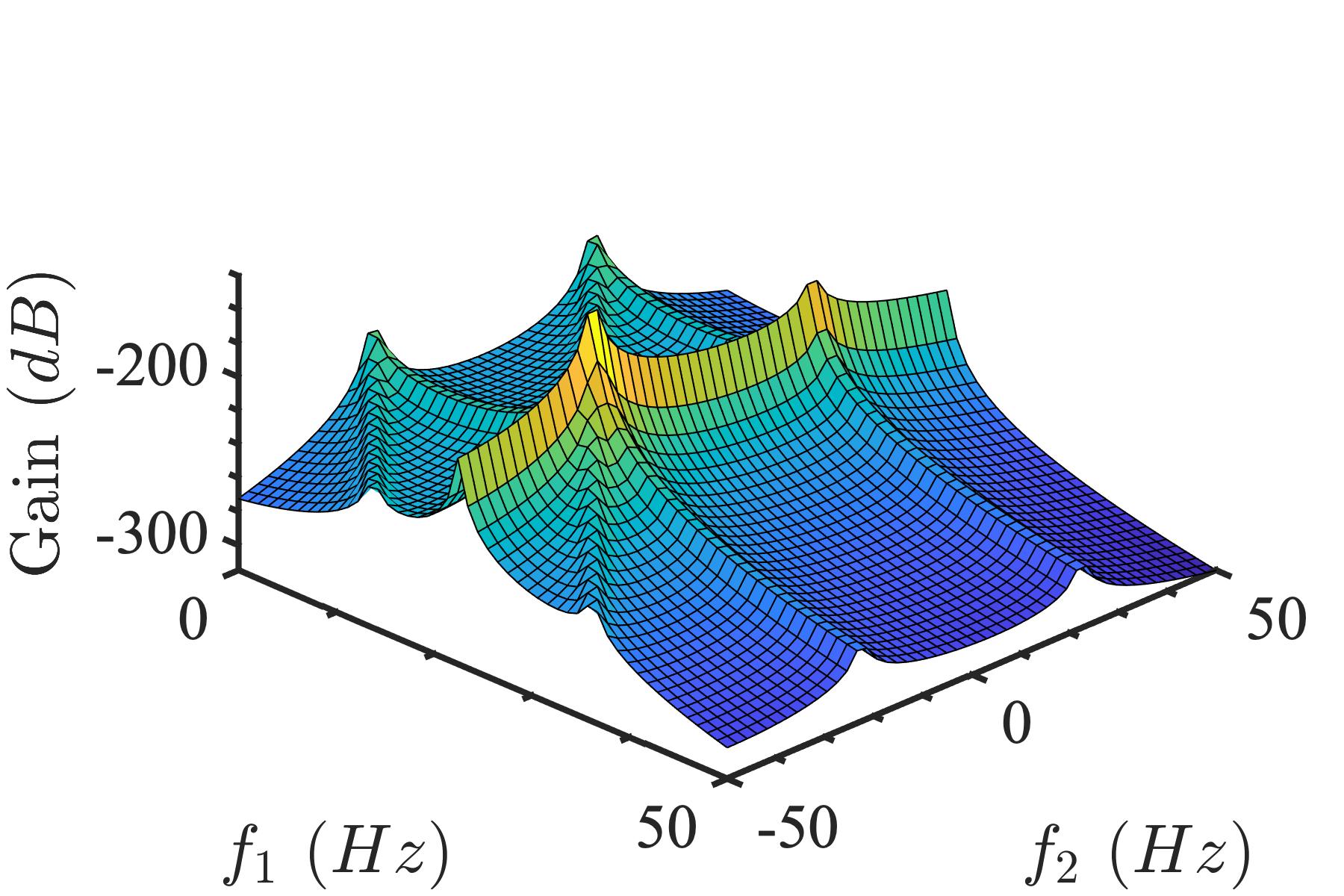}
  \caption{$\mathcal{M}_{D_1}$}
  \label{f:frfDuffD3M1}
\end{subfigure}
\begin{subfigure}{.38\textwidth}
  \centering
  \includegraphics[width=\textwidth]{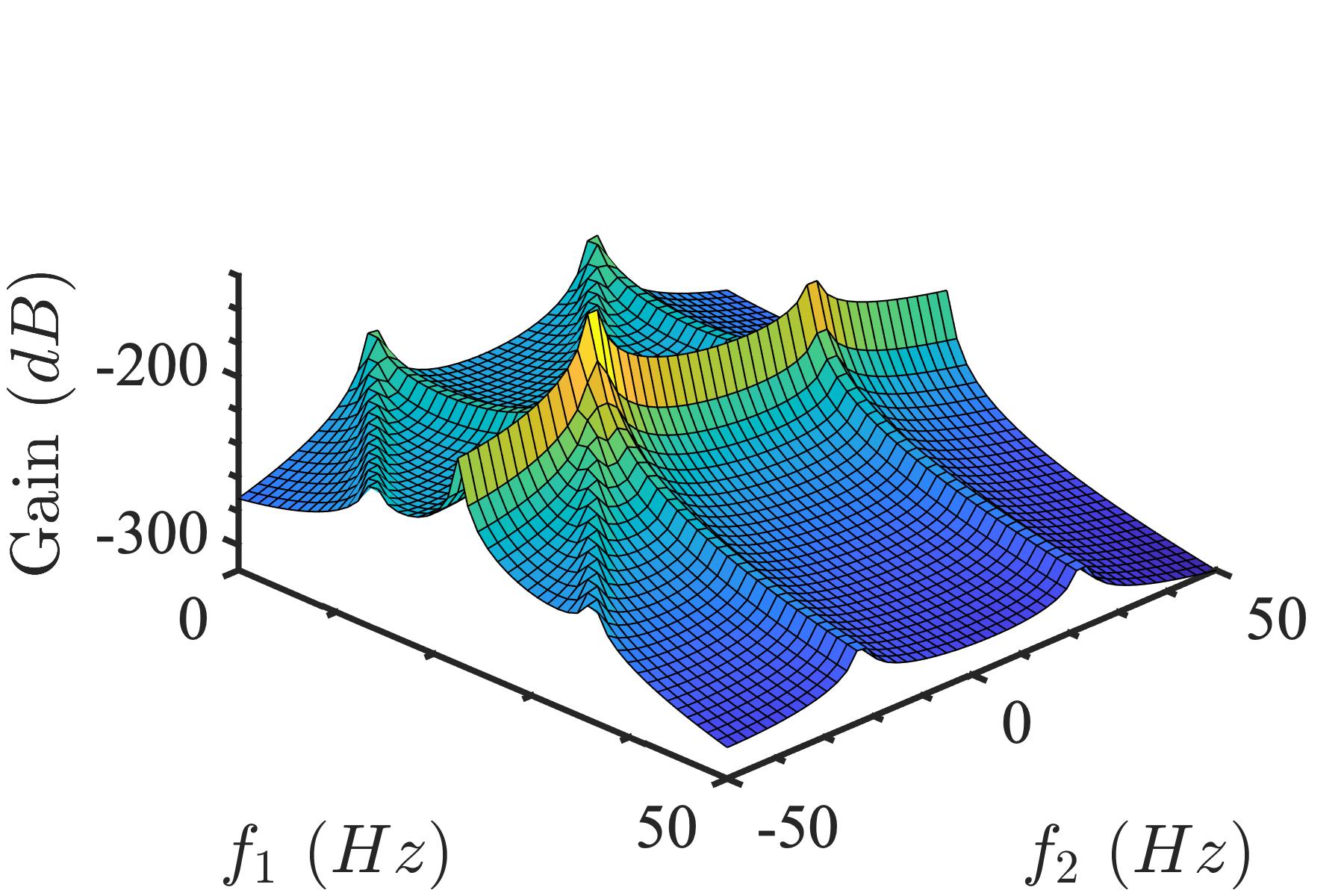}
  \caption{$\mathcal{M}_{D_2}$}
  \label{f:frfDuffD3M2}
\end{subfigure}%
\hspace{0.1\textwidth}
\begin{subfigure}{.38\textwidth}
  \centering
  \includegraphics[width=\textwidth]{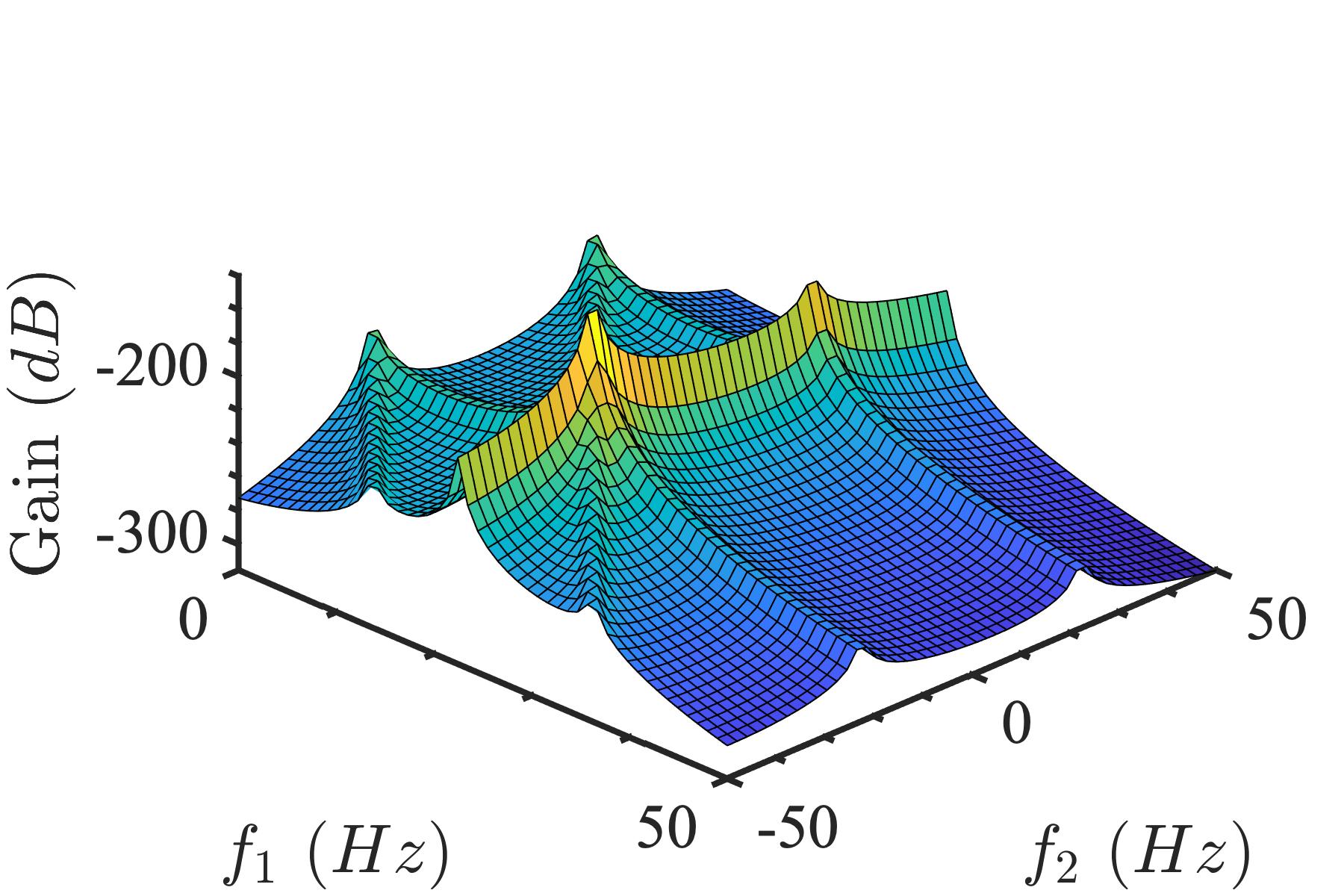}
  \caption{$\mathcal{M}_{D_3}$}
  \label{f:frfDuffD3M3}
\end{subfigure}
\caption{Third Order Frequency Response of the Duffing's oscillator and the identified models with $f_3=f_1$.}
\label{f:gfrfduff3}
\end{figure*}
\begin{figure*}[!t]
\centering
\begin{subfigure}{.33\textwidth}
  \centering
  \includegraphics[width=\textwidth]{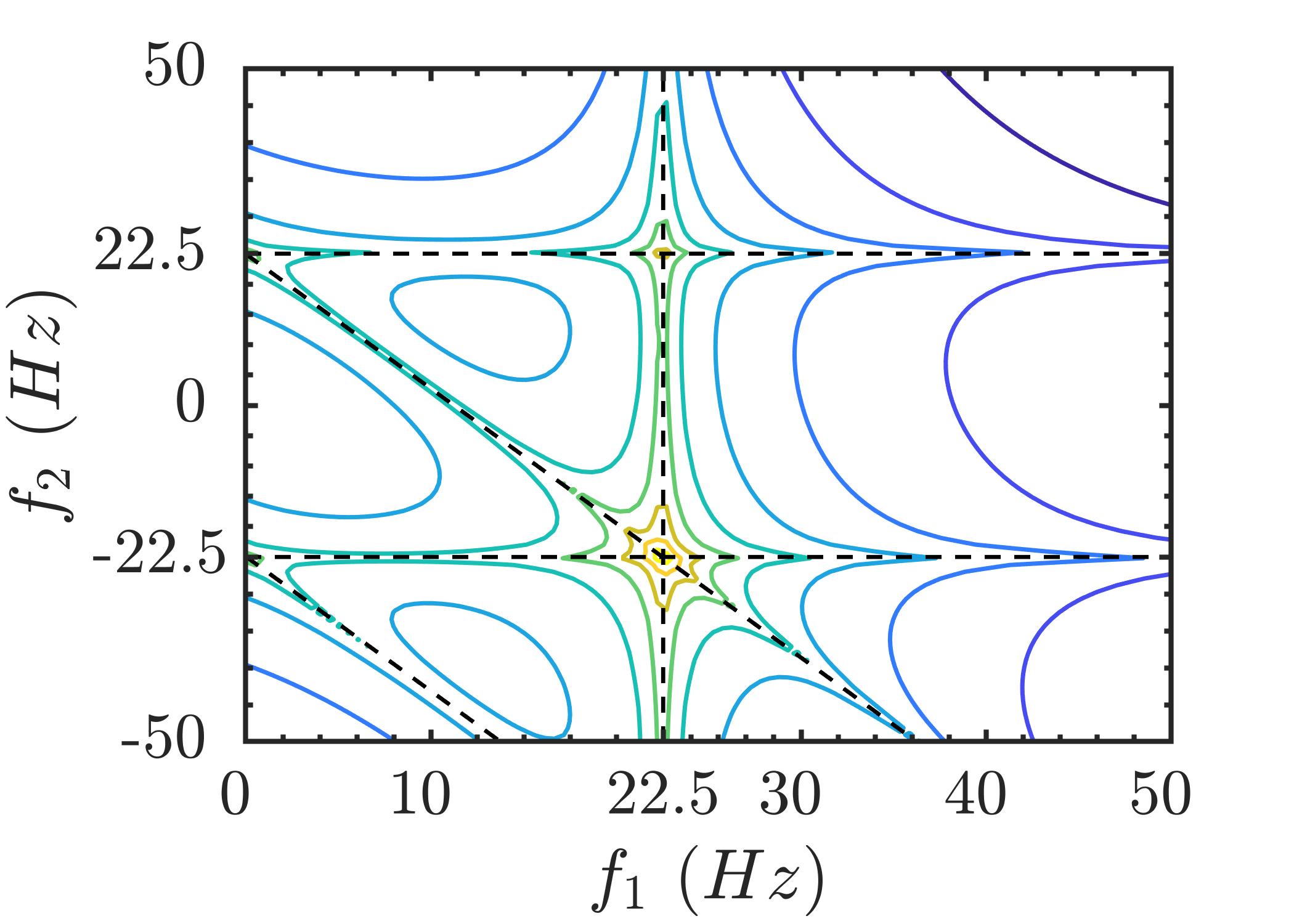}
  \caption{Duffing's Oscillator}
  \label{f:frfDuffC2}
\end{subfigure}%
\hspace{0.1\textwidth}
\begin{subfigure}{.33\textwidth}
  \centering
  \includegraphics[width=\textwidth]{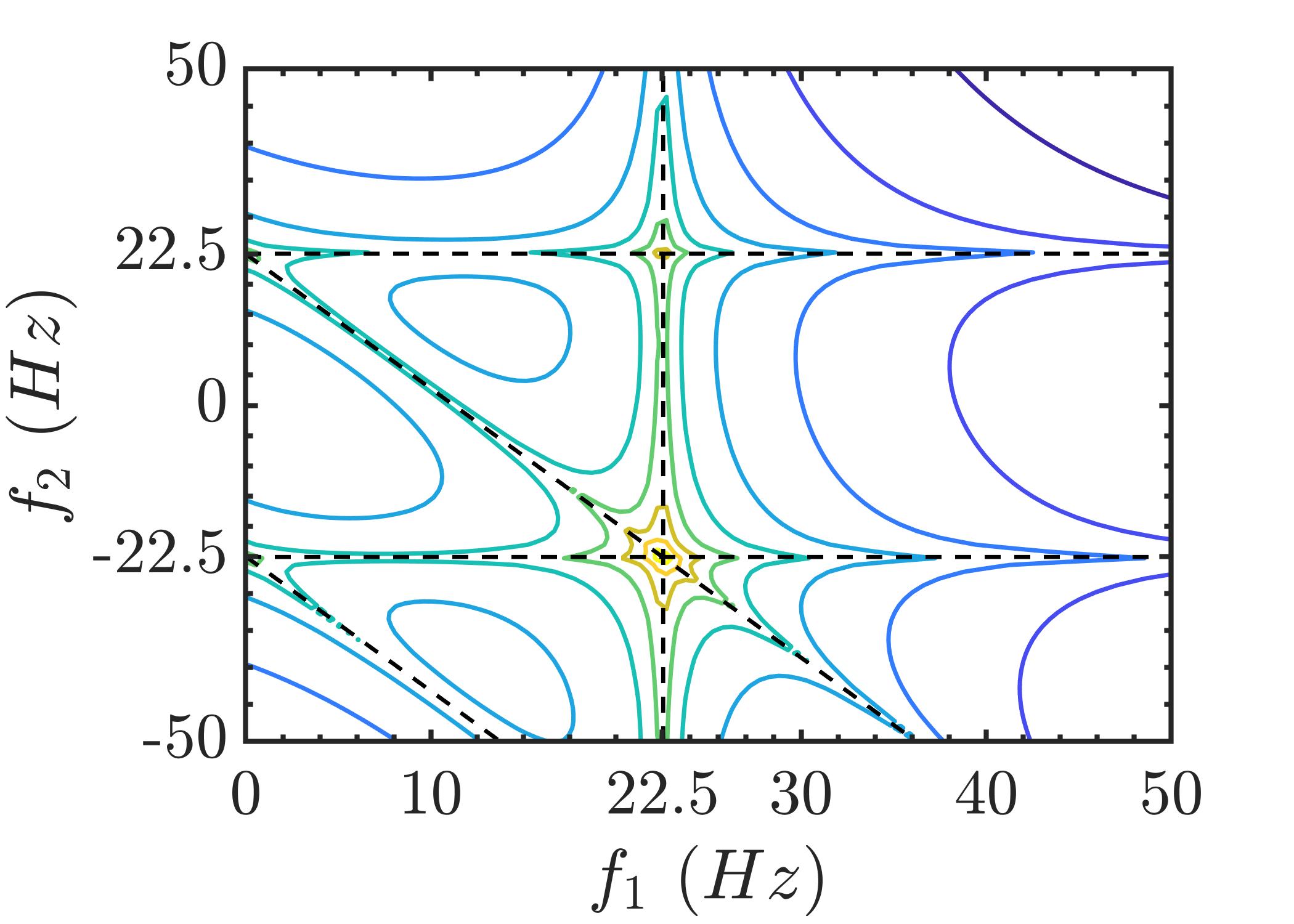}
  \caption{$\mathcal{M}_{D_1}$}
  \label{f:frfDuffD2M1}
\end{subfigure}
\begin{subfigure}{.33\textwidth}
  \centering
  \includegraphics[width=\textwidth]{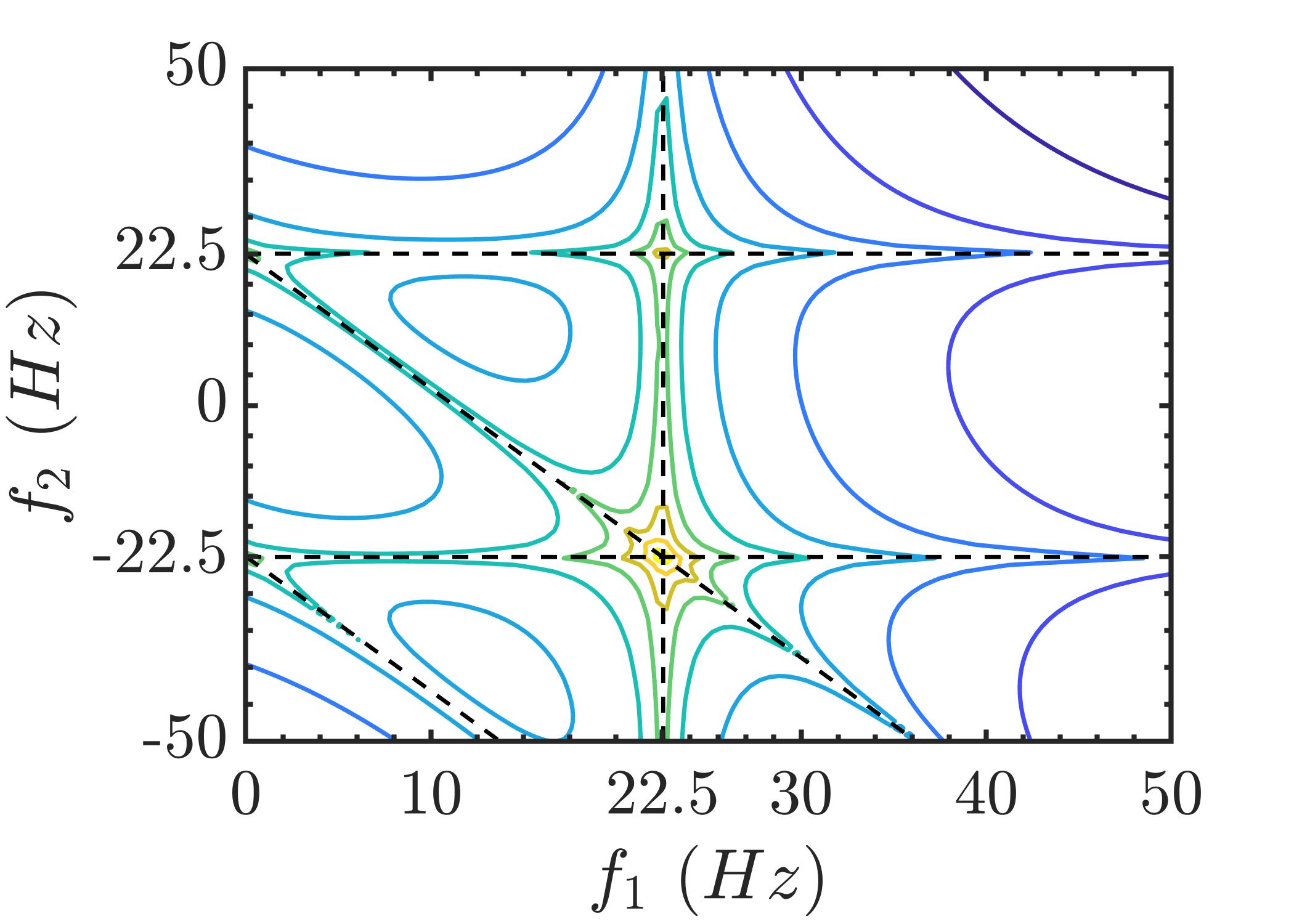}
  \caption{$\mathcal{M}_{D_2}$}
  \label{f:frfDuffD2M2}
\end{subfigure}%
\hspace{0.1\textwidth}
\begin{subfigure}{.33\textwidth}
  \centering
  \includegraphics[width=\textwidth]{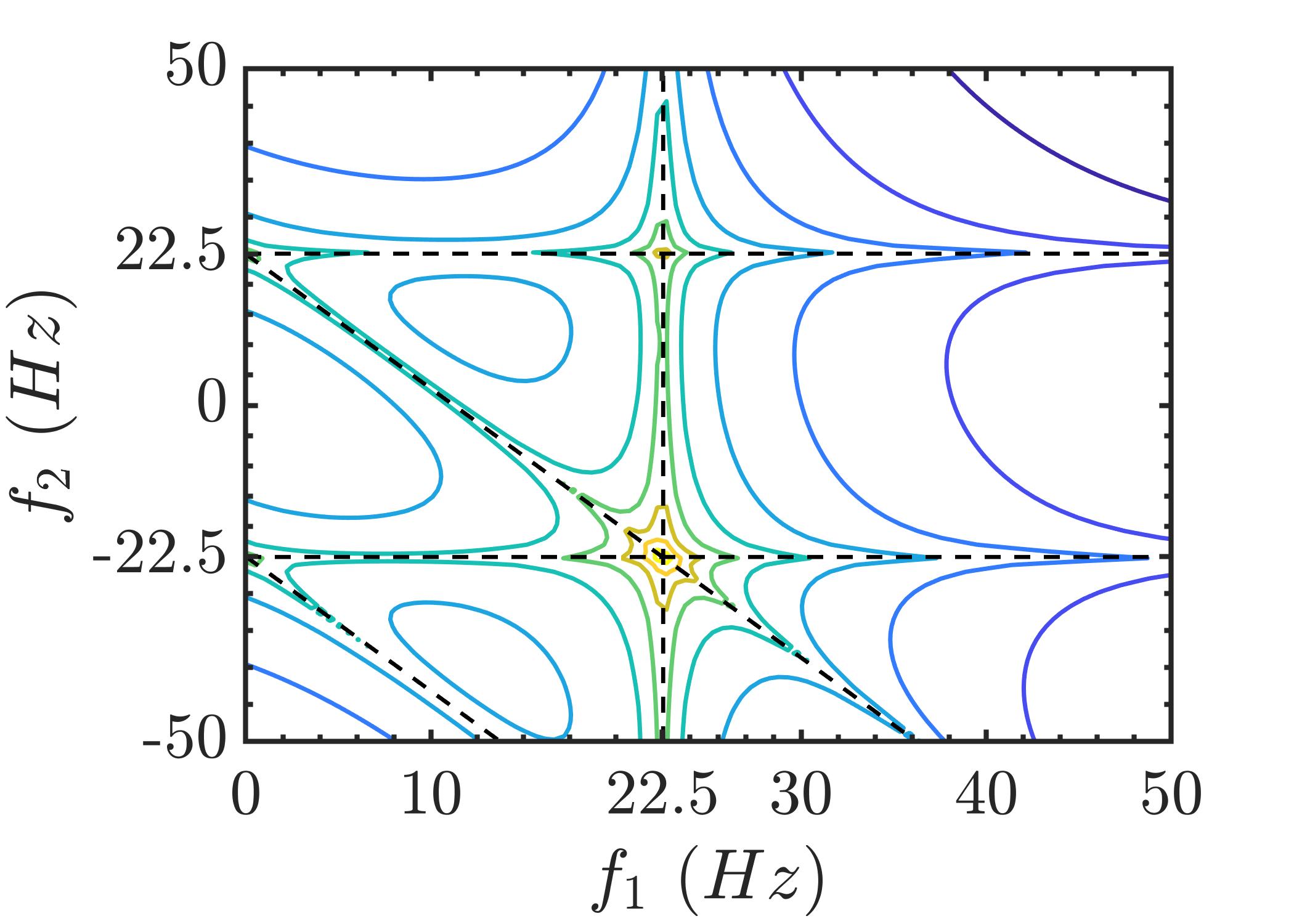}
  \caption{$\mathcal{M}_{D_3}$}
  \label{f:frfDuffD2M3}
\end{subfigure}
\caption{Third Order GFRF Contours of the Duffing's oscillator and the identified models with $f_3=f_1$. Note that the ridges align at $f_1+f_2+f_3= \pm 22.5$.}
\label{f:gfrfduff2}
\end{figure*}
It is clear that all the identified models could yield practically acceptable predictive performance and could be used to describe the dynamics of the Duffing's oscillator. This is expected as the discrete model for the continuous-time system is not unique~\cite{Billings:2013}. 

To investigate whether the identified models indeed describe the same dynamics, the Generalized Frequency Response Functions (GFRFs) are computed from these models by the harmonic probing procedure~\cite{Billings:Tsang:1989} and are compared with that of the continuous-time model. Note that the GFRFs are invariant descriptors of the system dynamics. This implies that if the identified models are correct the corresponding frequency response function should match with that of the continuous-time model.The GFRFs can easily be determined for differential equations models, the polynomial NARX models and the other system representations by mapping them into the frequency domain~\cite{Billings:2013,Billings:Tsang:1989,Billings:Peyton:1990}.


The linear and the third order GFRFs computed from the continuous-time model and the identified discrete models are shown in Fig.~\ref{f:gfrfduff1}-\ref{f:gfrfduff2}. We begin by identifying the unique frequency domain `\textit{traits}' of the Duffing's oscillator from the GFRFs shown in Fig.~\ref{f:frfDuffC1}, \ref{f:frfDuffC3} and~\ref{f:frfDuffC2}. As seen Fig.~\ref{f:frfDuffC1}, the linear resonant frequency of Duffing's oscillator is at $22.5 \ Hz$. Further, the third order GFRF, in Fig.~\ref{f:frfDuffC3} and~\ref{f:frfDuffC2}, is characterized by the presence of several \textit{peaks} and \textit{ridges}. The peaks occur whenever the system is excited by a three tone signal with $f_1=f_2=f_3=\pm 22.5 \ Hz$. Further, the ridges occur in either of the following conditions: $f_1$, $f_2$ or $f_3 = \pm 22.5 \ Hz$ or $f_1+f_2+f_3=\pm 22.5$.

It is interesting to see that all the identified models ($\mathcal{M}_{D_1}$, $\mathcal{M}_{D_2}$, $\mathcal{M}_{D_3}$) could replicate the aforementioned frequency domain behavior of the Duffing's oscillator (as seen in Fig.~\ref{f:gfrfduff1}-\ref{f:gfrfduff2}). Hence, this gives conclusive evidence that these apparently different models indeed describe the dynamics of the Duffing's oscillator. However, since the model $\mathcal{M}_{D_1}$ enables relatively compact system description, it is recommended following the principle of parsimony. 

\begin{figure*}[!t]
\centering
\small
\begin{subfigure}{.32\textwidth}
  \centering
  \includegraphics[width=\textwidth]{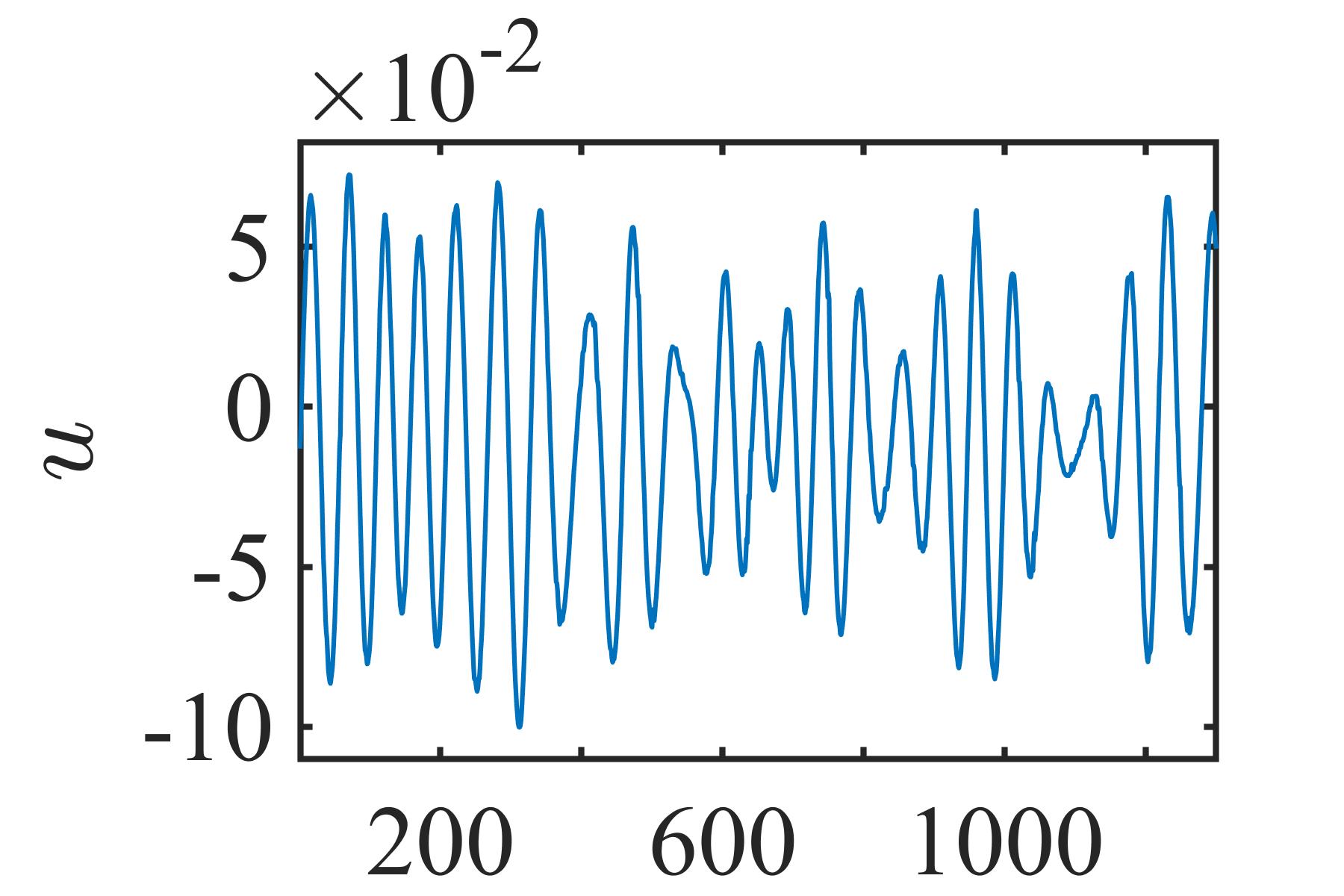}
  \caption{Flow velocity (input), $u$}
  \label{f:uwav}
\end{subfigure}%
\begin{subfigure}{.32\textwidth}
  \centering
  \includegraphics[width=\textwidth]{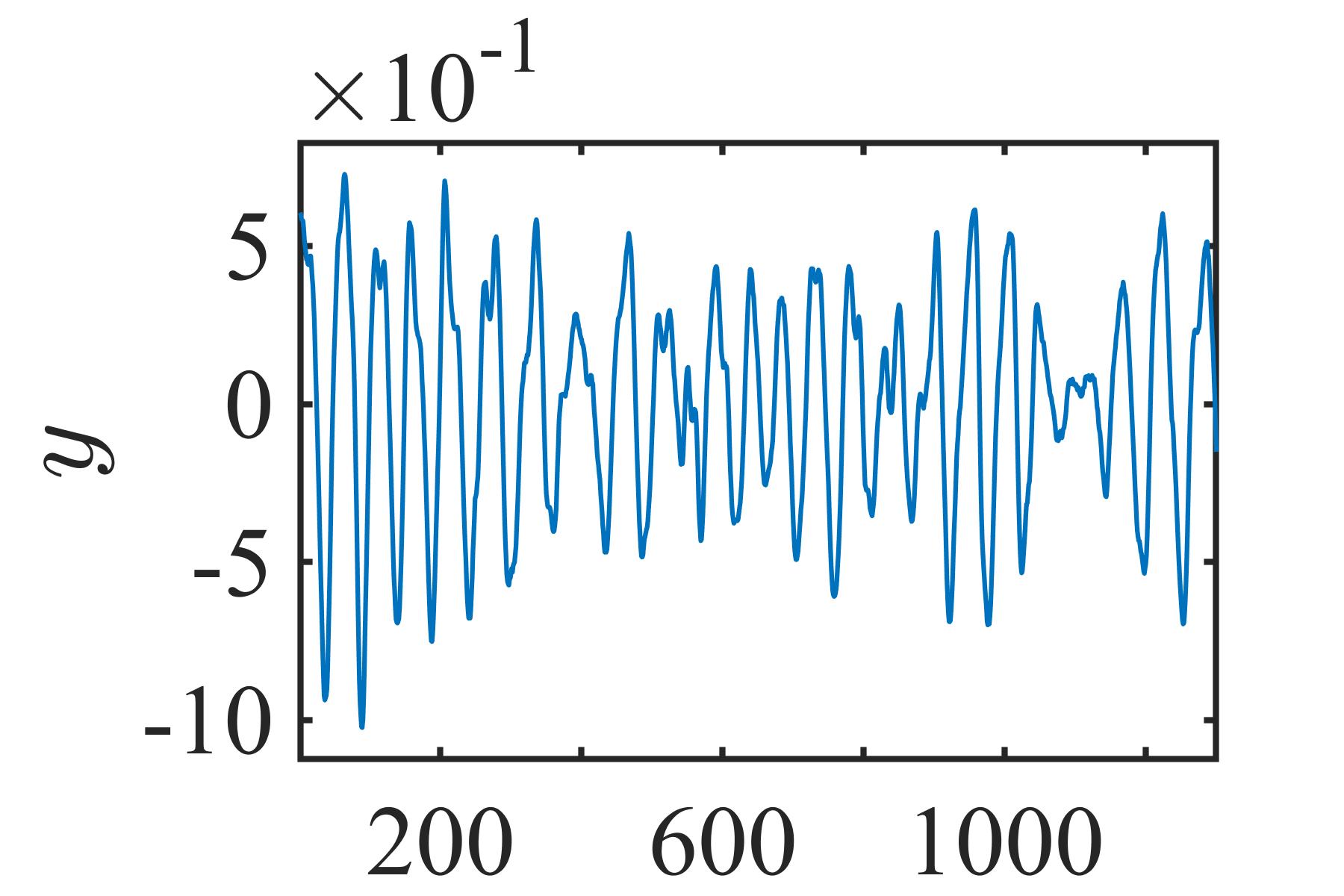}
  \caption{Wave force (output), $y$}
  \label{f:ywav}
\end{subfigure}
\begin{subfigure}{.32\textwidth}
  \centering
  \includegraphics[width=\textwidth]{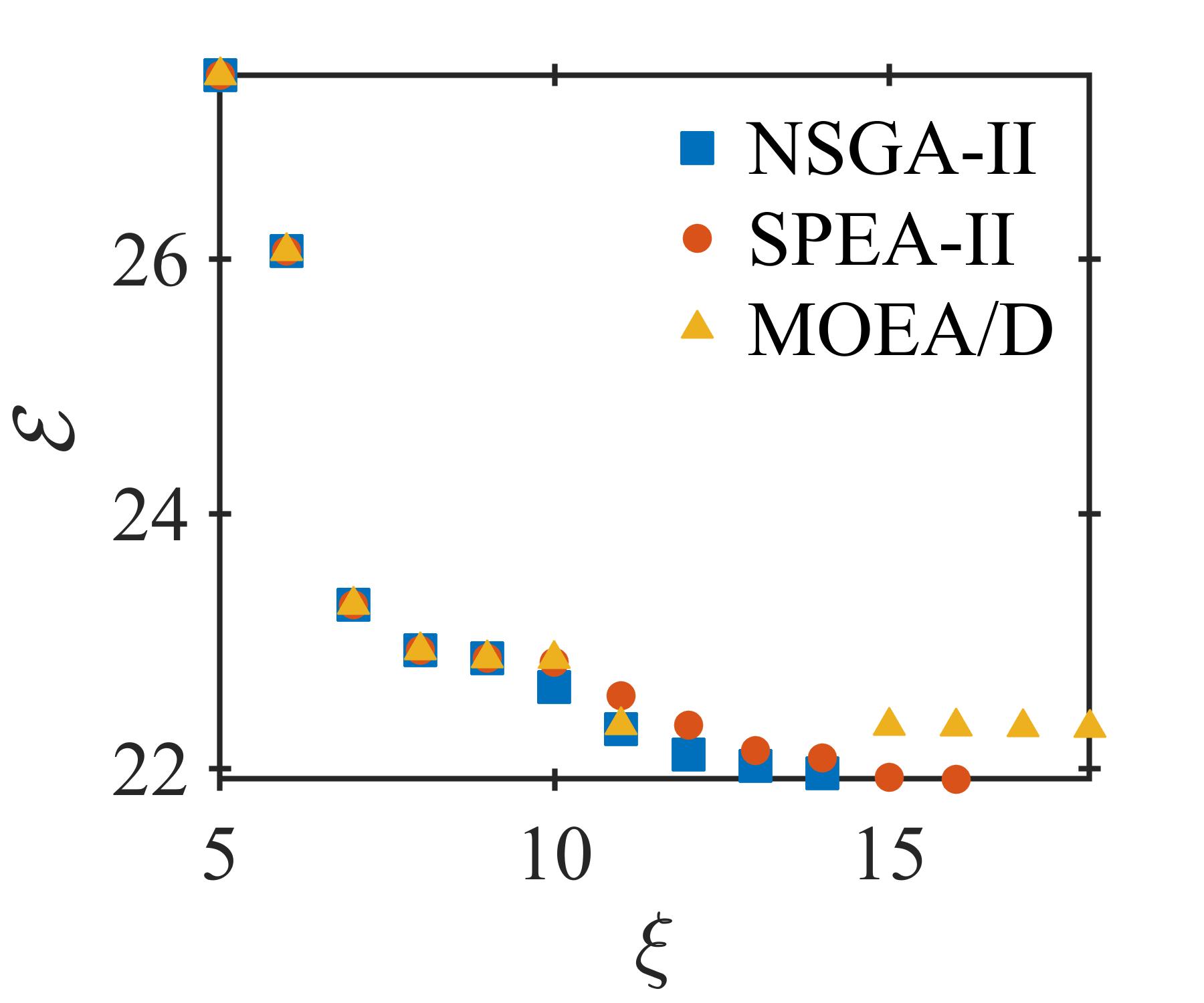}
  \caption{The approximate Pareto fronts}
  \label{f:frontwav}
\end{subfigure}
\caption{The nonlinear wave-force data.}
\label{f:wavdata}
\end{figure*}
\subsection{Identification of nonlinear wave forces}
\label{s:wav}

Next, we focus on a practical problem from the field of computational hydrodynamics which involves modelling of wave forces acting on fixed bodies in a viscous fluid. In particular, the objective is to capture the wave force dynamics from the flow velocity (input, $u$) and the wave force (output, $y$) histories~\cite{Swain:Billings:Stansby:1998}. A total of $1400$ data points are used for the identification purposes, which are shown in Fig.~\ref{f:wavdata}. Further details about this dataset can be found in~\cite{Swain:Billings:Stansby:1998}.

The models for wave-force dynamics are identified following the procedure outlined in Section~\ref{s:searchSetup} and Algorithm~\ref{al:moss}. To this end, the model set ($\mathcal{X}_{model}$) of $286$ NARX terms is generated by the following specification in~(\ref{eq:NARXmodel}): $[n_u,n_y,n_l]=[5,5,3]$. A total of $1000$ data points are being used for estimation purposes and the rest are used for validation. With these specifications, the approximate Pareto fronts (APF) are shown in Fig.~\ref{f:frontwav}. It is clear that NSGA-II could yield better structures for the wave-force data (\textit{to be discussed in} Section~\ref{s:CompMOEA}). Hence, only the results obtained by NSGA-II are discussed in the following. 
\begin{table}[!t]
\centering
\small
\caption{Selected Models for the Wave-force Dynamics}
\label{t:wavndsol}%
\begin{adjustbox}{max width=0.75\textwidth}
\begin{threeparttable}
    \begin{tabular}{cccccc}
    \toprule
    \multirow{2}{*}{\textbf{Model}} & \multirow{2}{*}{\makecell{$^\dagger$ \textbf{MMD} \\ \boldmath{$\mathcal{D}(\cdot)$}}} & \multirow{2}{*}{\makecell{$^\ddagger$ \textbf{MTD} \\ \boldmath{$\mathcal{R}(\cdot)$}}} & \multirow{2}{*}{\makecell{\textbf{Number of}\\ \textbf{Terms} (\boldmath$\xi$)}} & \multicolumn{2}{c}{\textbf{Prediction Error }} \\
    \cmidrule{5-6}          &       &       & & \textbf{NMSE} (\boldmath$\mathcal{E}$) & \textbf{MSE} \\
    \midrule
     $\mathcal{M}_{W_1}$    & 0.38   & 2.31 & 8     & 22.93 &  $1.36\times 10^{-2}$\\[1ex]
     $\mathcal{M}_{W_2}$    & 0.47   & 2.19 & 9     & 22.87 &  $1.35\times 10^{-2}$\\[1ex]
     $\mathcal{M}_{W_3}$    & 0.57   & 2.04 & 10    & 22.64 &  $1.32\times 10^{-2}$\\[1ex]
    \bottomrule
    \end{tabular}%
    \begin{tablenotes}
      \small
       \item $\dagger$ `$\mathcal{D}(\cdot)$' is determined following~(\ref{eq:MMD}) from the hypothetical ideal point: $\vec{J}^{\star}=\begin{Bmatrix} 5 & 21.96 \end{Bmatrix}$; $\ddagger$ `$\mathcal{R}(\cdot)$' is determined following~(\ref{eq:MTDRank}) with $\vec{w}=[0.83, \ 0.17]$
    \end{tablenotes}
 \end{threeparttable}
\end{adjustbox} 
\end{table}

\begin{figure*}[!t]
\centering
\small
\begin{subfigure}{.3\textwidth}
  \centering
  \includegraphics[width=\textwidth]{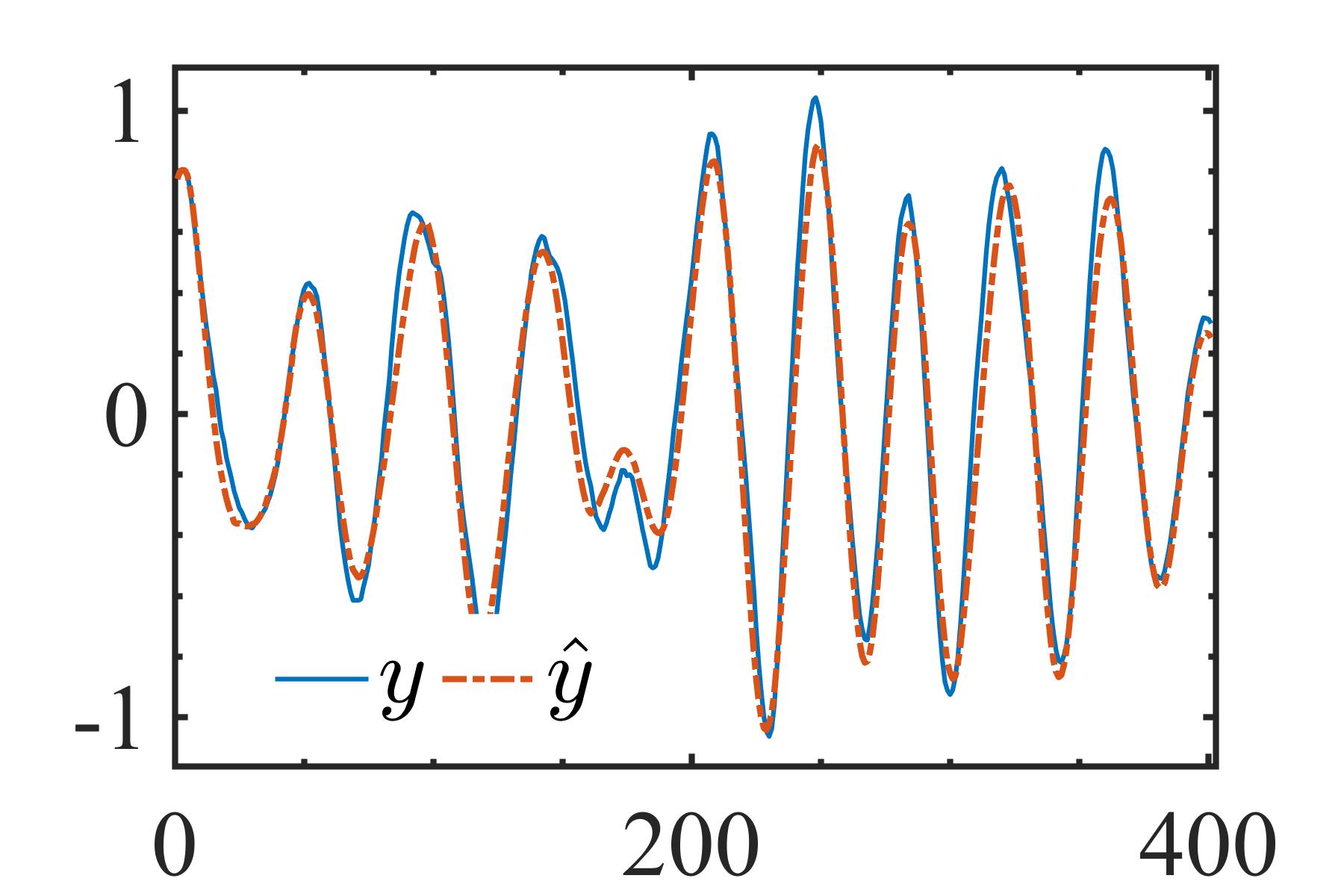}
  \caption{Validation ($y$); $\mathcal{M}_{W_1}$ ($\hat{y}$)}
  \label{f:f:wavmpo2}
\end{subfigure}
\hfill
\begin{subfigure}{.3\textwidth}
  \centering
  \includegraphics[width=\textwidth]{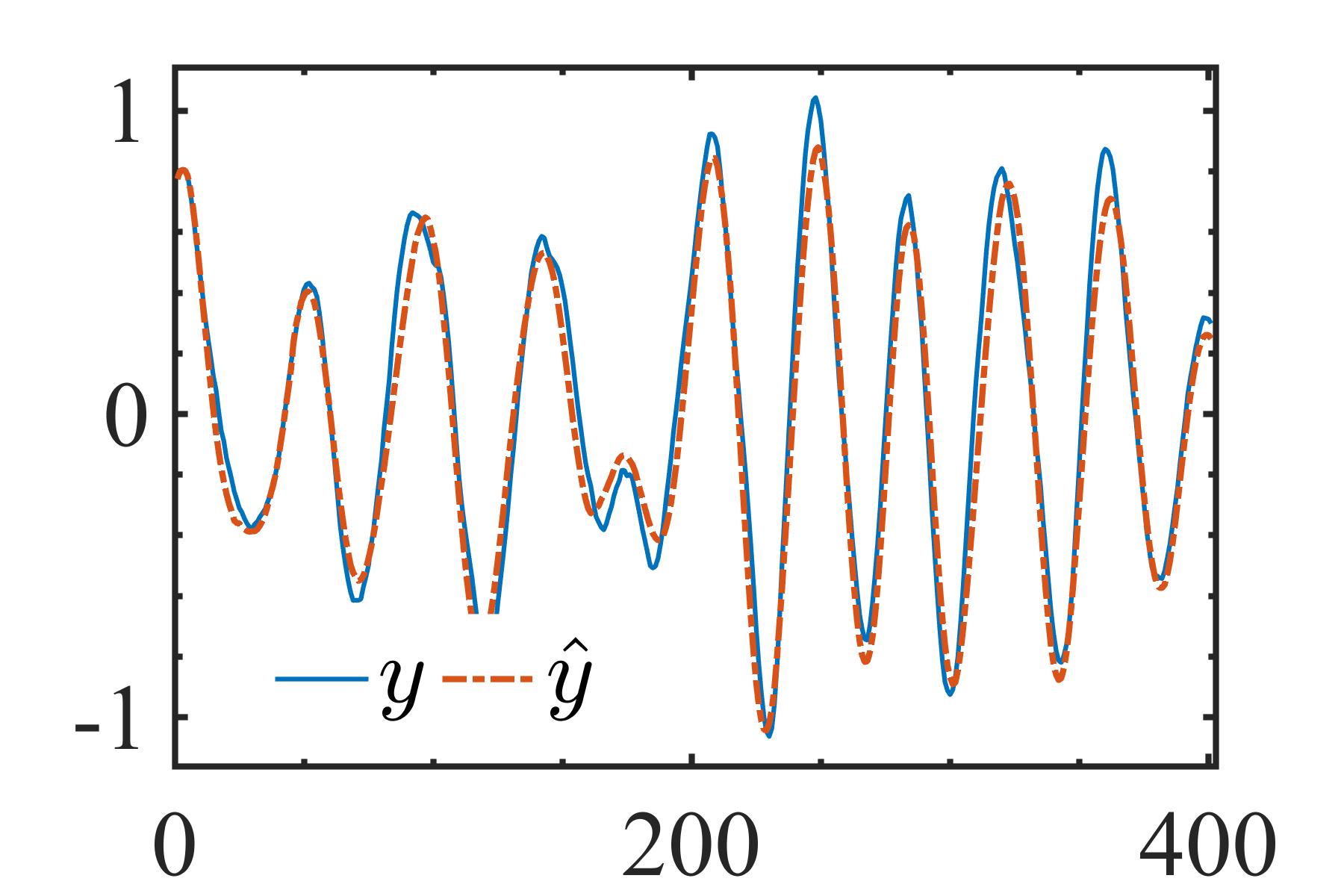}
  \caption{Validation ($y$); $\mathcal{M}_{W_2}$ ($\hat{y}$)}
  \label{f:f:wavmpo3}
\end{subfigure}
\hfill
\begin{subfigure}{.3\textwidth}
  \centering
  \includegraphics[width=\textwidth]{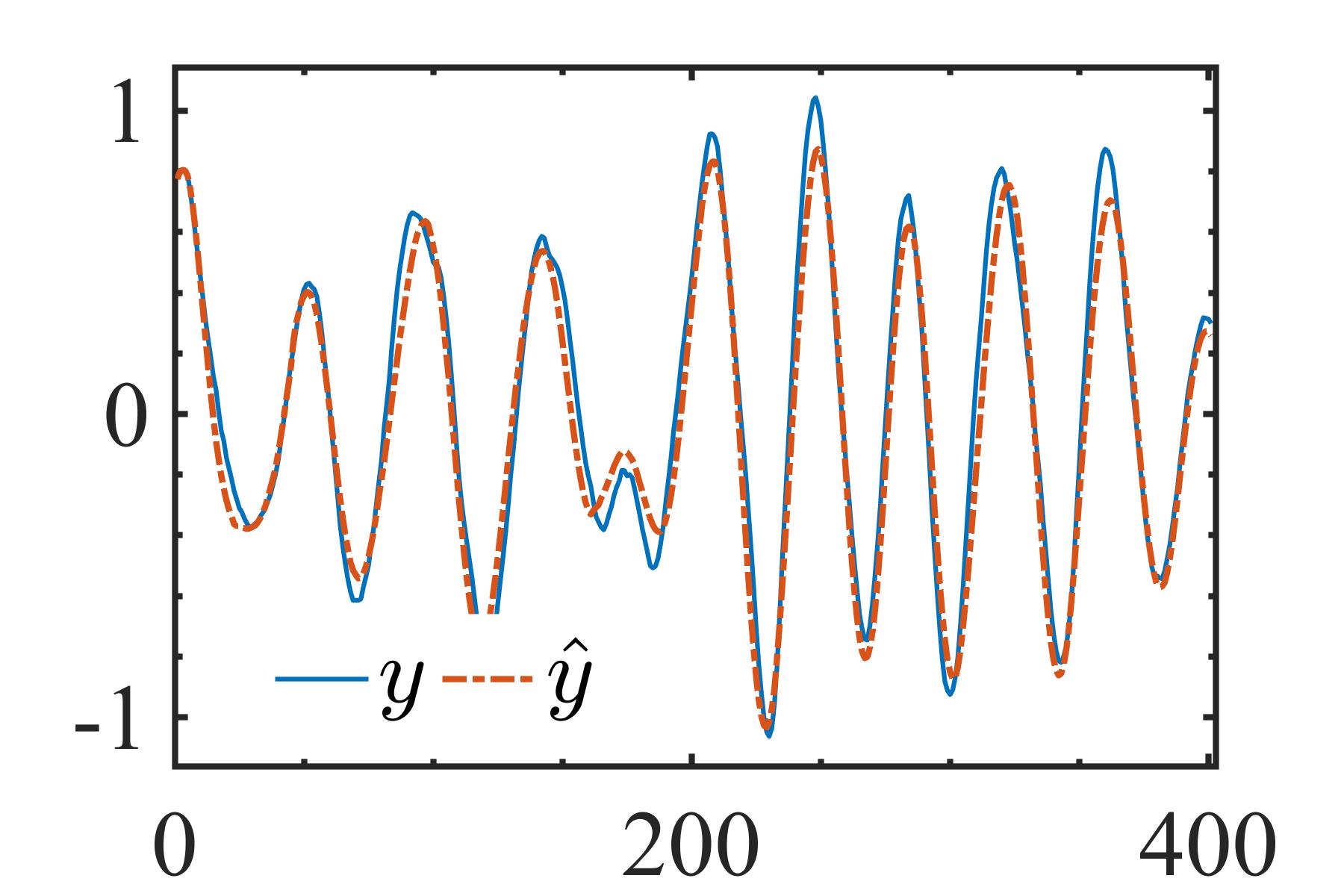}
  \caption{Validation ($y$); $\mathcal{M}_{W_3}$ ($\hat{y}$)}
  \label{f:f:wavmpo4}
\end{subfigure}
\caption{The model predicted output ($\hat{y}$) obtained with the identified models (\ref{eq:MW1})-(\ref{eq:MW3}) over the validation data.}
\label{f:wavmpo}
\end{figure*}

A total of $10$ non-dominated structures are identified by NSGA-II. These structures are ranked based on Manhattan distance `$\mathcal{D}(\cdot)$' (see Section~\ref{s:MMD}) and the global rank `$\mathcal{R}(\cdot)$' (see Section~\ref{s:MTD}). Based on these rankings, following three models are selected for further analysis:
\begin{linenomath*}
\begin{small}                                     
\begin{align}
   \label{eq:MW1}                                  
   \mathcal{M}_{W_1}: \ y(k) = & \ 1.411 \, y(k-1) - 0.49494 \, y(k-3) + 3.4176 \, u(k-1) - 3.3247 \, u(k-2) + 0.0072609 \, y(k-4)^3 \nonumber \\
                    & + 0.7105 \, u(k-5)y(k-1)^2 - 4.1731 \, u(k-5)y(k-4)y(k-2) + 3.5563 \, u(k-5)y(k-4)y(k-3)\\
   \label{eq:MW2}                                  
   \mathcal{M}_{W_2}: \ y(k) = & \ 1.41 \, y(k-1) -0.49533 \, y(k-3) + 3.6004 \, u(k-1) -                           3.4881 \, u(k-2) + 0.01363 \, y(k-4)^3 \nonumber \\ 
                    & + 0.73643 \, u(k-5)y(k-1)^2 - 0.27807 \, u(k-3)y(k-4)y(k-3) + 3.6831 \, u(k-5)y(k-4)y(k-3) \nonumber \\ 
                    & - 4.0926 \, u(k-5)y(k-4)y(k-2)\\
    \label{eq:MW3}                                  
    \mathcal{M}_{W_3}: \ y(k) = & \ 1.41 \, y(k-1) - 0.49371 \, y(k-3) + 3.4953 \, u(k-1) -3.4115 \, u(k-2) + 0.0064625 \, y(k-4)^3 \nonumber\\
                    &  + 0.8456 \, u(k-5)y(k-1)^2 - 7.1498 \, u(k-4)y(k-4)y(k-2) + 5.843 \, u(k-4)y(k-4)y(k-3) \nonumber\\ 
                    & - 0.52694 \, u(k-1)y(k-4)^2 + 1.1227 \, u(k-2)y(k-4)y(k-1)
\end{align}                                
\end{small} 
\end{linenomath*}
Table~\ref{t:wavndsol} gives the objective functions of the selected non-dominated structures along with the corresponding Manhattan distance and the global rank. The prediction errors (NMSE and MSE) in Table~\ref{t:wavndsol} are determined over the model predicted output with the validation data-points, which are shown in Fig.~\ref{f:wavmpo}. It is clear that there is no significant difference in the predictive performance of the identified models. 

\begin{figure*}[!t]
\centering
\small
\begin{subfigure}{.28\textwidth}
  \centering
  \includegraphics[width=\textwidth]{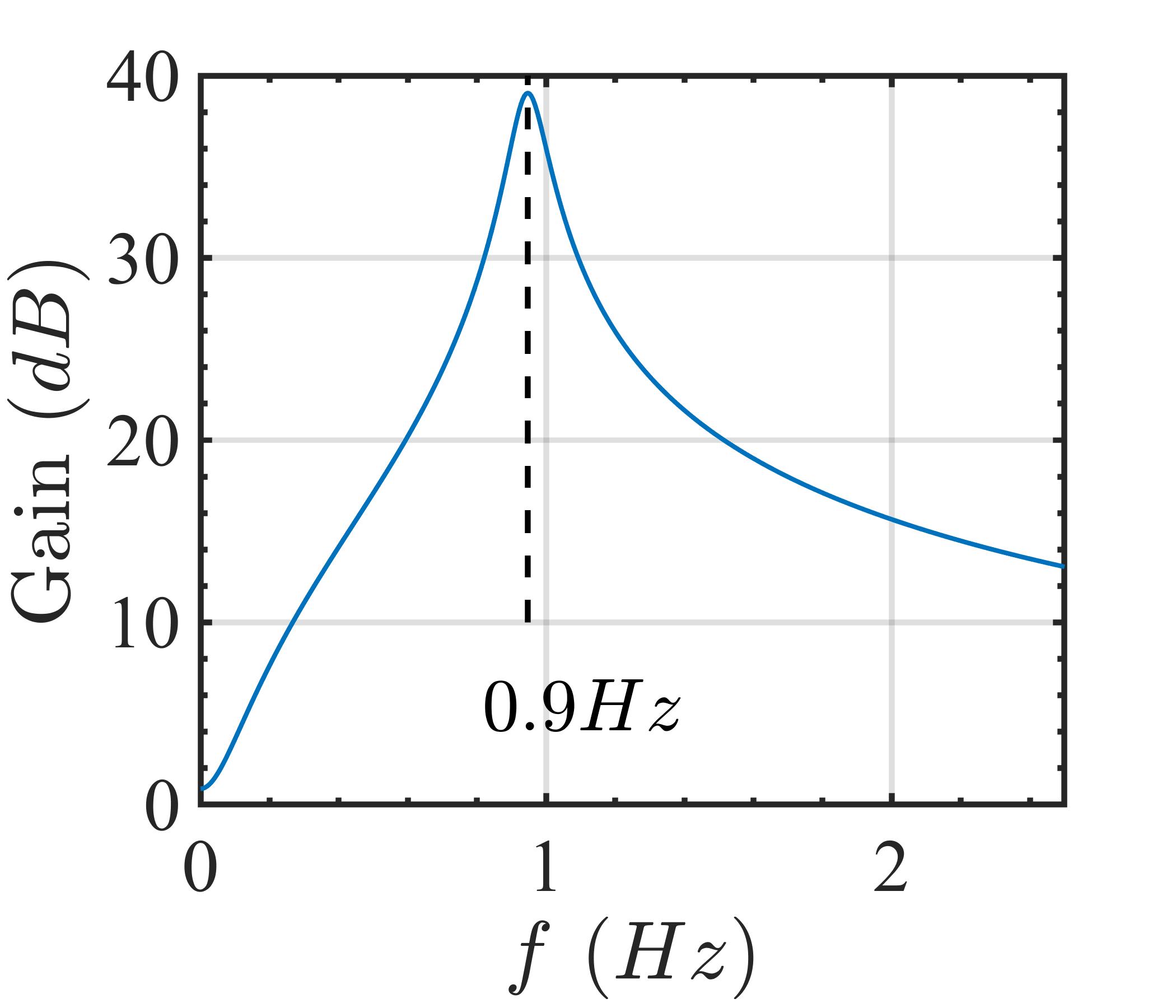}
  \caption{$\mathcal{M}_{W_1}$}
  \label{f:wavfrf2}
\end{subfigure}
\hfill
\begin{subfigure}{.28\textwidth}
  \centering
  \includegraphics[width=\textwidth]{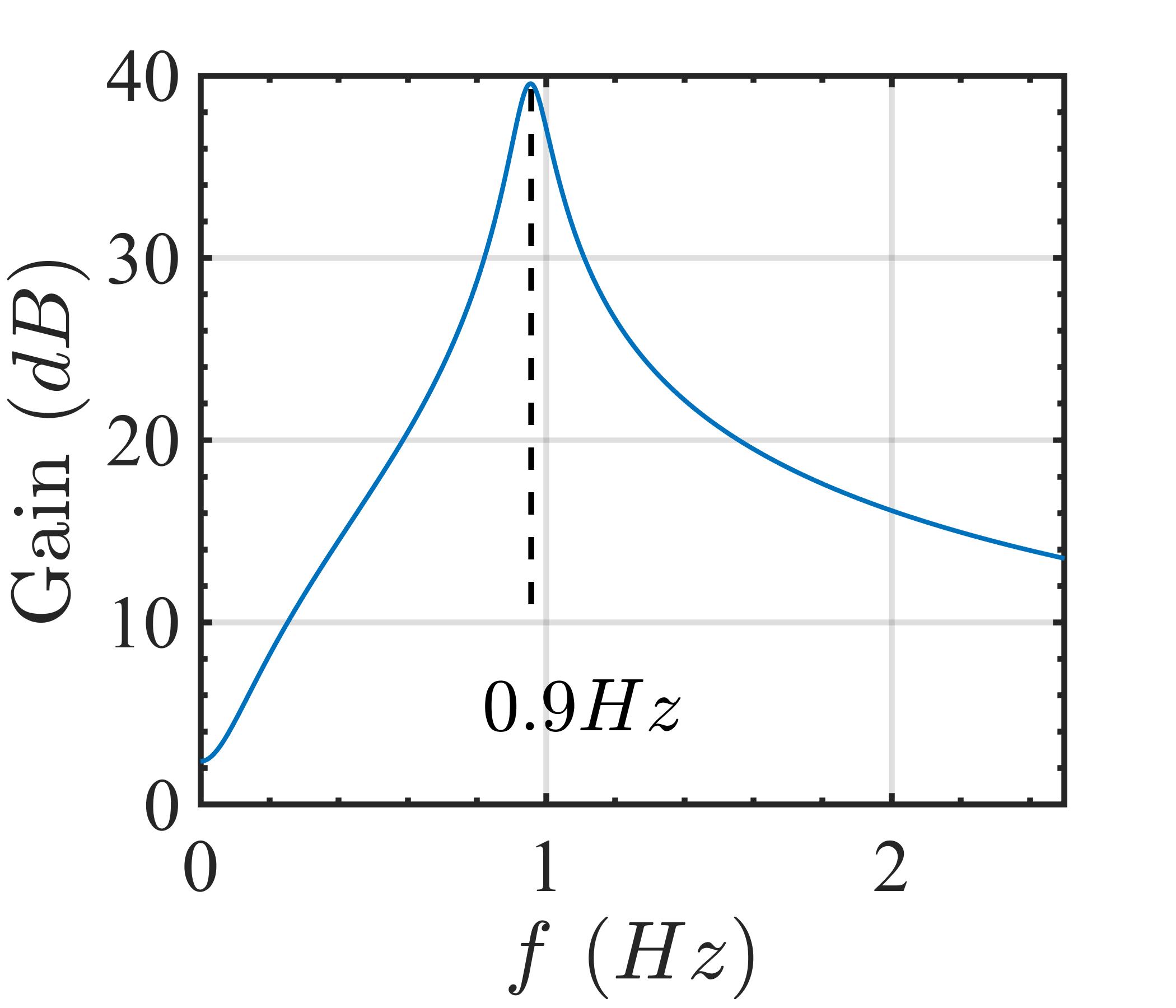}
  \caption{$\mathcal{M}_{W_2}$}
  \label{f:f:wavfrf3}
\end{subfigure}%
\hfill
\begin{subfigure}{.28\textwidth}
  \centering
  \includegraphics[width=\textwidth]{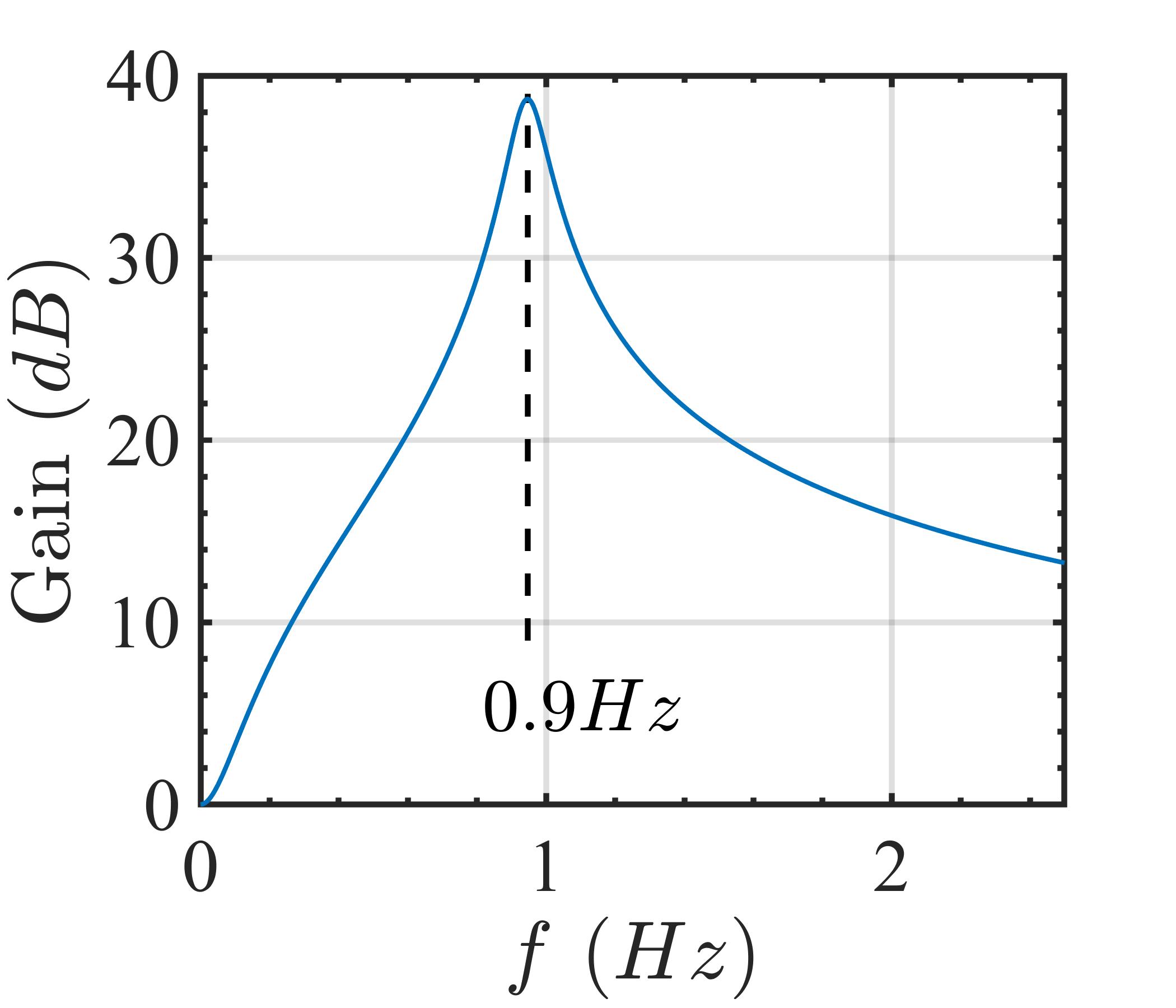}
  \caption{$\mathcal{M}_{W_3}$}
  \label{f:f:wavfrf4}
\end{subfigure}%
\caption{The linear frequency response functions (magnitude) of the models identified for the wave-force data.}
\label{f:wavgfrf}
\end{figure*}

For further validation, the frequency domain behavior of the identified models is compared using GFRFs. Note that even though the system dynamics may be explained by distinct discrete-time models, they are uniquely defined in the frequency domain~\cite{Billings:2013}. Hence, if the apparently different models have captured the wave force dynamics then the corresponding GFRF should match. To investigate this, GFRFs of the identified models are determined by mapping them to frequency domain, and are shown in Fig.~\ref{f:wavgfrf}. These results do not show any significant difference in the linear resonant frequency or the peak magnitude. 

Further, it is interesting to note that the similar frequency response behavior was observed in the distinct NARX model identified for the same wave force data in our previous investigation~\cite{Hafiz:Swain:Floating:2019}.   

\section{Results: Search Perspective}
\label{s:CompMOEA}

The objective of this part of the study is to determine whether there exists any significant difference in the search performance of the MOEAs. To this end, in the following, MOEAs are first compared using the quantitative performance metrics, drawing on the recommendations in~\cite{Zitzler:Thiele:2003}. Subsequently, a rigorous statistical analysis is carried out to determine the significance of the results. 

For this test, in addition to the seven discrete-time systems ($\mathcal{S}_1$-$\mathcal{S}_7$), a nonlinear continuous-time system (Duffing's oscillator, Section~\ref{s:ResDuff}) and a practical case study (wave-force identification, Section~\ref{s:wav}) are included. For each test system, the non-dominated structures are identified over $40$ independent runs of MOEAs as per the search environment described in Section~\ref{s:searchSetup} and Algorithm~\ref{al:moss}. The goal point for the system $\mathcal{S}_7$ is updated to $\{ \xi_{lim}, \ \mathcal{E}_{lim} \} = \{ 50, \ 30 \}$, as discussed in Section~\ref{s:commentPH}. The set of non-dominated structures which are obtained over $40$ runs is treated as the approximate Pareto set (APS). Further, for a particular system, the APS identified by NSGA-II, SPEA-II and MOEA/D are respectively denoted by $\Gamma_A$, $\Gamma_B$ and $\Gamma_C$.

\subsection{Dominance based comparison of MOEAs}
\label{s:CompMOEA1}


First, the set coverage metric $\mathcal{C}(\cdot)$ is considered which allows the pairwise comparison of the given APS. Note that, since $\mathcal{C}(\cdot)$ is not symmetric, the comparison of given APS (\textit{say}, $\Gamma_A$ and $\Gamma_B$) requires evaluation of both $\mathcal{C}(\Gamma_A,\Gamma_B)$ and $\mathcal{C}(\Gamma_B,\Gamma_A)$. Hence, for sake of clarity, the difference in these two metrics $\Delta \mathcal{C}(\cdot)$ is reported in Table~\ref{t:setcoverage}. This is defined as:
\begin{align}
    \label{eq:DeltaCmetric}
    \Delta \mathcal{C}(\Gamma_A,\Gamma_B) & = \mathcal{C}(\Gamma_A,\Gamma_B) - \mathcal{C}(\Gamma_B,\Gamma_A)
\end{align}
A positive value of $\Delta \mathcal{C}(\Gamma_A,\Gamma_B)$ implies that $\Gamma_A$ weakly dominates $\Gamma_B$ whereas a negative value implies otherwise. This metric is determined for all the test systems and it is shown in Table~\ref{t:setcoverage}. The results are shown for three pair-wise comparisons among NSGA-II, SPEA-II and MOEA/D. Based on these pair-wise comparisons, for each system, the dominance ordering for MOEAs is determined and is shown in the last column of Table~\ref{t:setcoverage}. For instance, for the system $\mathcal{S}_1$, when NSGA-II is compared to SPEA-II and MOEA/D, a positive value of $\Delta \mathcal{C}(\cdot)$ is obtained, \textit{i.e.}, $\Gamma_A \preceq \Gamma_B$ and $\Gamma_A \preceq \Gamma_C$. Further, for the same system, when SPEA-II is compared to MOEA/D, a negative value of $\Delta \mathcal{C}(\cdot)$ is obtained, \textit{i.e.}, $\Gamma_C \preceq \Gamma_B$. Hence, based on these pairwise comparisons, the following dominance ordering of MOEAs can be inferred for $\mathcal{S}_1$ : $\Gamma_A \preceq \Gamma_C \preceq \Gamma_B$.

By inspecting the dominance relations shown in Table~\ref{t:setcoverage}, a clear pattern emerges. For instance, NSGA-II outperforms both SPEA-II and MOEA/D on most of the test systems (six out of nine). Further, SPEA-II exhibits better performance than MOEA/D on six test systems. Hence, considering all the test systems, the performance of MOEAs can be ranked as: NSGA-II$>$SPEA-II$>$MOEA/D.

\begin{table}[!t]
  \centering
  \small
  \caption{Comparative Evaluation of MOEAs: Set Coverage Metric$^\dagger$}
  \label{t:setcoverage}%
  \begin{adjustbox}{max width=0.99\textwidth}
    \begin{threeparttable}
    \begin{tabular}{ccccc}
    \toprule
    \multirow{2}{*}{\textbf{System}} & {\makecell{\textbf{NSGA-II vs.} \\ \textbf{SPEA-II}}} & {\makecell{\textbf{SPEA-II vs.} \\ \textbf{MOEA/D}}} & {\makecell{\textbf{NSGA-II vs.} \\ \textbf{MOEA/D}}} & \multirow{2}{*}{\textbf{Remarks}} \\ [0.8ex]
     \cmidrule{2-4}     & \boldmath{$\Delta \mathcal{C}(\Gamma_A,\Gamma_B)$} & \boldmath{$\Delta \mathcal{C}(\Gamma_B,\Gamma_C)$} & \boldmath{$\Delta \mathcal{C}(\Gamma_A,\Gamma_C)$} &  \\ [0.8ex]
    \midrule
    $\mathcal{S}_1$    & 0.38  & -0.19 & 0.38 & \boldmath$\Gamma_A \preceq \Gamma_C \preceq \Gamma_B$ \\ [1.1ex]
    $\mathcal{S}_2$    & 0.24  & -0.17 & -0.04  & \boldmath$\Gamma_C \preceq \Gamma_A \preceq \Gamma_B$ \\ [0.9ex]
    $\mathcal{S}_3$    & -0.28 & 0.39  & 0.33 & \boldmath$\Gamma_B \preceq \Gamma_A \preceq \Gamma_C$ \\ [0.9ex]
    $\mathcal{S}_4$    & 0.35  & 0     & 0.12 & \makecell{\boldmath$\Gamma_A \preceq \Gamma_B$ \\ \boldmath$ \Gamma_A \preceq \Gamma_C$} \\ [2.2ex]
    $\mathcal{S}_5$    & 0.12  & 0.29  & 0.29 & \boldmath$\Gamma_A \preceq \Gamma_B \preceq \Gamma_C$ \\ [0.9ex]
    $\mathcal{S}_6$    & 0.33  & 0.44  & 0.44 & \boldmath$\Gamma_A \preceq \Gamma_B \preceq \Gamma_C$ \\ [0.9ex]
    $\mathcal{S}_7$    & 0.58  & 0.45  & 0.70 & \boldmath$\Gamma_A \preceq \Gamma_B \preceq \Gamma_C$ \\ [1.2ex]
    \makecell{Duffing's \\ Oscillator}  & -0.35 & 0.59  & 0.59 & \boldmath$\Gamma_B \preceq \Gamma_A \preceq \Gamma_C$ \\ [2ex]
    \makecell{Wave-force}    & 0.42  & 0.46  & 0.64 & \boldmath$\Gamma_A \preceq \Gamma_B \preceq \Gamma_C$ \\ [0.9ex]
    \bottomrule
    \end{tabular}%
    \begin{tablenotes}
      \small
       \item $^\dagger$ $\Gamma_A$, $\Gamma_B$ and $\Gamma_C$ respectively denote the approximate Pareto set identified by NSGA-II, SPEA-II and MOEA/D
    \end{tablenotes}
    \end{threeparttable}
 \end{adjustbox} 
\end{table}%

\subsection{Non-parametric statistical comparison of MOEAs}
\label{s:CompMOEA2}

Owing to the stochastic nature of MOEAs, further statistical analysis is carried out to determine the significance of the comparative analysis. In particular, the objective is to test the hypothesis that NSGA-II is significantly better than the other two MOEAs, SPEA-II and MOEA/D. For this purpose, non-parametric statistical tests are considered; as in practice the results obtained by MOEAs usually do not satisfy the necessary conditions of parametric tests~\cite{Derrac:Salvador:2011}. Therefore, in the following, multiple non-parametric statistical comparisons are carried out based on the recommendations in~\cite{Sheskin:2003,Derrac:Salvador:2011}.

Note that while the coverage metric $\mathcal{C}(\cdot)$ determines the dominance ordering of the given approximate Pareto set (APS), it does not provide any information on the degree of improvement. Hence, for the statistical analysis, the hyper-volume (HV) metric is considered (see Section~\ref{s:QuantMetric}). The reference point necessary for HV calculation is determined following the guidelines in~\cite{Ishibuchi:Imada:2017}. Prior to the determination of HV, the objective values are normalized and bounded in $[0,1]$. Following this procedure, HV is determined for all the test systems and are shown in Table~\ref{t:hv}. The significance of these results is determined in two steps. 

\begin{table}[!t]
  \centering
  \small
  \caption{Comparative Evaluation of MOEAs: Hyper-Volume (HV)}
  \label{t:hv}
  \begin{adjustbox}{max width=0.95\textwidth}
    \begin{threeparttable}
    \begin{tabular}{cccc}
    \toprule
    \textbf{System} & \textbf{NSGA-II} & \textbf{SPEA-II} & \textbf{MOEA/D} \\
    \midrule
    $\mathcal{S}_1$    & 0.7310 & 0.7260 & 0.7269 \\[0.5ex]
    $\mathcal{S}_2$    & 0.9280 & 0.9279 & 0.9280 \\[0.5ex]
    $\mathcal{S}_3$    & 0.8768 & 0.8771 & 0.8760 \\[0.5ex]
    $\mathcal{S}_4$    & 0.9316 & 0.9315 & 0.9315 \\[0.5ex]
    $\mathcal{S}_5$    & 0.6388 & 0.6382 & 0.6354 \\[0.5ex]
    $\mathcal{S}_6$    & 0.9389 & 0.9389 & 0.9388 \\[0.8ex]
    $\mathcal{S}_7$    & 0.8745 & 0.8686 & 0.8634 \\[0.8ex]
    \makecell{Duffing's\\Oscillator} & 0.9353 & 0.9353 & 0.9353 \\[1.5ex]
    Wave-force  & 0.7609 & 0.7462 & 0.7353 \\
    \midrule
    \makecell{\textbf{Average}\\ \textbf{Rank}} & 1.2   & 2.1   & 2.7 \\
    \midrule
    \makecell{\textbf{Friedman}\\ \textbf{Static}} & {9.56} & \boldmath{$p$}\textbf{-value} & {0.0084}\\
    \bottomrule
    \end{tabular}%
    \end{threeparttable}
 \end{adjustbox}  
\end{table}%
\begin{table}[!t]
\centering
\small
\caption{Outcome of the Hommel's Post-hoc Procedure for $95\%$ Confidence Interval}
\label{t:frt}
\begin{adjustbox}{max width=0.9\textwidth}
    \begin{threeparttable}
    \begin{tabular}{ccccc}
    \toprule
    \multirow{2}{*}{\textbf{Algorithm}} & \multicolumn{4}{c}{\textbf{HV}} \\
    \cmidrule{2-5}  & \textbf{Test Static} & \boldmath{$p$}\textbf{-value} & \textbf{APV} & \boldmath{$H_0$}\\
    \midrule
    SPEA  & 1.88 & 0.0593 & 0.050 & \xmark \\
    MOEA/D & 3.06 & 0.0022 & 0.025 & \xmark \\
    \bottomrule
    \end{tabular}%
    \begin{tablenotes}
      \scriptsize
       \item $^\dagger$ $H_0$ denotes \textit{null-hypothesis}
    \end{tablenotes}
    \end{threeparttable}
\end{adjustbox}
\end{table}

First, the Friedman Two-way Analysis of Variance by Ranks~\cite{Sheskin:2003,Derrac:Salvador:2011} is applied to test the null hypothesis that \textit{median value of HV obtained with MOEAs are equal}. For this purpose, the hyper-volume values are ranked from `$1$' (\textit{highest}) to `$3$' (\textit{lowest}). The procedure is repeated for all the test systems. Subsequently, the average value of ranks, the test statistic and the corresponding \textit{p-value} are shown in Table~\ref{t:hv}. These results indicate that the null-hypothesis can easily be rejected for $95\%$ confidence interval. This implies that there exists a significant difference in the performance of MOEAs.

Next, multiple comparisons of NSGA-II with the other MOEAs is carried out through a set of interconnected hypotheses. In particular, the determination of two interconnected null hypotheses is required to compare NSGA-II with SPEA-II and MOEA/D. Each null hypothesis ($H_0$) implies that the compared MOEA is significantly better than NSGA-II. The test statistic and \textit{$p-$value} needed to evaluate these hypotheses are determined following the guidelines in~\cite{Derrac:Salvador:2011,Sheskin:2003}. The \textit{$p-$value} are adjusted for multiple comparisons following the Hommel's post-hoc procedure~\cite{Derrac:Salvador:2011}, and denoted as Adjusted p-values (APV). The hypotheses are evaluated at $95\%$ confidence interval. The outcomes of this test are shown in Table~\ref{t:frt}. This test disproves both the \textit{null-hypotheses} and thereby confirms that NSGA-II is significantly better than SPEA-II and MOEA/D.

\section{Robustness to Control Parameters}
\label{s:resRobust}


A judicious tuning of search parameters is crucial to the performance of any evolutionary algorithm (EA)~\cite{Eiben:Hinterding:1999}. For instance, the \textit{cross-over} ($p_c$) and \textit{mutation probabilities} ($p_m$), the choice of \textit{recombination} mechanism such as \textit{single-point} or \textit{uniform crossover} can have significant effect on the search performance. These parameters can be categorized into \textit{quantitative} or \textit{qualitative} parameters. 


Although there exist several parameter tuning approaches, \textit{e.g.} see~\cite{Eiben:Hinterding:1999}, in most of the existing investigations, the parameters are selected either based on the prevalent practice or based on limited trial-and-error approach~\cite{Fonseca:Fleming:1996,Rodriguez:Fonseca:Fleming:1997,Rodriguez:Fleming:1998,Rodriguez:Fleming:2005,Rodriguez:Fonseca:2004,Zakaria:Jamaluddin:2012,Lavinia:Patelli:2009a,Lavinia:Patelli:2009b}. Even though MOEAs considered in this study have been developed over past two decades, their search performance has received relatively less attention from the identification perspective. It is therefore crucial to evaluate the \textit{robustness} of these algorithms with respect to \textit{quantitative} and \textit{qualitative} search parameters. This has been the motivation behind the empirical investigation which is discussed in the following. 

The \textit{quantitative} control parameters include but not limited to the \textit{size} of the \textit{population} and external archive (if any), the \textit{cross-over} ($p_c$) and \textit{mutation} probabilities ($p_m$), and other algorithm specific parameters, \textit{e.g.}, \textit{neighborhood size} `$T$' in MOEA/D. In this study, the effects of the cross-over ($p_c$) and mutation probabilities ($p_m$) on the search performance are of particular interest. For this purpose, the search performance of MOEAs is determined over a \textit{grid} in the parameter space. Each \textit{point} on this \textit{grid} denotes a particular combination of parameters $\{ p_c, \ p_m \}$. For each parameter combination, the search performance is quantified using the \textit{hyper-volume} metric. This will determine the region of better performance over the \textit{grid-map}. Such promising regions in the parameter space are often referred to as `\textit{sweet-spot}'~\cite{Goldberg:1999,Purshouse:Fleming:2007}. It is easier to follow that the \textit{robustness} of the given MOEA, with respect to the search parameter, can indirectly be estimated from the \textit{sweet-spot} in the parameter space. More details and the results of this experiment are respectively discussed in Section~\ref{s:ExpSetupPcPm} and~\ref{s:SweetSpot}.

The \textit{qualitative} parameters mainly reflect the choice of algorithm design, \textit{e.g.}, choice of \textit{selection}, \textit{recombination} and \textit{mutation} mechanisms. In MOEAs, the \textit{proximity} and \textit{diversity} preserving mechanisms are closely tied to the \textit{selection} process. The choice of \textit{selection} mechanism is thus dependent on the framework of MOEA, \textit{e.g.}, \textit{crowded tournament selection} in NSGA-II. Hence, for each MOEA, the \textit{selection} mechanism is chosen in the original research is considered. In contrast, the choice of the \textit{recombination} and the \textit{mutation} mechanisms is not dependent on the framework of MOEA. Hence, it is possible to incorporate suitable mechanism. In particular, we are interested in evaluating the effects of \textit{recombination} mechanism, which will be discussed in Section~\ref{s:SingleOrUniform}.

\begin{algorithm}[!t]
    \small
    \SetKwInOut{Input}{Input}
    \SetKwInOut{Output}{Output}
    \SetKwComment{Comment}{*/ \ \ \ }{}
    \Input{Input-output Data, $(u,y)$}
    \Output{Hyper-volume Performance Metric, $\overline{PM}$}
    \BlankLine
    Specify the search grid with `$n_{comb}$' number of parameter combinations\\
    Select the search algorithm: NSGA-II, SPEA-II or MOEA/D\\
    Select the recombination mechanism: \textit{Single Point} or \textit{Uniform} crossover\\
    \BlankLine
    \BlankLine
    \Comment*[h] {Stage-1: Search for Non-dominated Structures}\\
    \BlankLine
    \For{i = 1 to $n_{sys}$\nllabel{line:S1start}}
    {   \BlankLine
        \For{j = 1 to $n_{comb}$}
        {   
             \BlankLine
             Set the \textit{crossover rate} to $p_c^{\ j}$\\
             Set the \textit{mutation rate} to $p_m^{\ j}$
            \BlankLine
            \Comment*[h] {Perform `$R$' runs of MOEA with $\{ p_c^{\ j}, p_m^{\ j} \}$}\\
            \BlankLine
            $\Gamma_{run} \leftarrow \varnothing$, $\Lambda_{run} \leftarrow \varnothing$\\
            \For{k = 1 to $R$} 
            {   \BlankLine
                Identify and store the \textit{non-dominated} structures, \textit{i.e.},\\
                \BlankLine
                $\Gamma_{run} \leftarrow \Gamma_{run} \cup \begin{Bmatrix} \mathcal{X}_1 & \mathcal{X}_2 & \dots \end{Bmatrix}$\\
                $\Lambda_{run} \leftarrow \Lambda_{run} \cup \begin{Bmatrix} \vec{J}(\mathcal{X}_1) & \vec{J}(\mathcal{X}_2) & \dots \end{Bmatrix}$
                \BlankLine
            } 
        \BlankLine
        \Comment*[h]{Approximate Pareto Set}\\
        $\Gamma_{i,j} \preceq \Gamma_{run}, \qquad \Lambda_{i,j} \preceq \Lambda_{run} $ \\
        \BlankLine
        }
    }\nllabel{line:S1end}
    \BlankLine
    \Comment*[h] {Stage-2: Evaluation of Performance Metric}\\
    \BlankLine
    \Comment*[h] {`\textit{ideal}' Pareto Front}\\
    \For{i = 1 to $n_{sys}$\nllabel{line:S2start}}
    {   
        \BlankLine
        $\Lambda_{i}^\star \preceq \big\{ \Lambda_{i,1} \cup \Lambda_{i,2} \cup \dots \cup \Lambda_{i,n_{comb}} \big\} $
    }\nllabel{line:S2end1}
    \BlankLine
    \Comment*[h] {Metric Evaluation}\\
    \For{j = 1 to $n_{comb}$}
    {   \BlankLine
        \For{i = 1 to $n_{sys}$}
        {
            \BlankLine
            Determine the Hyper-Volume ratio with respect to $\Lambda_{i}^\star$, \textit{i.e.},\\
            $PM_{i,j} = \ \text{HV Ratio} (\Lambda_{i,j},\Lambda^{\ast}_i)$
            \BlankLine
        }
        $\overline{PM_j} = \frac{1}{n_{sys}} \sum \limits_{i=1}^{n_{sys}} PM_{i,j}$\nllabel{line:S2PM}
    }\nllabel{line:S2end2}
\caption{Robustness Analysis}
\label{al:pcpm}
\end{algorithm}
\subsection{Experimental Setup}
\label{s:ExpSetupPcPm}

The search setup for the robustness analysis is outlined in Algorithm~\ref{al:pcpm}. For this test, the \textit{grid} is designed in the parameter space using $10$ distinct values of $p_c$ and $p_m$, where $p_c\in[0.1, 1]$ and $p_m\in [0.001,0.02)$. This gives a total of $100$ distinct combinations of $\{p_c,p_m\}$. The search performance is evaluated on all the discrete time systems ($\mathcal{S}_1-\mathcal{S}_7$). In Algorithm~\ref{al:pcpm}, the total number of parameter combinations and systems are respectively denoted by `$n_{comb}$' and `$n_{sys}$', \textit{i.e.}, $n_{comb}=100$ and $n_{sys}=7$. 

The goal of this test is to determine the unary performance metric corresponding to each parameter combination on the test grid. To this end, the test begins by selecting the MOEA and the crossover mechanism, and it is further carried out in the following two stages: 1) Search for Approximate Pareto Set (APS) and 2) Evaluation of Performance Metric.

In the first stage, for each test system, APS is identified corresponding to each parameter combination. The steps involved in this stage are outlined in Line~\ref{line:S1start}-\ref{line:S1end}, Algorithm~\ref{al:pcpm}. Due to stochastic nature of MOEAs, the non-dominated structures identified over `$R$' independent runs ($R=20$) are determined to be APS. For instance, in Algorithm~\ref{al:pcpm}, `$\Gamma_{i,j}$' and `$\Lambda_{i,j}$' respectively denote APS and the corresponding front identified for the $i^{th}$ system with the $j^{th}$ parameter combination in MOEA. This procedure is repeated for each system across all the combination of parameters. 

In the second stage, the identified APS are quantified using unary performance metric. The overall procedure involved in this stage is outlined in Line~\ref{line:S2start}-\ref{line:S2end2}, Algorithm~\ref{al:pcpm}. In particular, for the given combination of parameters, the goal is to determine an aggregated value of performance metric across all the test systems. For this purpose, the hyper-volume ratio (HVR) is appropriate as it yields normalized values which are bounded in $[0,1]$. As the name suggests, HVR is in essence ratio of hyper-volume values of the identified APF and the `\textit{ideal}' Pareto front. Note that since the `\textit{ideal}' Pareto front is not known, it is approximated from the union set of APF which are identified with different parameter combinations, as outlined in Line~\ref{line:S2start}-\ref{line:S2end1}, Algorithm~\ref{al:pcpm}. Subsequently, HVR is determined for each system across all the parameter combination from the corresponding APS. For example, in Algorithm~\ref{al:pcpm}, `$PM_{i,j}$' denotes the HVR for the $i^{th}$ system with the $j^{th}$ parameter combination. Finally, for each parameter combination, the aggregated performance metric ($\overline{PM}$) is determined as shown in Line~\ref{line:S2PM}, Algorithm~\ref{al:pcpm}.

The overall procedure outlined in Algorithm~\ref{al:pcpm} is repeated for all combinations of MOEAs (NSGA-II, SPEA-II and MOEA/D) and recombination mechanisms (\textit{Single Point} and \textit{Uniform}). 

\subsection{Sweet-spots in the Parameter Space}
\label{s:SweetSpot}

The objective of this part of the study is to determine the \textit{robustness} of MOEAs to parameter specification. As discussed earlier, this can indirectly be determined by observing the region of `\textit{good}' performance in the search parameter space, \textit{i.e.}, the bigger the region of high performance, the higher is the \textit{robustness}. To this end, an aggregated performance metric ($\overline{PM}$) across all the test systems is determined for each parameter combination, as per the procedure outlined in Algorithm~\ref{al:pcpm}. Following this procedure a \textit{contour map} of performance($\overline{PM}$) is determined in the parameter space; as each parameter combination is essentially a point on the parameter grid (\textit{as discussed in the previous section}). Consequently, the contours are used to highlight the region of high performance, which are often referred to as `\textit{sweet-spots}'. Note that this approach is somewhat similar to earlier investigation of Goldberg~\cite{Goldberg:1999} and Purshouse and Fleming~\cite{Purshouse:Fleming:2007}, where the objective was to determine the performance \textit{sweet-spots} over the  \textit{grid} of some \textit{quantitative} or \textit{qualitative} search parameters. 

\begin{figure*}[!t]
\centering
\small
\begin{subfigure}{.37\textwidth}
  \centering
  \includegraphics[width=\textwidth]{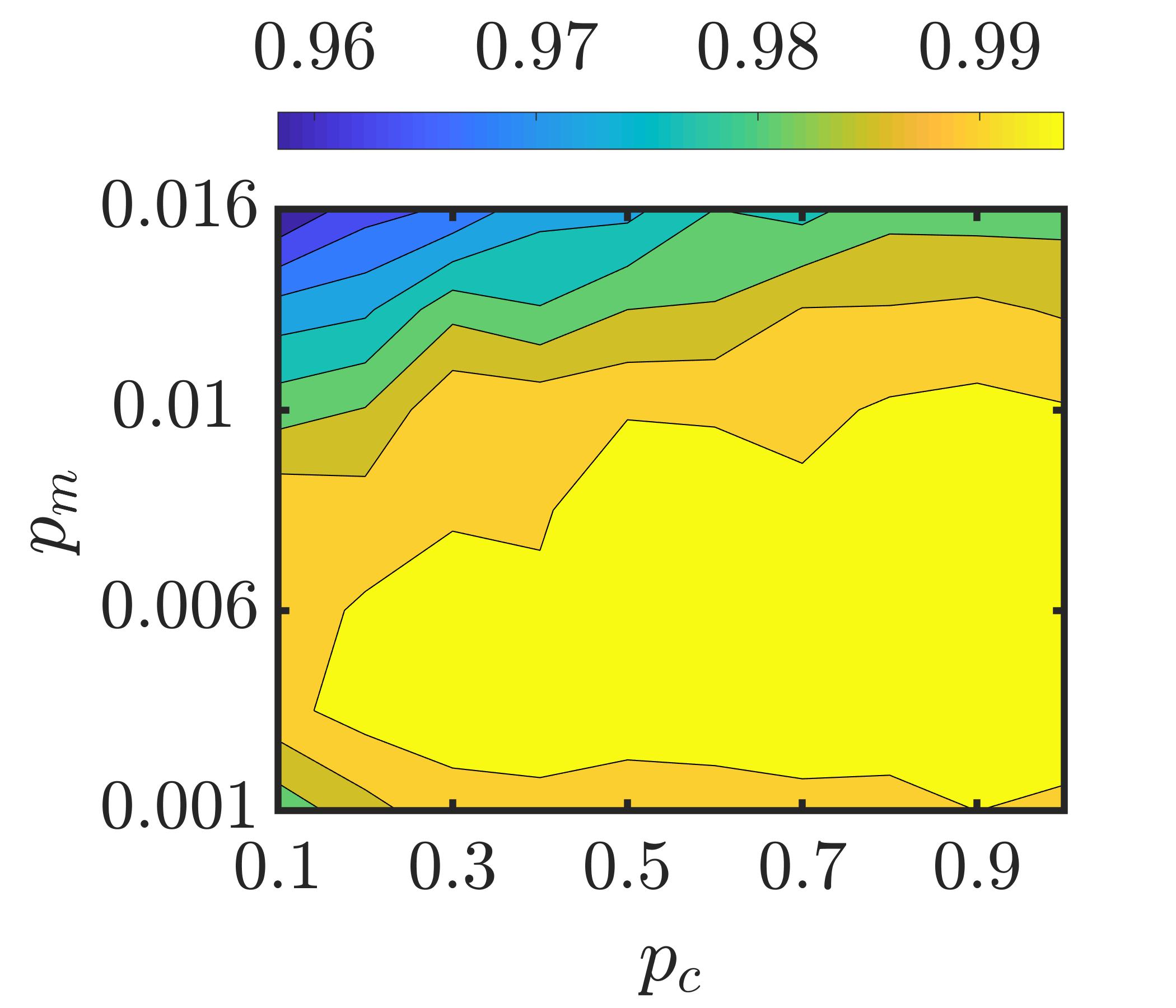}
  \caption{NSGA-II + Single Point Crossover}
  \label{f:pcpmnsga1}
\end{subfigure}%
\hspace{.1\textwidth}
\begin{subfigure}{.37\textwidth}
  \centering
  \includegraphics[width=\textwidth]{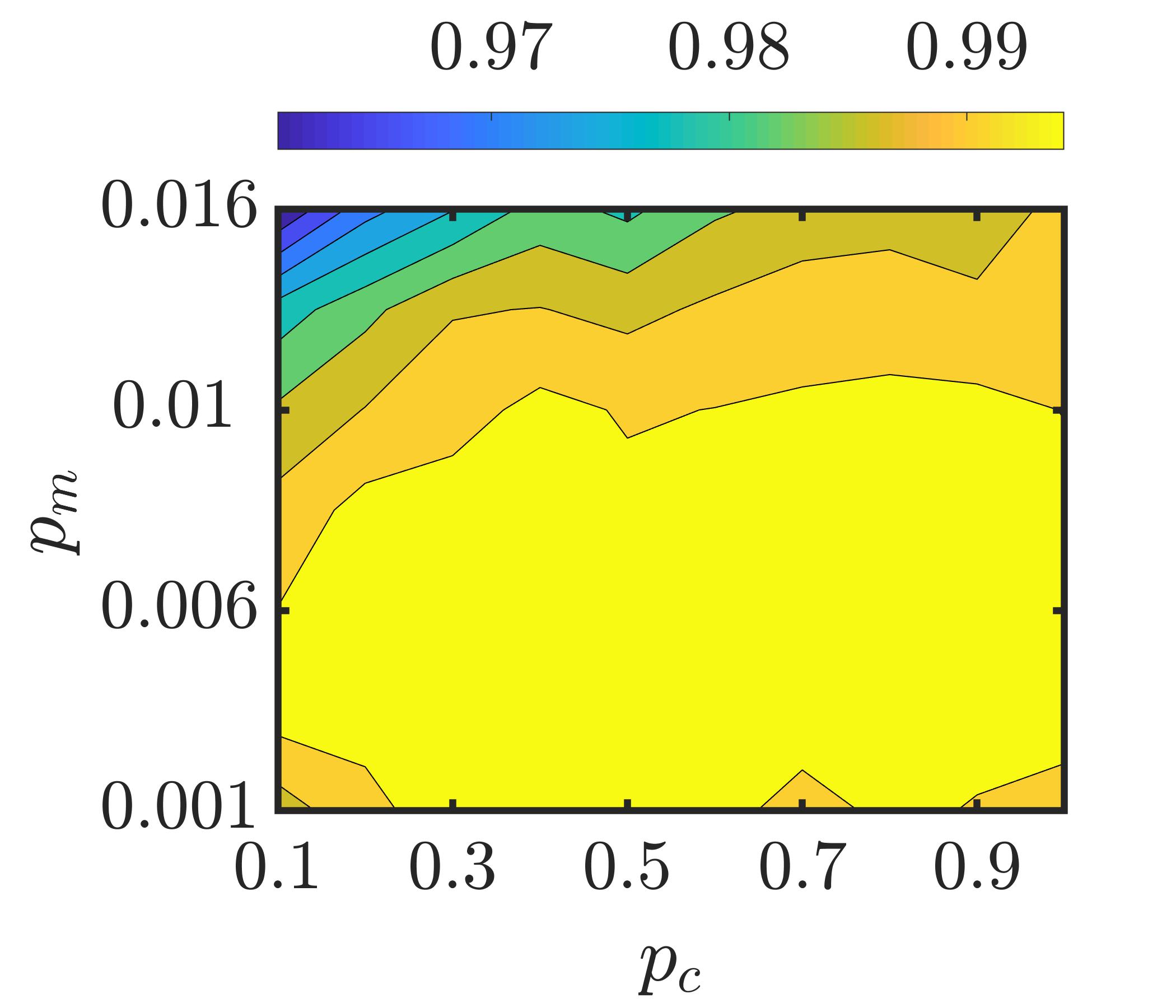}
  \caption{NSGA-II + Uniform Crossover}
  \label{f:pcpmnsga2}
\end{subfigure}
\begin{subfigure}{.37\textwidth}
  \centering
  \includegraphics[width=\textwidth]{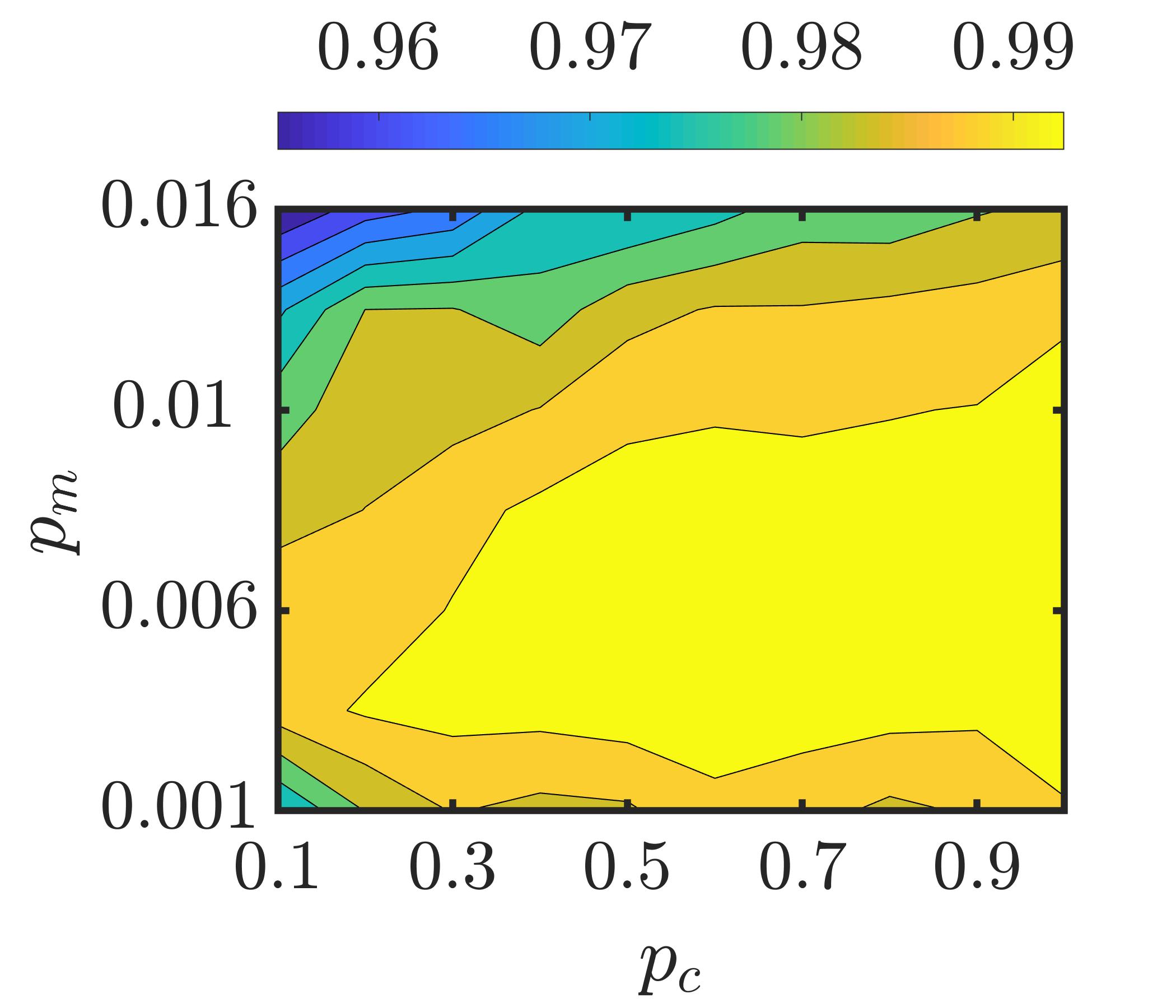}
  \caption{SPEA-II + Single Point Crossover}
  \label{f:pcpmspea1}
\end{subfigure}%
\hspace{.1\textwidth}
\begin{subfigure}{.37\textwidth}
  \centering
  \includegraphics[width=\textwidth]{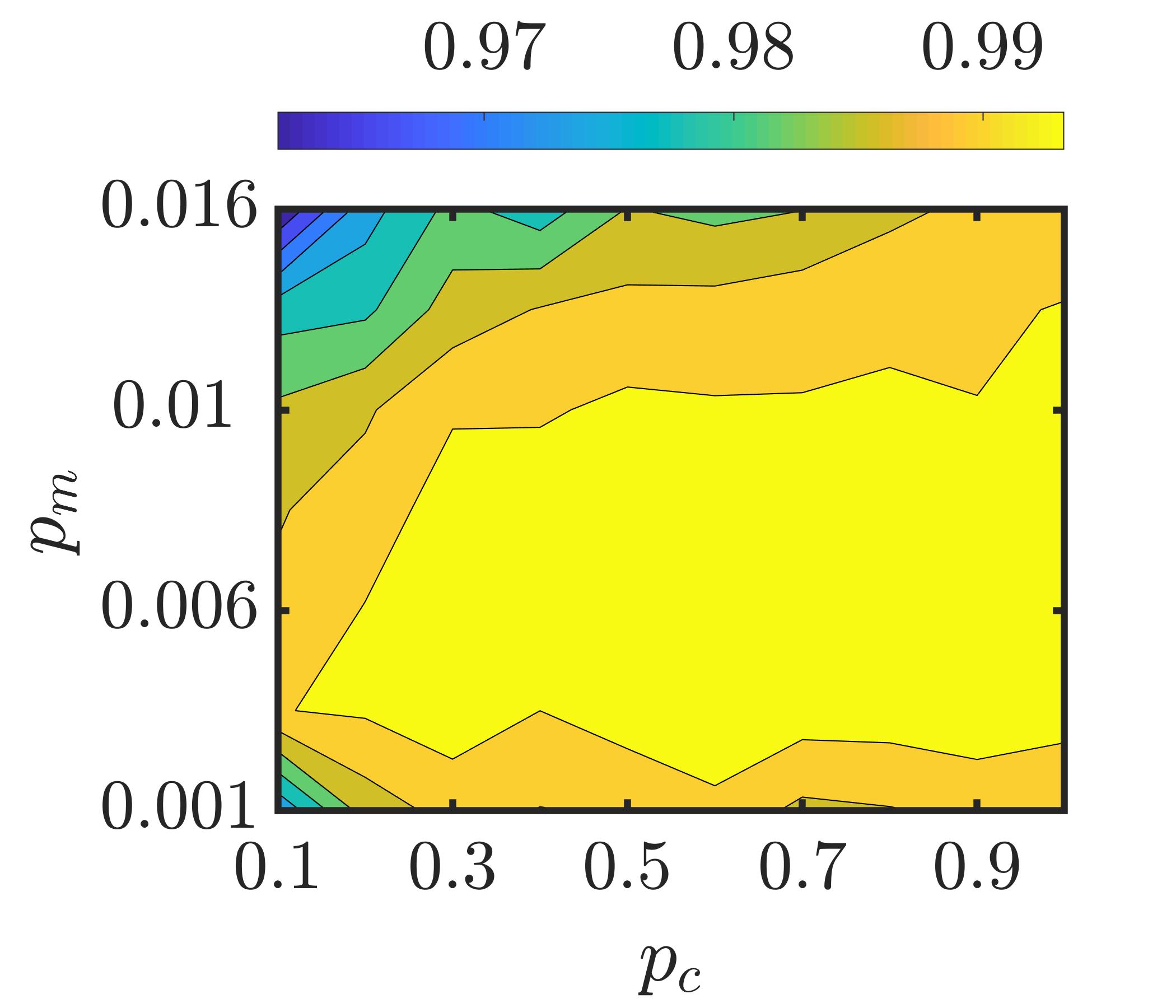}
  \caption{SPEA-II + Uniform Crossover}
  \label{f:pcpmspea2}
\end{subfigure}
\begin{subfigure}{.37\textwidth}
  \centering
  \includegraphics[width=\textwidth]{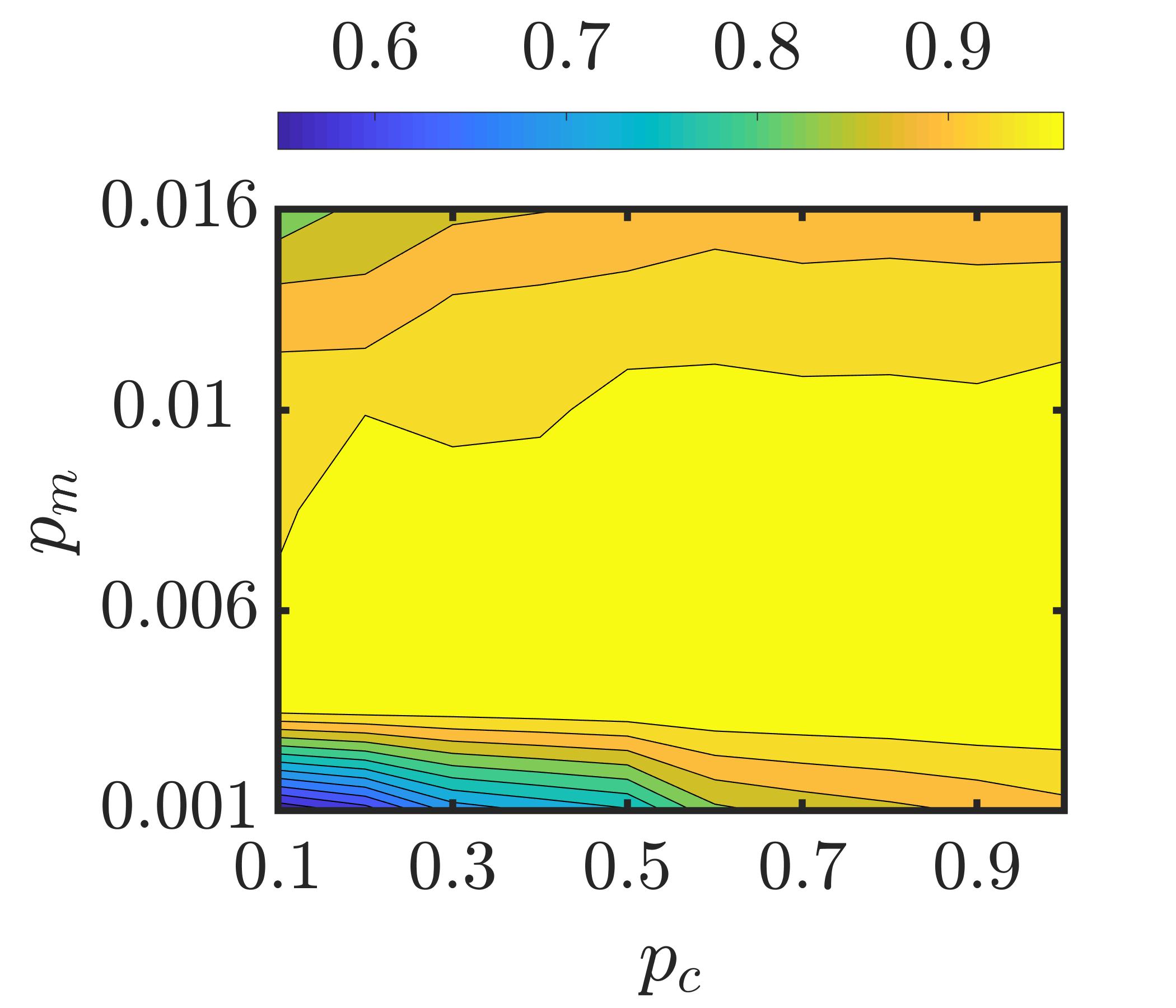}
  \caption{MOEA/D + Single Point Crossover}
  \label{f:pcpmmoea1}
\end{subfigure}%
\hspace{.1\textwidth}
\begin{subfigure}{.37\textwidth}
  \centering
  \includegraphics[width=\textwidth]{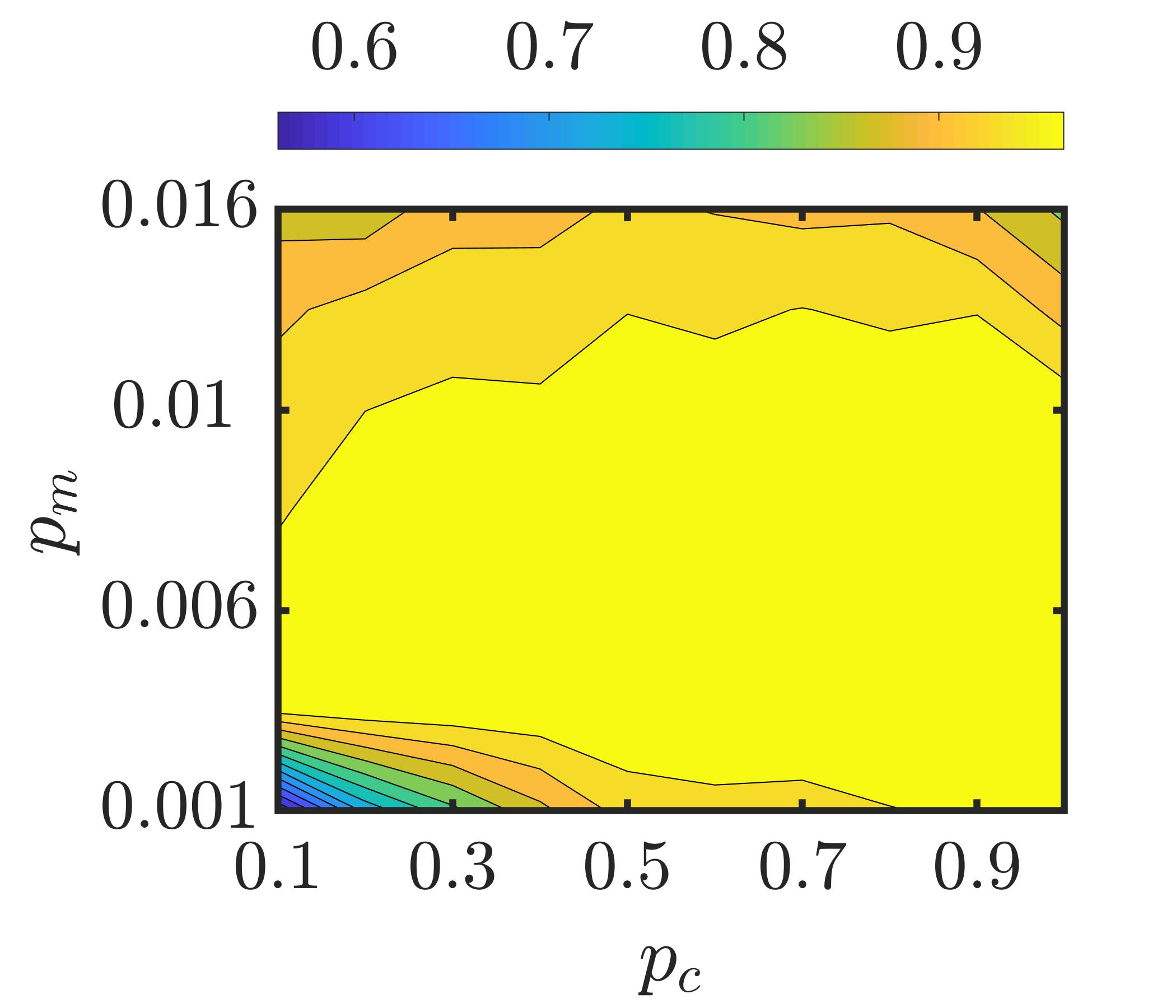}
  \caption{MOEA/D + Uniform Crossover}
  \label{f:pcpmmoea2}
\end{subfigure}
\caption{Average value of the HV ratio over all the test systems ($\mathcal{S}_1-\mathcal{S}_6$) with different combinations of $\{p_c, p_m \}$. A higher value of HV ratio is desirable.}
\label{f:pcpmoa}
\end{figure*}

The results of this investigation are shown in Fig.~\ref{f:pcpmoa}. For each contour-map the corresponding color-bar indicates the scale of $\overline{PM}$. A higher value of $\overline{PM}$ is desirable which is represented by a lighter shade of yellow. Hence, the \textit{sweet-spot} is highlighted by a yellow bounded area in each contour-map. It is interesting to see that the \textit{sweet-spots} of NSGA-II and MOEA/D inflate/grow when uniform crossover is used (see Fig~\ref{f:pcpmnsga1}-\ref{f:pcpmnsga2}, NSGA-II and Fig.~\ref{f:pcpmmoea1}-\ref{f:pcpmmoea2}, MOEA/D), which implies that the robustness of NSGA-II and MOEA/D is better with uniform crossover (\textit{to be discussed in the next section}). Hence, we focus only on the results obtained with the uniform crossover, \textit{i.e.}, Fig.~\ref{f:pcpmnsga2} (NSGA-II), Fig.\ref{f:pcpmspea2} (SPEA-II) and Fig.\ref{f:pcpmmoea2} (MOEA/D).

It is clear that the variance in $\overline{PM}$ with NSGA-II and SPEA-II is relatively low. Hence, the search performance will not significantly deteriorate even with improper choice of parameters. In contrast, the parameter selection is a critical issue in MOEA/D which is indicated by a high variance in $\overline{PM}$. From the contours, the \textit{sweet-spots} of the MOEAs are determined as follows: $p_c \in (0.2,1]$ and $p_m\in [0.002,0.009]$ (NSGA-II, Fig.~\ref{f:pcpmnsga2}); $p_c \in [0.3,1]$ and $p_m\in [0.003,0.01]$ (SPEA-II, Fig.\ref{f:pcpmspea2}) and $p_c \in [0.5,1]$ and $p_m\in [0.003,0.013]$ (MOEA/D, Fig.\ref{f:pcpmmoea2}). Note that these performance \textit{sweet-spots} can also be used as \textit{rule of thumb} or a starting point to select the parameters for the other identification problems. 

The comparative analysis suggests that the search performance of MOEA/D is poorer compared to the other dominance based MOEAs such as NSGA-II and SPEA-II. This could be ascribed to the choice of the control parameters. In this study, a detailed investigation is carried out to select the crossover and mutation probabilities as well as the recombination mechanism. However, further investigation is required to determine the sensitivity of MOEA/D to the other control parameters such as the choice of aggregation function, neighborhood size, `$T$', and replacement neighbor size,`$nr$'.

\subsection{Recombination: Single Point or Uniform?}
\label{s:SingleOrUniform}

The objective of this part of the study is to determine if there exist any significant effects of the recombination mechanism on the search performance of MOEAs. 

The motivation for this test is in part based on the previous investigation by the authors in which Simple Genetic Algorithm (SGA) was found to be more susceptible to the selection of spurious terms~\cite{Hafiz:Swain:CEC:2018}. While the cause of this poor performance is still to be thoroughly investigated, in theory, it can be ascribed to poor \textit{selection} and/or \textit{recombination} mechanism. Especially, the single point crossover in SGA is designed to preserve low-order schemas which are considered to be the building blocks of the string~\cite{Goldberg:Book:1989}. This assumption, however, may not be valid in the structure selection, since the \textit{significant/system} terms may not be positioned close to each other in the binary string representation, \textit{i.e.}, the desired schemas may not necessarily be of low-order. Further, it has been argued that the pursuit to preserve low-order schemas may lead to selection of bits that are not part of the desired schema~\cite{Mitchell:1998}. This could explain the selection of spurious terms (\textit{bits}) in our previous investigations. It has been argued that the \textit{parameterized uniform crossover} can overcome these issues~\cite{Mitchell:1998,Spears:DeJong:1995}. Since MOEAs being used are in essence based on genetic algorithms, it is pertinent to evaluate the effects of recombination mechanisms on the search performance of MOEAs. 

\begin{table}[!t]
  \centering
  \small
  \caption{Performance Metric ($PM$) for System $\mathcal{S}_1$}
  \label{t:WilcoxPS1}%
    \begin{tabular}{ccccccc}
    \toprule
    \multirow{3}{*}{\makecell{\textbf{Parameter} \\ \textbf{Combination}\\\boldmath{$\{p_c^{ \ j},\ p_m^{ \ j}\}$} }} &  \multicolumn{2}{c}{\textbf{NSGA-II}} & \multicolumn{2}{c}{\textbf{SPEA-II}} & \multicolumn{2}{c}{\textbf{MOEA/D}} \\
    \cmidrule{2-7} & \makecell{\textbf{Single}\\\textbf{Point}} & \textbf{Uniform} & \makecell{\textbf{Single}\\\textbf{Point}} & \textbf{Uniform} & \makecell{\textbf{Single}\\\textbf{Point}} & \textbf{Uniform} \\
    \midrule
    $\{0.1, 0.001\}$ & 0.9696 & 0.9764 & 0.9649 & 0.9419 & 0.5240 & 0.5573 \\[0.5ex]
    $\{0.1,0.0035\}$ & 0.9836 & 0.9867 & 0.9774 & 0.9883 & 0.9830 & 0.9820 \\[0.5ex]
    \vdots & \vdots & \vdots & \vdots & \vdots & \vdots & \vdots \\[0.5ex]
    $\{0.8,0.001\}$ & 0.9812 & 0.9890 & 0.9542 & 0.9704 & 0.8668 & 0.9203 \\[0.5ex]
    \vdots & \vdots & \vdots & \vdots & \vdots & \vdots & \vdots \\[0.5ex]
    \bottomrule
    \end{tabular}%
\end{table}%

\begin{table}[!t]
  \centering
  \small
  \caption{Statistical Comparison of Crossover Mechanisms}
  \label{t:WilcoxPS2}%
    \begin{tabular}{cccc}
    \toprule
    \multirow{2}{*}{\textbf{System}} & \multicolumn{3}{c}{\textbf{Null Hypothesis} (\boldmath{$H_0$})} \\
    \cmidrule{2-4} & \textbf{NSGA-II} & \textbf{SPEA-II} & \textbf{MOEA/D} \\
    \midrule
    $\mathcal{S}_1$    & \xmark     & \cmark     & \xmark \\
    $\mathcal{S}_2$    & \cmark     & \cmark     & \xmark \\
    $\mathcal{S}_3$    & \cmark     & \xmark     & \xmark \\
    $\mathcal{S}_4$    & \cmark     & \cmark     & \xmark \\
    $\mathcal{S}_5$    & \xmark     & \xmark     & \xmark \\
    $\mathcal{S}_6$    & \xmark     & \cmark     & \xmark \\
    $\mathcal{S}_7$    & \xmark     & \xmark     & \xmark \\
    \bottomrule
    \end{tabular}%
\end{table}%

In particular, the goal here is to test the hypothesis that the uniform crossover can \textit{significantly} improve the search performance of MOEAs in comparison to the single point crossover. To this end, the test outlined in Algorithm~\ref{al:pcpm} is carried out with all the MOEAs and in conjunction with both Single Point and Uniform crossovers. Following these steps, the performance metric ($PM$) is obtained for all possible \textit{instances}, where each instance denotes a unique grouping of MOEA, recombination mechanism and parameter combination. For sake of clarity, the results with three parameter combinations (out of 100) for $\mathcal{S}_1$ are shown in Table~\ref{t:WilcoxPS1}. These results can be interpreted as follows: Consider the first row in Table~\ref{t:WilcoxPS1}. These results indicate that when the crossover and mutation probabilities are set to $\{ 0.1, 0.001\}$, NSGA-II can yield better results with uniform crossover; since the corresponding $PM$ value ($0.9764$, Table~\ref{t:WilcoxPS1}) is higher than that of the single point crossover (\textit{i.e.}, $0.9696$, Table~\ref{t:WilcoxPS1}). Following the same arguments, local conclusions can be inferred for other parameter combinations. It can be argued that the crossover mechanism, which yields better performance across higher number of parameter combination, is desirable. Hence, for each MOEA, the following hypotheses are evaluated using the Wilcoxon matched-pairs signed-ranks test~\cite{Sheskin:2003,Derrac:Salvador:2011}:
\medskip
\begin{itemize}
    \item \textit{Null Hypothesis} ($H_0$): The median value of performance metrics from single point and uniform crossover are \textit{equal}, \textit{i.e.}, $H_0 : \mu_D=0$
    \smallskip
    \item \text{Alternative Hypothesis} ($H_1$): The median value of performance metric from uniform crossover is \textit{higher} than that of the single point crossover, \textit{i.e.}, $H_1 : \mu_D>0$
\end{itemize}
\medskip
These hypotheses are determined for each test system using the performance metric ($PM$) values across all the parameter combinations. The outcome of the hypotheses evaluation at $95\%$ confidence interval are shown in Table~\ref{t:WilcoxPS2}. It is interesting to see that, with MOEA/D, the \textit{null} or \textit{equivalence} hypothesis is rejected for all the test systems, \textit{i.e.}, there is enough evidence to support the argument that MOEA/D would yield better search performance with the \textit{uniform} crossover. In contrast, the null hypothesis is rejected for few systems with the other MOEAs (4 with NSGA-II and 3 with SPEA-II). This indicates that with NSGA-II and SPEA-II, \textit{uniform} crossover will provide better or equivalent performance to \textit{single point} crossover. Hence, on the basis of these observations, the \textit{uniform} crossover is selected in this study.

\section{Conclusions}
\label{s:conclusion}

A Multi-Objective Structure Selection (MOSS) framework for the identification of nonlinear system is proposed. In particular, the structure selection is formulated as a Multi-Criteria Decision Making (MCDM) problem to balance the \textit{bias-variance} dilemma. The proposed framework is critically analyzed both from the perspective of \textit{system identification} and \textit{search algorithm} through comparative analysis on several benchmark nonlinear systems and a practical identification problem. A thorough parameter sensitivity analysis has also been carried out to determine the robustness of MOEAs. These results show that the region of high performance or \textit{sweet spots} in the parameter space is very large, especially for the dominance based MOEAs. This implies that the choice of the control parameters are not critical to the search performance of MOEAs, provided these lie in the identified \textit{sweet spots}. 

The key benefit of MOSS is the ability to identify multiple valid discrete-time models for continuous-time systems. This has been demonstrated by validating the identified models in frequency domain using generalized frequency response functions. 



The proposed MOSS framework can easily accommodate distinct MOEA architectures. In this investigation, this is demonstrated by \textit{dominance} based and  \textit{decomposition} based architectures. In addition, it is possible to incorporate distinct preferences to focus the search in the region of interest to the decision maker. The proposed approach therefore provides a flexible generic identification framework.


\linespread{1}
\section*{Acknowledgement}
Faizal Hafiz is grateful to Education New Zealand for supporting this research through New Zealand International Doctoral Research Scholarship (NZIDRS).

\appendix
\small

\bibliographystyle{elsarticle-num}

\end{document}